\documentclass[a4paper,11pt]{article}

\pdfoutput=1 

\usepackage{jcappub} 

\usepackage[T1]{fontenc} 
\usepackage{graphicx}
\usepackage{subcaption}
\usepackage{cleveref}
\usepackage[normalem]{ulem}
\usepackage{changes}

\def\ln{{\rm ln}}

\renewcommand{\vec}[1]{\boldsymbol{#1}}

\title{\boldmath Probing Parity Violation with Weak Lensing Trispectrum}

\author[a,b]{Sijin  Chen,}
\author[a,b,c]{Zhengyangguang Gong,}
\author[a,b,d,e]{Jiamin Hou}

\affiliation[a]{Universit\"ats-Sternwarte, Fakult\"at f\"ur Physik, Ludwig-Maximilians-Universit\"at M\"unchen, Scheinerstra{\ss}e~1, 81679~M\"unchen, Germany}
\affiliation[b]{Max Planck Institute for Extraterrestrial Physics, Giessenbachstra{\ss}e~1, 85748~Garching, Germany}
\affiliation[c]{Steward Observatory, University of Arizona, 933 North Cherry Avenue, Tucson, AZ 85721, USA}
\affiliation[d]{Institute of Astronomy, University of Cambridge, Madingley Rd, Cambridge CB3 0HA, UK}
\affiliation[e]{Kavli Institute for Cosmology Cambridge, Madingley Road, Cambridge CB3 0HA, UK}

\emailAdd{sijin@usm.lmu.de}
\emailAdd{zgong@arizona.edu}
\emailAdd{jiamin.hou@physik.uni-muenchen.de}

\abstract{We establish the weak gravitational lensing convergence trispectrum as an independent probe of cosmological parity violation in the late-time Large-Scale Structure (LSS). To map three-dimensional primordial symmetries into two-dimensional observables, we derive a generalized, compact projection formalism for the reduced angular trispectrum applicable
to two classes of factorizable primordial curvature trispectra considered in this work. Applying this framework, we compute the parity-odd signal and forecast the expected Signal-to-Noise Ratio (SNR) using two 
phenomenological parity-violating trispectrum templates. These two templates encode the initial conditions for the late-time weak lensing observables. One template peaks at the squeezed limit, and the other template peaks at the collapsed limit. Our analysis evaluates idealized Dirac-delta source redshift distributions alongside actual tomographic profiles from the Dark Energy Survey Year 3 (DES Y3) and forecasted profiles for the Legacy Survey of Space and Time Year 10 (LSST Y10). We map the geometric sensitivity of these templates, providing physical intuition into how the resulting SNR is strongly modulated by the specific source galaxy redshift distributions and the underlying 
geometry of the configurations. Furthermore, we rigorously test the validity of the Limber approximation for higher-order angular statistics, demonstrating the necessity of exact line-of-sight numerical integration to capture the parity-violating signal accurately. By demonstrating the theoretical detectability of parity-breaking signatures through weak lensing, this work serves as a proof of principle, showing how upcoming weak lensing surveys can be leveraged to 
probe the fundamental symmetries of the early universe. }

\begin{document}
\maketitle
\flushbottom

\section{Introduction}
\label{sec:intro}

A parity transformation flips the spatial coordinates, taking $\boldsymbol{x} \rightarrow -\boldsymbol{x}$, and the violation of this symmetry indicates that physical laws differentiate between left- and right-handed configurations. 
The discovery of parity violation in the weak interaction in the late 1950s was a landmark result in particle physics~\cite{Lee:1956qn,Wu:1957my,Garwin:1957hc}. It revealed that fundamental interactions can distinguish left- from right-handed states, a feature that became central to the  chiral structure of the weak force. 
The later discovery of charge-conjugation and parity (CP) violation further demonstrated that discrete symmetry breaking can reveal fundamental structures of a theory, such as complex phases. These examples motivate the broader question of whether parity symmetry may also  be violated on cosmological scales~\cite{Komatsu:2022nvu}. 
Probing parity violation across the Universe would therefore provide a potential signature of new physics in the primordial cosmos.

Several cosmological observables have been proposed as laboratories to investigate parity-violating signals~\cite{Lue:1998mq,Komatsu:2022nvu,Philcox:2023uor,Shim:2024tue,Zhu:2024wme,Hou:2024udn}. Historically, a key probe has been the polarization of the Cosmic Microwave Background (CMB) \cite{Komatsu:2022nvu,Lue:1998mq,Kamionkowski:2010rb,Shiraishi:2010kd,Minami:2020odp,Bartolo:2015dga,Shiraishi:2016mok}. The CMB polarization field is a spin-$2$ field on the celestial sphere and can be decomposed into parity eigenstates known as $E$ and $B$ modes~\cite{Zaldarriaga:1996xe,Kamionkowski:1996ks}. Because the $B$ mode possesses parity-odd properties, correlation functions involving an odd number of $B$ components naturally yield parity-odd statistics. Because the $B$ mode is parity odd, correlation functions involving an odd number of $B$ modes vanish if parity is conserved, making them sensitive probes of parity-violating physics.  There are already intriguing hints of parity violation in the CMB in the form of cosmic birefringence~\cite{Minami:2020odp,Diego-Palazuelos:2022dsq,Eskilt:2022cff,Diego-Palazuelos:2025dmh}, which is often motivated by theoretical scenarios involving axion-like particles and Chern-Simons couplings~\cite{Carroll:1989vb,Carroll:1991zs,Harari:1992ea,Carroll:1998zi,Lue:1998mq,Alexander:2004wk,Lyth:2005jf,Takahashi:2009wc,Alexander:2009tp,Satoh:2010ep,Soda:2011am,Shiraishi:2011st,Dyda:2012rj,Wang:2012fi,Zhu:2013fja}.

Higher-order statistics of primordial curvature perturbations provide a powerful window into inflationary physics and primordial symmetries~\cite{Bartolo:2004if}. Beyond constraining the amplitude and shape of primordial non-Gaussianity, their transformation properties under parity allow us to search for possible parity-violating physics during inflation. For statistically isotropic scalar perturbations, both the power spectrum and the bispectrum remain parity-even, implying that parity-odd information first appears at the four-point level. The trispectrum therefore represents the lowest-order scalar statistics capable of carrying parity-violating information~\cite{Cahn:2021ltp,Shiraishi:2016mok,Jeong:2012df}. Parity-violating physics during inflation can generate parity-odd primordial trispectra~\cite{Sorbo:2011rz,Niu:2022fki,Fujita:2023inz,Reinhard:2024evr,Lee:2023jby,Stefanyszyn:2023qov,Cabass:2022oap,Thavanesan:2025kyc,Cabass:2022rhr,Jazayeri:2023kji} as a leading-order signature. Since standard gravitational evolution is parity-conserving, any statistically significant parity-odd signal detected in the late-time matter distribution would provide compelling evidence for primordial parity-violating physics.

While the CMB provides a two-dimensional snapshot, the LSS provides access to parity violation in full three-dimensional space. The three-dimensional distribution of galaxies contains a vastly larger number of observable modes, making it a powerful statistical arena for detecting subtle primordial signatures. Geometrically, the relevant four-point configuration forms a tetrahedron, the lowest-order three-dimensional shape that cannot be rotated into its mirror image. Recent observational analyses have successfully placed constraints on cosmological parity violation using the parity-odd 3D galaxy 4PCF \cite{Philcox:2022hkh,Hou:2022wfj,Krolewski:2024paz,Philcox:2024mmz,Slepian:2025kbb,Cabass:2022oap,Bao:2025onc,Jeong:2012df,Hou:2025cey}. Besides the galaxy distribution, the observed shapes of distant galaxies offer another powerful probe of the LSS. Recent studies have been proposed leveraging the intrinsic alignments of these galaxy shapes as a spin-$2$ probe of parity violation \cite{Kurita:2025hmp}.

Complementary to galaxy clustering, weak lensing directly traces the projected matter distribution through coherent distortions of background galaxy shapes, bypassing uncertainties associated with galaxy bias. Like the CMB observables, weak lensing is an integrated line-of-sight observable, but it traces the late-time large-scale structure rather than the primordial surface of last scattering, and its source redshift dependence provides tomographic access to the evolution of the matter field.  As the corresponding projected scalar field, weak lensing convergence therefore provides a natural arena for searching for parity-violating signatures in higher-order statistics. Current Stage-III weak lensing surveys, including Hyper Suprime-Cam Survey (HSC) \cite{Aihara_2017}, the Dark Energy Survey (DES) \cite{DES:2015gax,DES:2016jjg,DES:2020aks,DES:2020ekd}  and the Kilo-Degree Survey (KiDS) \cite{Kuijken:2015vca,Wright:2024qvd}, have already delivered competitive cosmological constraints using two-point statistics. With the rapidly increasing power of ongoing and upcoming surveys, it is becoming feasible to extend weak lensing analyses beyond the power spectrum and explore parity-sensitive trispectrum measurements.

With these ideas in mind, this work focuses on the weak lensing trispectrum of the LSS. Previous works have shown that parity violation can be probed with the CMB statistics such as its temperature field \cite{Hu:2001fa,Okamoto:2002ik} and lensing trispectrum \cite{Greco:2025xtt}. We build on this foundation to demonstrate that weak lensing trispectrum in the late Universe naturally yields a parity-odd sensitive statistics, offering a direct probe of the projected matter distribution that complements both 2D CMB projections and 3D galaxy clustering.

This paper is organized as follows: In \Cref{section: weak lensing projection and angular power spectrum}, we define the weak lensing convergence and compute its angular power spectrum to establish our notation and ensure our formalism is sensible. In \Cref{sec: weak lensing convergence trispectrum}, we present the trispectrum calculation, with a particular emphasis on identifying parity violation across different primordial trispectrum templates. In \Cref{sec: numerical result}, we estimate the signal-to-noise ratio for the parity-odd weak lensing trispectrum for two specific templates and compare the sensitivity of various geometric configurations. We conclude in \Cref{sec: conclusion} with a summary of our main findings and suggestions for future work. Finally, the appendices provide additional technical details, including the harmonic space derivation of the weak lensing trispectrum, the Limber approximation, the numerical integration pipeline and its validation, as well as an investigation of the cumulative SNR saturation problem.


Throughout this work, we adopt the following Fourier transform convention:
\begin{equation*}
    \tilde{f}\left( \boldsymbol{k} \right) \equiv \int{\mathrm{d}\boldsymbol{x}\,\,f\left( \boldsymbol{x} \right)}e^{-i\boldsymbol{k}\cdot \boldsymbol{x}} \qquad f\left( \boldsymbol{x} \right) \equiv \int{\frac{\mathrm{d}\boldsymbol{k}}{\left( 2\pi \right) ^3}\tilde{f}\left( \boldsymbol{k} \right) e^{i\boldsymbol{k}\cdot \boldsymbol{x}}} .
\end{equation*}
where a tilde denotes the Fourier-space representation of a quantity.. We further use the shorthand notation $\int_{\boldsymbol{k}}\equiv \int{\frac{\mathrm{d}\boldsymbol{k}}{\left( 2\pi \right) ^3}}$ and $\boldsymbol{k}_{12\dots n} = \boldsymbol{k}_1 + \boldsymbol{k}_2 + \dots + \boldsymbol{k}_n$. The notations $\boldsymbol{k}_{1\dots n}$ and $\sum_{i=1}^{n}\boldsymbol{k}_i$ will be used interchangeably throughout this work.


\section{Theoretical Formalism}\label{section: weak lensing projection and angular power spectrum}

In this section, we briefly review the basic theory of weak gravitational lensing. We begin by summarizing the formalism of the weak lensing convergence, which will be used extensively throughout this paper. 
For comprehensive reviews of weak gravitational lensing, we refer the interested reader to Refs.~\cite{Bartelmann:1999yn, Mandelbaum:2017jpr}. Throughout this work, to evaluate our theoretical expressions and generate illustrative results, we assume a spatially flat $\Lambda$CDM cosmology consistent with the Planck 2018 constraints \cite{Planck:2018vyg}: $\Omega _m = 0.315$, $\Omega_b = 0.049$, $H_0 = 67.40 \ \text{km} \ \text{s}^{-1} \text{Mpc}^{-1}$, $A_s = 2.10 \times 10^{-9}$, and $\sigma_8 = 0.811$.

\subsection{Gravitational Lensing}




Weak gravitational lensing is described by the convergence $\kappa$ and shear $\gamma$, both derived from the projected lensing potential $\psi$. Since this work focuses on the convergence trispectrum, we briefly summarize the relevant projection formalism below. 

For a source located at a comoving distance $\chi$, this lensing potential can be expressed as a weighted line-of-sight integral of the three-dimensional Newtonian gravitational potential $\Phi$, between the observer and the source \cite{Narayan:1996ba}:
\begin{equation}\label{eq: 2D potential projection from 3D potential}
    \psi\left(\boldsymbol{\hat{\theta}}, \chi\right) = \frac{2}{c^2} \int _{0} ^{\chi} \frac{f_{K}\left(\chi - \chi^{\prime}\right)}{f_{K}\left(\chi\right) f_{K}\left(\chi^{\prime}\right)} \Phi\left(\chi^{\prime} \boldsymbol{\hat{\theta}}, \chi^{\prime}\right) \mathrm{d} \chi^{\prime},
\end{equation}
where $\chi'$ is the comoving distance of the lensing matter continuously distributed along the line of sight. The function $f_{K}(\chi)$ is the comoving angular diameter distance, with $K$ denoting the spatial curvature of the Universe. Throughout this work, we assume a spatially flat universe, for which $f_{K}(\chi) = \chi$. The quantity $\Phi$ is the three-dimensional gravitational potential, and $c$ is the speed of light. The gravitational potential $\Phi$ is related to the matter density field through the Poisson equation 
\begin{equation}\label{eq:3D poisson equation}
    \nabla^{2} \Phi = 4\pi G a^{2} \bar{\rho}\,\delta_m ,
\end{equation}
where $G$ is the gravitational constant, $a$ is the scale factor, $\rho$ is the matter density, $\bar{\rho}$ is the mean value, and $\delta_m \equiv \rho/\bar{\rho} - 1$ is the matter density contrast.

The weak lensing convergence is related to the lensing potential through its second-order derivatives $\kappa = \frac{1}{2}\nabla^{2}_{\boldsymbol{\theta}} \psi $ \cite{Schneider:2005ka}. Thus, combining Eqs.~\eqref{eq: 2D potential projection from 3D potential} and \eqref{eq:3D poisson equation}, the convergence can be expressed as a line-of-sight projection of the three-dimensional matter density contrast \cite{Bartelmann:1999yn},
\begin{equation}\label{eq: kappa single plane}
    \kappa(\boldsymbol{\hat{\theta}}, \chi) = \frac{3 H_{0}^{2} \Omega_{m}}{2 c^{2}} \int_{0}^{\chi} \frac{f_{K}(\chi - \chi')\, f_{K}(\chi')}{f_{K}(\chi)} \frac{\delta_m(\chi' \boldsymbol{\hat{\theta}}, \chi')}{a(\chi')} \,\mathrm{d}\chi' .
\end{equation}

The above derivation assumes a fixed source plane. However, in real weak lensing observations, the catalogue of source galaxies follows a normalized probability distribution $n\left(\chi\right)$ which spans a range of $\chi$ values, and Eq.~\eqref{eq: kappa single plane} is then weighted by this distribution to obtain the projected convergence as a function of direction \cite{Bartelmann:1999yn}:
\begin{align}\label{eq: kappa projection expression}
    \begin{split}
        \kappa \left( \boldsymbol{\hat{\theta} } \right) =\frac{3H_{0}^{2}\Omega _m}{2c^2}\int_0^{\chi_{H}}{}g\left( \chi \right) f_K\left( \chi \right) \frac{\delta_m \left( \chi \boldsymbol{\hat{\theta} },\chi  \right)}{a\left( \chi  \right)} \mathrm{d}\chi,
    \end{split}
\end{align}
where $\chi_{H}$ denotes the maximum comoving distance of the source distribution, and $g\left(\chi\right)$ is the lensing efficiency given by:
\begin{equation}\label{eq: lensing efficiency}
    g\left(\chi\right) = \int _{\chi} ^{\chi_{H}} n\left(\chi^{\prime}\right) \frac{f_{K}\left(\chi^{\prime} - \chi\right)}{f_{K}\left(\chi^{\prime}\right)} \mathrm{d}\chi^{\prime}
\end{equation}
Then we could express the lensing projection kernel using Eqs.~\eqref{eq: kappa projection expression} and \eqref{eq: lensing efficiency} as:
\begin{equation}\label{eq: lensing kernel}
    q\left(\chi\right) = \frac{3H_0^2 \Omega_m}{2c^2} \frac{f_{K}\left(\chi\right)}{a\left(\chi\right)}g\left(\chi\right).
\end{equation}
Thus, in the following derivations we will write the relation between weak lensing convergence and 3D matter density contrast as:
\begin{equation}\label{eq: convergence lensing kernel delta}
    \kappa \left( \boldsymbol{\hat{\theta}} \right) =\int_0^{\chi _{H}}{}q\left( \chi ^{\prime} \right) \delta _m \left( \chi ^{\prime}\boldsymbol{\hat{\theta} },\chi ^{\prime} \right) \mathrm{d}\chi ^{\prime}.
\end{equation}


To illustrate why a full-sky analysis is required, we first consider the corresponding two-dimensional flat-space case. In a two-dimensional flat space, a parity transformation (flipping the spatial coordinates) is equivalent to a simple spatial rotation of $180^{\circ}$. Because the background cosmology is statistically isotropic, projecting a genuine 3D parity-violating signature to a 2D flat space gives a vanishing signal. To extract the parity-odd signatures, we must analyze the convergence field in harmonic space, where the geometric symmetries and parity transformations are naturally governed by the angular momenta (as will be explicitly detailed in \Cref{sec: weak lensing convergence trispectrum}). We define the harmonic coefficients as $\kappa_{\ell m} = \int \mathrm{d} ^{2}\boldsymbol{\hat{\theta}} \kappa \left(\boldsymbol{\hat{\theta}}\right) Y^{*}_{\ell m} \left(\boldsymbol{\hat{\theta}}\right)$.  
Substituting the Fourier representation of the matter density contrast into the harmonic transform and integrating over the angular coordinates $\boldsymbol{\hat{\theta}}$ yields:
\begin{equation}\label{eq: weak lensing convergence projection for one point}
    \kappa _{\ell m}=4\pi i^{\ell}\int_0^{\chi _H}{\mathrm{d}\chi ^{\prime}q\left( \chi ^{\prime} \right)}\int_{\boldsymbol{k}}{\tilde{\delta}_m\left( \boldsymbol{k} \right) j_{\ell}\left( k\chi ^{\prime} \right) Y_{\ell m}^{*}\left( \hat{\boldsymbol{k}} \right)} , 
\end{equation}
where $\tilde{\delta}_{m} (\boldsymbol{k})$ is the matter density contrast in Fourier space, $j_{\ell}(k\chi^{\prime})$ is the spherical Bessel function of order $\ell$.

\subsubsection{Lensing Kernel}

\begin{figure}[htbp]
    \centering

    \begin{subfigure}[b]{0.495\textwidth}
        \centering
        \includegraphics[width=\textwidth]{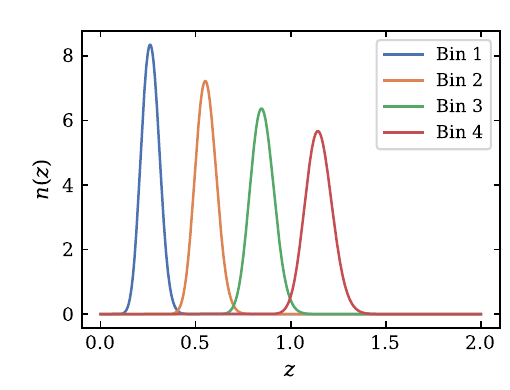}
        \caption{LSST-like Y10 redshift distribution}
        \label{fig: LSST redshift distribution}
    \end{subfigure}
    \hfill
    \begin{subfigure}[b]{0.495\textwidth}
        \centering
        \includegraphics[width=\textwidth]{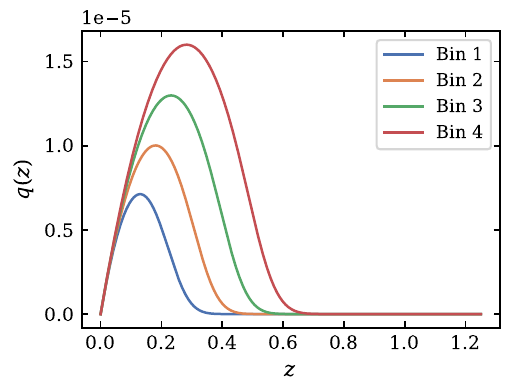}
        \caption{LSST-like Y10 lensing kernel}
        \label{fig: LSST lensing kernel}
    \end{subfigure}

    \caption{The normalized tomographic redshift distributions and corresponding lensing kernels for the LSST-like Y10 analysis. The panels show four representative tomographic bins selected from the full LSST-like $10$-bin setup, corresponding to the $1$st, $4$th, $7$th and $10$th bins, and relabeled as Bins $1$-$4$ for simplicity.  
    }
    \label{fig: LSST redshift distribution and lensing kernel}
\end{figure}

The functional form of the lensing kernel, $q\left(z\right)$, is determined by the line-of-sight distribution of source galaxies, $n\left(z\right)$. To illustrate the impact of different source distributions and to generate realistic theoretical predictions, we consider both idealized and realistic source distributions:
\begin{itemize}
    \item Idealized Single Source Plane (Dirac delta): First, we consider an idealized population of sources located at a single redshift, corresponding to $n\left(\chi \right) = \delta_{D}\left(\chi - \chi_s\right)$. In this limit, the projection integral simplifies to the following form
    \begin{equation}
        q\left( \chi \right) =\frac{3H_{0}^{2}\Omega _m}{2c^2}\frac{1}{a\left( \chi \right)}\frac{\chi \left( \chi _s-\chi \right)}{\chi _s}.
    \end{equation}
    As shown in \Cref{fig: lensing kernel under dirac delta distribution}, this results in a kernel that rises to a peak roughly halfway between the observer and the source, before reaching zero at $z=z_s$. We utilize this setup to validate the geometric properties of our formalism.

\begin{figure}[htbp]
    \centering
    \includegraphics[width=0.8\linewidth]{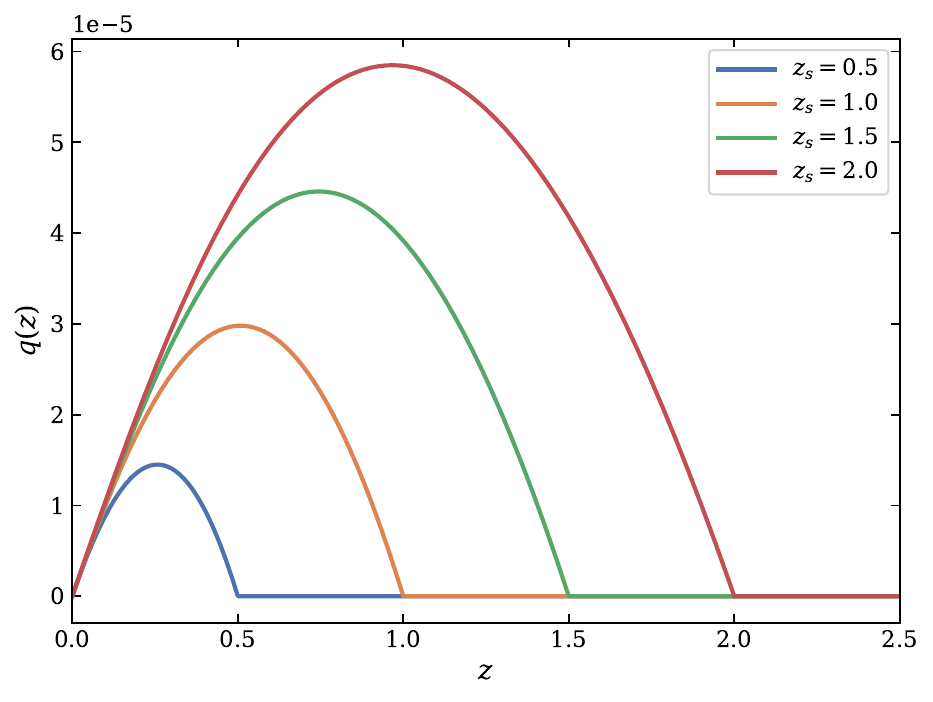}
    \caption{The weak lensing convergence kernel $q\left(z\right)$ as a function of redshift, computed for idealized source populations located at a single redshifit (Dirac delta distribution). The colored curves correspond to source planes fixed at $z_s=0.5, 1.0, 1.5$ and $2.0$. As described in Eq.~\eqref{eq: convergence lensing kernel delta}, the kernel quantifies the projection weight with which matter fluctuations at redshift $z$ lens the background sources; the sensitivity drops to zero for matter located behind the source $z > z_s$. }
    \label{fig: lensing kernel under dirac delta distribution}
\end{figure}
    

    \item Realistic Survey Distributions (DES Year 3 and LSST-like Year 10): To obtain more realistic predictions, we consider representative source redshift distributions from both DES Y3 and LSST-like Y10 surveys. Unlike the idealized case, real surveys contain galaxies distributed over a broad range of redshifts, which smooths the projection kernel. \\

    For DES Y3 \cite{DES:2020ebm}, we adopt the observed source redshift distributions and tomographic binning. \Cref{fig: DES redshift distribution} shows the normalized distributions for the four tomographic bins, while \Cref{fig: DES lensing kernel} presents the corresponding effective lensing kernels obtained by integrating Eq.~\eqref{eq: lensing efficiency} and \eqref{eq: lensing kernel} over these distributions. \\

\begin{figure}[htbp]
    \centering

    \begin{subfigure}[b]{0.495\textwidth}
        \centering
        \includegraphics[width=\textwidth]{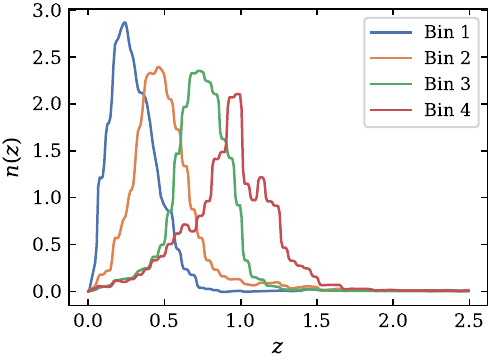}
        \caption{DES Y3 redshift distribution}
        \label{fig: DES redshift distribution}
    \end{subfigure}
    \hfill
    \begin{subfigure}[b]{0.495\textwidth}
        \centering
        \includegraphics[width=\textwidth]{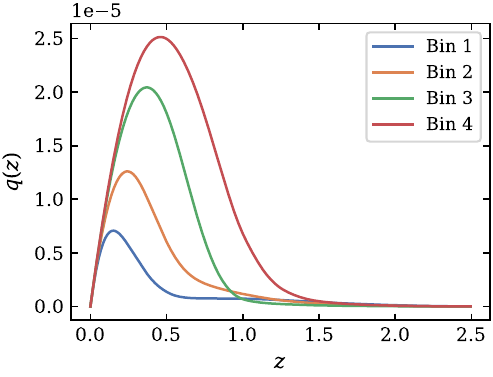}
        \caption{DES Y3 lensing kernel}
        \label{fig: DES lensing kernel}
    \end{subfigure}

    \caption{The normalized tomographic redshift distributions and resulting lensing kernels for the DES Y3 analysis. 
    }
    \label{fig: DES redshift distribution and lensing kernel}
\end{figure}

For the LSST-like case, we adopt an analytic model for the source redshift distribution \cite{LSSTDarkEnergyScience:2018jkl}, 
    \begin{equation}\label{eq: LSST total probability distribution}
        n\left( z \right) \propto z^2\exp \left[ -\left( \frac{z}{0.28} \right) ^{0.90} \right] ,
    \end{equation}
which provides a representative description of future deep photometric surveys. For the LSST-like Y10 construction, galaxies are binned according to their observed photometric redshifts ($z_{\text{ph}}$) rather than their true redshifts ($z$), leading to overlapping true redshift distributions between bins. The true redshift distribution of galaxies in the $i$th photometric bin, defined by $z^{(i)}_{\text{ph}} < z_{\text{ph}} < z^{(i+1)}_{\text{ph}}$, is given by  \cite{Ma:2005rc}:
\begin{equation}
    n_i\left( z \right) =\int_{z_{\text{ph}}^{\left( i \right)}}^{z_{\text{ph}}^{\left( i+1 \right)}}{\mathrm{d}z_{\text{ph}}\,\,n\left( z \right) p\left( z_{\text{ph}}|z \right)}
\end{equation}
where the photometric redshift uncertainty is modeled as a Gaussian distribution,
\begin{equation}
    p\left( z_{\text{ph}}|z \right) =\frac{1}{\sqrt{2\pi}\sigma _z}\exp \left[ -\frac{\left( z-z_{\text{ph}} \right) ^2}{2\sigma _{z}^{2}} \right] 
\end{equation}
with $\sigma _z = 0.03(1+z)$ \cite{Sinde:2026adb}. The full survey is typically divided into $10$ tomographic bins with a characteristic bin width of $\Delta z \simeq 0.1$ \cite{Sinde:2026adb}. For simplicity and clarity of presentation, we select four representative bins with boundaries $[0.2, 0.3]$, $[0.5, 0.6]$, $[0.8, 0.9]$ and $[1.1, 1.2]$. These bins are chosen to span a broad redshift range while keeping a uniform bin width. The resulting redshift distributions and corresponding lensing kernels are shown in \Cref{fig: LSST redshift distribution and lensing kernel}. 

    \end{itemize}

\subsection{Weak Lensing Convergence Power Spectrum}

With the formalism of the projection of one point prepared, we can now compute the angular power spectrum
\begin{align}\label{eq: power spectrum by multiplying two point}
    \begin{split}
        \left< \kappa _{\ell _1m_1}^{*}\kappa _{\ell _2m_2} \right> =\left( 4\pi \right) ^2i^{\ell _2-\ell _1}\int_{\boldsymbol{k}_1}{}\int_0^{\chi _H}{}\mathrm{d}\chi _{1}^{\prime}q\left( \chi _{1}^{\prime} \right) j_{\ell _1}\left( k_1\chi _{1}^{\prime} \right) Y_{\ell _1m_1}\left( \hat{\boldsymbol{k}}_1 \right) 
\\
\times \int_{\boldsymbol{k}_2}{}\int_0^{\chi _H}{}\mathrm{d}\chi _{2}^{\prime}q\left( \chi _{2}^{\prime} \right) j_{\ell _2}\left( k_2\chi _{2}^{\prime} \right) Y_{\ell _2m_2}^{*}\left( \hat{\boldsymbol{k}}_2 \right) \left< \,\,\tilde{\delta}_{m}^{*}\left( \boldsymbol{k}_1,\chi _{1}^{\prime} \right) \,\,\tilde{\delta}_m\left( \boldsymbol{k},\chi _{2}^{\prime} \right) \right> , 
    \end{split}
\end{align}
We can identify the matter power spectrum in the second line. Assuming linear evolution\footnote{For simplicity, we assume throughout this work linear evolution for the matter power spectrum and later on trispectrum calculation. Readers can refer to Ref.~\cite{Azyzy2025} for more complex nonlinear evolution of matter trispectrum in the context of primordial parity violation.}, the matter power spectrum could be written as~\cite{Eisenstein:1997ik, Eisenstein:1997jh, Borges:2007bh, Slepian:2015zra, Dodelson:2020bqr}, 
\begin{align}\label{eq: dirac delta power point correlation function}
    \begin{split}
        \left< \tilde{\delta}_{m}^{*}\left( \boldsymbol{k}_1,\chi _{1}^{\prime} \right) \tilde{\delta}_m\left( \boldsymbol{k}_2,\chi _{2}^{\prime} \right) \right> &=\left[ \frac{2k_{1}^{2}\mathcal{T} \left( k_1 \right)}{5\Omega _mH_{0}^{2}} \right] ^2D\left( \chi _{1}^{\prime} \right) D\left( \chi _{2}^{\prime} \right) 
\\
&\times \left( 2\pi \right) ^3\delta _{D}^{\left( 3 \right)}\left( \boldsymbol{k}_1-\boldsymbol{k}_2 \right) \mathcal{P} _{\mathcal{R}}\left( k_1 \right) .
    \end{split}
\end{align}
Here $\mathcal{T}\left(k\right)$ is the transfer function, $D\left(\chi\right)$ is the growth factor, $\mathcal{P}_{\mathcal{R}}\left(k\right)$ denotes the primordial power spectrum of the comoving curvature perturbations and $\delta ^{\left(3\right)} _{D}$ is the Dirac delta function. 

By substituting Eq.~\eqref{eq: dirac delta power point correlation function} into Eq.\eqref{eq: power spectrum by multiplying two point}, and integrating $\boldsymbol{k}$ over the Dirac delta function and over the spherical harmonics, we can get the expression for the weak lensing convergence power spectrum:
\begin{align}\label{eq: weak lensing convergence angular power spectrum}
    \begin{split}
        \left< \kappa _{\ell _1m_1}^{*}\kappa _{\ell _2m_2} \right> &=\frac{8\delta _{\ell _1\ell _2}^{K}\delta _{m_1m_2}^{K}}{25\pi \Omega _{m}^{2}H_{0}^{4}}\int{}\mathrm{d}k_1\,\,k_{1}^{6}\mathcal{T} ^2\left( k_1 \right) \mathcal{P} _{\mathcal{R}}\left( k_1 \right) \int_0^{\chi _H}{}\mathrm{d}\chi _{1}^{\prime}q\left( \chi _{1}^{\prime} \right) j_{\ell _1}\left( k_1\chi _{1}^{\prime} \right) 
\\
&\times \int_0^{\chi _H}{}\mathrm{d}\chi _{2}^{\prime}q\left( \chi _{2}^{\prime} \right) j_{\ell _1}\left( k_1\chi _{2}^{\prime} \right) D\left( \chi _{1}^{\prime} \right) D\left( \chi _{2}^{\prime} \right) .
    \end{split}
\end{align}
where $\delta ^{K}$ is the Kronecker delta function. Physically, the multipole $\ell$ sets the angular scale of fluctuations, while the magnetic quantum number $m$ labels different orientations (or azimuthal phases) at fixed $\ell$. Statistical isotropy implies that the two-point function is diagonal in harmonic space~\cite{Hu:2001fa}, 
\begin{equation}
    \left<\kappa ^{*} _{\ell_1 m_1} \kappa_{\ell_2 m_2}\right> = C_{\ell_1} \delta^{K}_{\ell_1 \ell_2} \delta^{K}_{m_1 m_2} .
\end{equation}
To evaluate Eq.~\eqref{eq: weak lensing convergence angular power spectrum}, and proceed to higher-order statistics, we must specify the primordial power spectrum. Following the assumption of adiabatic initial conditions in \cite{Greco:2025xtt}, the primordial curvature power spectrum is parameterized as a power-law:
\begin{equation}\label{eq: primordial power spectrum}
    P_{\mathcal{R}}\left(k\right) = \frac{2\pi^2}{k^3} A_{s}\left(k_p\right) \left(\frac{k}{k_p}\right)^{n_s - 1}, 
\end{equation}
where $A_s$ is the primordial scalar amplitude and $n_s$ is the scalar spectral tilt defined at the pivot scale $k_p$. This parameterization defines the initial conditions for the transfer function $\mathcal{T}\left(k\right)$ and allows us to compute the theoretical prediction for the convergence power spectrum. For our numerical calculations, we adopt $n_s \simeq 1$ for $k_p = 0.05 \, \text{Mpc}^{-1}$~\cite{Planck:2018jri}. 

\begin{figure}
    \centering
    \includegraphics[width=1.0\linewidth]{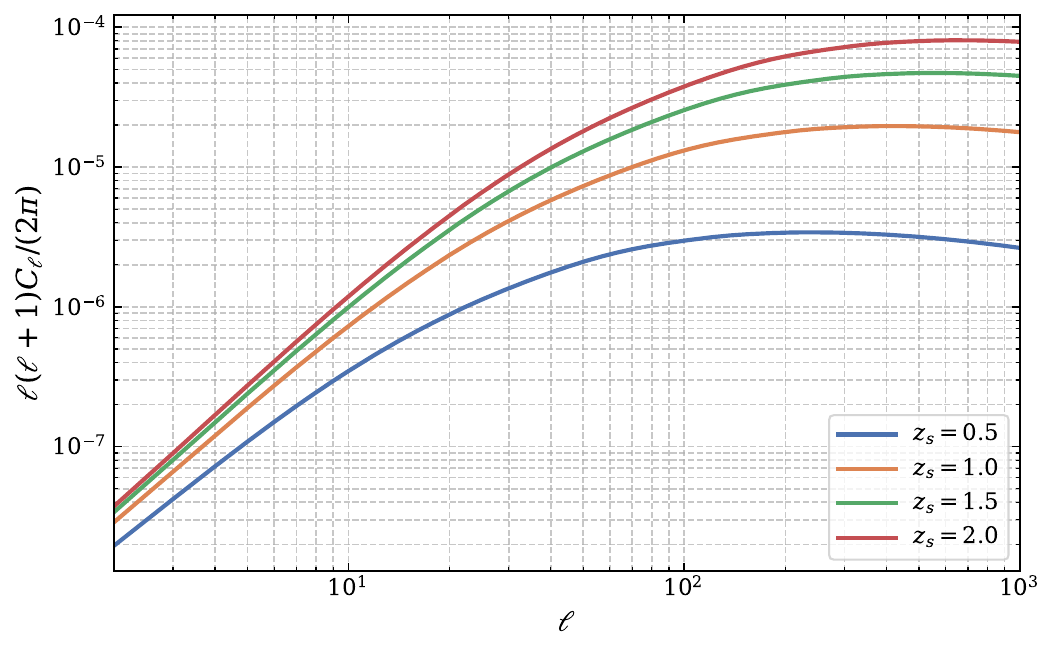}
    \caption{The weak lensing convergence angular power spectrum for idealized Dirac delta source distributions located at distinct redshifts $z_s$. The theoretical curves of Eq.~\eqref{eq: weak lensing convergence angular power spectrum} are computed under FFTLog and adopt the Eisenstein \& Hu transfer function~\cite{Eisenstein:1997ik} and linear growth factor, evaluated using the \texttt{Colossus}~\cite{Diemer:2017bwl} cosmology package. }
    \label{fig: angular power spectrum limber}
\end{figure}

We then compute the weak lensing convergence angular power spectrum for several idealized single-source-plane configurations. \Cref{fig: angular power spectrum limber} shows the resulting spectra for different source redshifts. As expected, higher-redshift source planes lead to larger convergence power spectrum amplitudes due to the increased lensing efficiency along the line of sight. 

\section{Weak Lensing and Trispectrum}
\label{sec: weak lensing convergence trispectrum}

While the angular power spectrum discussed in Section \ref{section: weak lensing projection and angular power spectrum} captures Gaussian information of the convergence field, the signature of parity violation arising from early-universe physics is inherently non-Gaussian. To probe such parity-violating signals, we must go beyond two-point statistics. Since the bispectrum of a scalar field vanishes for parity-odd configurations under the assumption of statistical isotropy, the trispectrum represents the lowest-order statistics \cite{Shiraishi:2016mok} capable of preserving parity-violating signatures in the weak lensing convergence field. 

This section outlines the calculation of the weak lensing convergence angular trispectrum. We first introduce the three-dimensional parity-odd trispectrum and isolate the parity-violating signature contained in its imaginary part. We then establish the geometric intuition underlying the line-of-sight projection and illustrate representative tomographic setups.  Building on the projection formalism, we present the general integral expression for the weak lensing angular trispectrum. To investigate the observational signatures of parity violation, we evaluate this expression explicitly for two characteristic templates, corresponding to the squeezed and collapsed limits.

\subsection{The Three-Dimensional Parity-Odd Trispectrum}

The trispectrum, defined as the connected part of the four-point function in Fourier space, is written as
\begin{equation}
    \left< \phi \left( \boldsymbol{k}_1 \right) \phi \left( \boldsymbol{k}_2 \right) \phi \left( \boldsymbol{k}_3 \right) \phi \left( \boldsymbol{k}_4 \right) \right> =\left( 2\pi \right) ^3\delta _{D}^{\left( 3 \right)}\left( \boldsymbol{k}_{1234} \right) T\left( \boldsymbol{k}_1,\boldsymbol{k}_2,\boldsymbol{k}_3,\boldsymbol{k}_4 \right)  .
\end{equation}
where the Dirac delta enforces closure of the four wavevectors into a tetrahedron.

Under parity, $\mathbf{k} \to -\mathbf{k}$, and for a real scalar field one has $\phi^*(\mathbf{k})=\phi(-\mathbf{k})$, implying
\begin{equation}\label{eq: trispectrum reality condition}
    \mathbb{P} \left[ T\left( \boldsymbol{k}_1,\boldsymbol{k}_2,\boldsymbol{k}_3,\boldsymbol{k}_4 \right) \right] =T\left( -\boldsymbol{k}_1,-\boldsymbol{k}_2,-\boldsymbol{k}_3,-\boldsymbol{k}_4 \right) =T^*\left( \boldsymbol{k}_1,\boldsymbol{k}_2,\boldsymbol{k}_3,\boldsymbol{k}_4 \right) .
\end{equation}
This allows a decomposition into parity-even and parity-odd components,
\begin{equation}
T\left( \boldsymbol{k}_1,\boldsymbol{k}_2,\boldsymbol{k}_3,\boldsymbol{k}_4 \right) =T_+\left( \boldsymbol{k}_1,\boldsymbol{k}_2,\boldsymbol{k}_3,\boldsymbol{k}_4 \right) +i\,T_-\left( \boldsymbol{k}_1,\boldsymbol{k}_2,\boldsymbol{k}_3,\boldsymbol{k}_4 \right) \,,
\end{equation}
where $T_+$ is invariant under parity and encodes parity-even correlations, while $T_-$ changes sign and captures parity-violating information.

The parity-odd signal is associated with the handedness of the tetrahedral configuration, which is quantified by the scalar triple product
\begin{equation}
    \sigma \equiv \text{sgn} \left[\boldsymbol{\hat{k}}_1 \cdot \left(\boldsymbol{\hat{k}}_2 \times \boldsymbol{\hat{k}}_3\right)\right].
\end{equation}
Rotational invariance then implies that the imaginary part of the trispectrum can be written as
\begin{equation}\label{eq: trispectrum into triple product and tau}
T_-(\boldsymbol{k}_1,\boldsymbol{k}_2,\boldsymbol{k}_3,\boldsymbol{k}_4)
= \big[\boldsymbol{\hat{k}}_1 \cdot (\boldsymbol{\hat{k}}_2 \times \boldsymbol{\hat{k}}_3)\big]\,
\tau_-(k_i, |\boldsymbol{k}_1+\boldsymbol{k}_2|, |\boldsymbol{k}_1+\boldsymbol{k}_4|)\,,
\end{equation}
where $\tau_{-}$ is a scalar function of the side lengths and diagonals that describe the specific shape of the trispectrum and index $i$ goes from 1 to 4.

\subsection{Geometric Projection Setup}

To illustrate the geometric structure underlying the projected weak lensing angular trispectrum, we consider two representative tomographic configurations: auto-correlations, where all source galaxies are located within the same tomographic bin, and cross-correlations, where source galaxies are distributed across different redshift bins. 

\begin{figure}[htbp]
    \centering
    \begin{subfigure}[b]{0.495\textwidth}
        \centering
        \includegraphics[width=\textwidth]{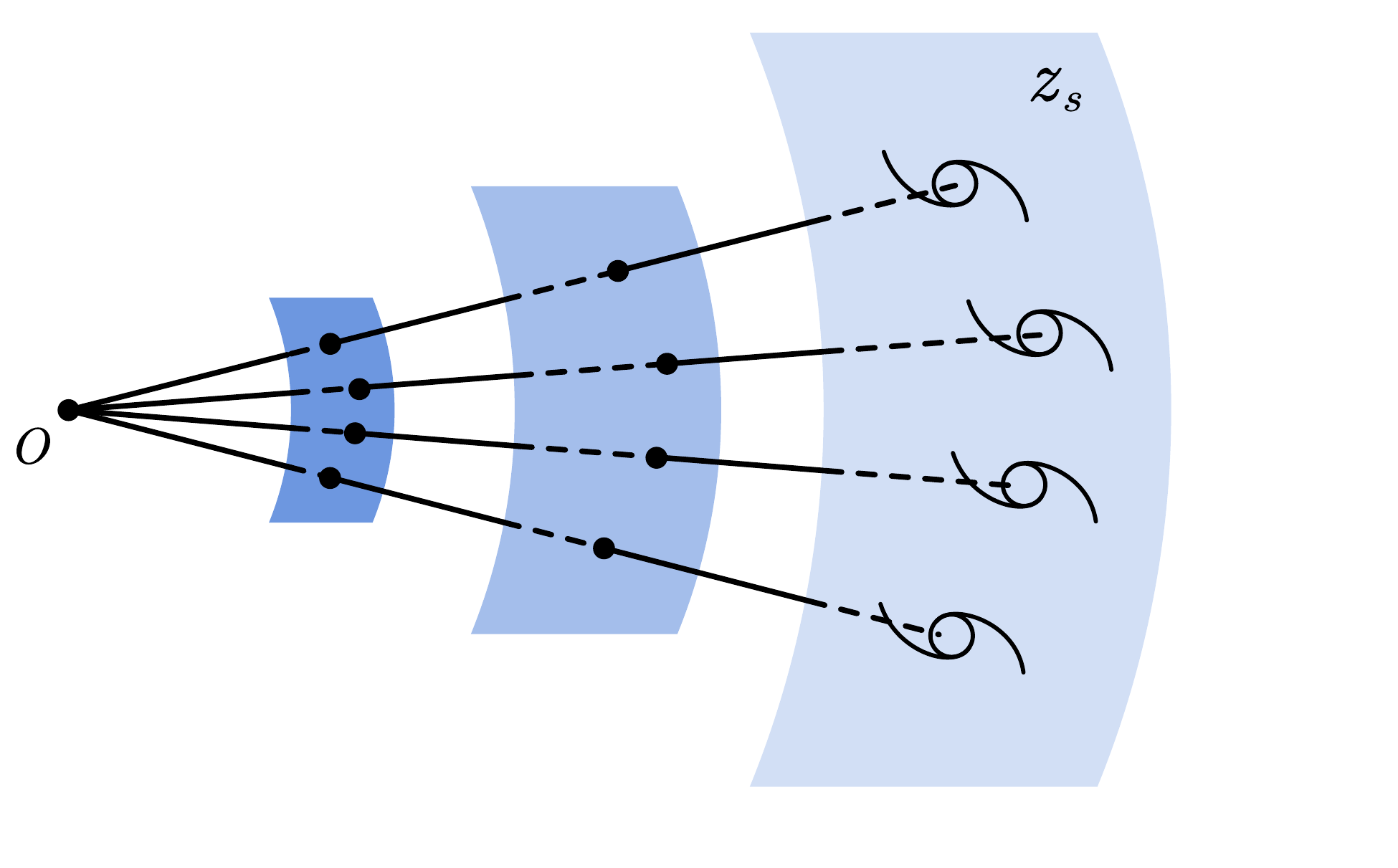}
        \caption{Same tomographic bin.}
        \label{fig: schematic diagram same tomo}
    \end{subfigure}
    \hfill
    \begin{subfigure}[b]{0.495\textwidth}
        \centering
        \includegraphics[width=\textwidth]{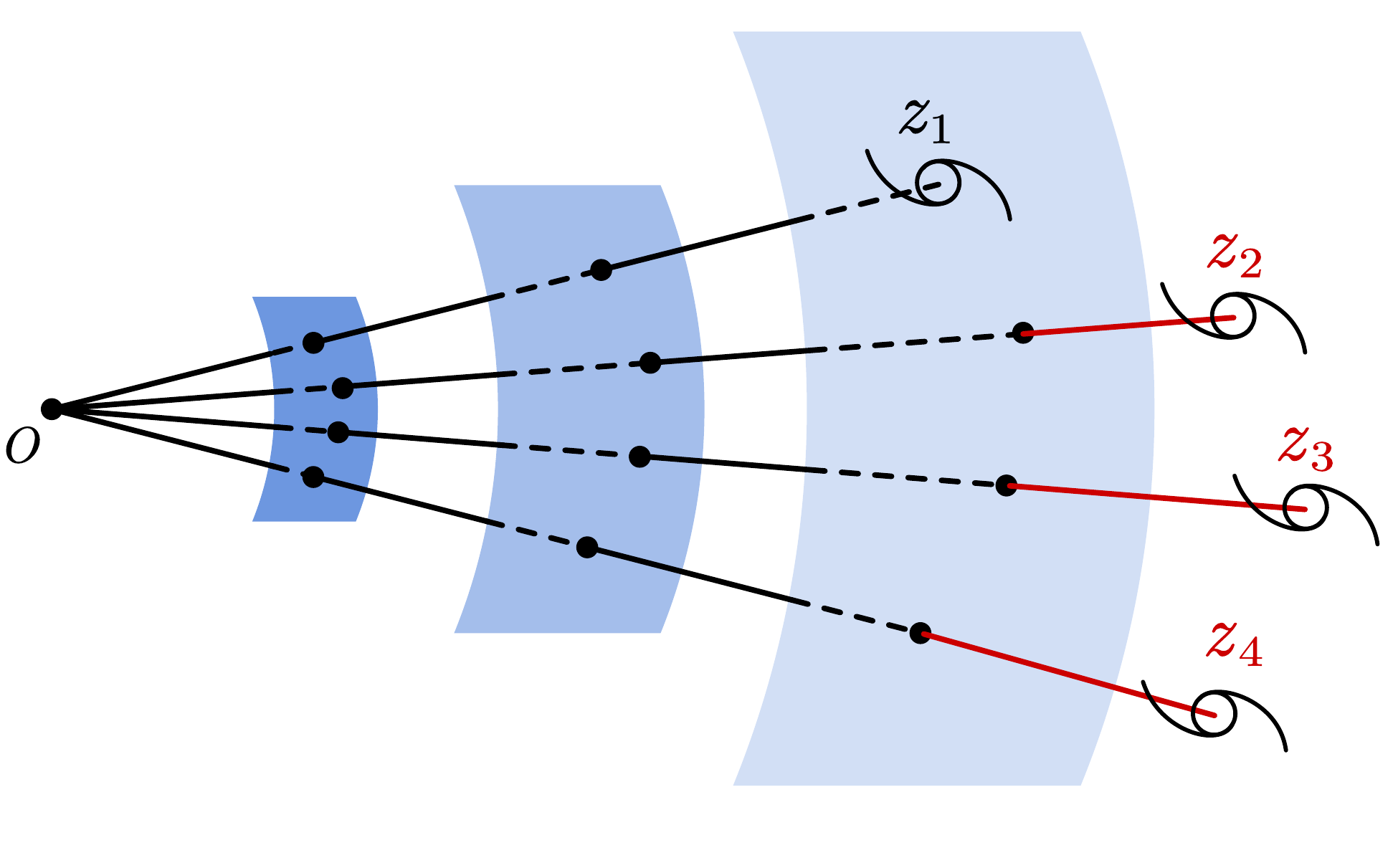}
        \caption{Different tomographic bins.}
        \label{fig: schematic diagram diff tomo}
    \end{subfigure}

    \caption{Schematic illustration of the geometric projection and line-of-sight integration for weak lensing four-point correlation. Panel (\subref{fig: schematic diagram same tomo}) shows the auto-correlation configuration, where all source galaxies reside within the same tomographic bin at redshift $z_s$. Panel (\subref{fig: schematic diagram diff tomo}) shows the cross-correlation configuration, where the source galaxies are distributed across different tomographic bins $(z_1 < z_2, z_3, z_4)$. The shaded blue regions schematically represent onion-like spherical shells at different comoving distances from the observer. The figure highlights that the projected signal is primarily accumulated from structures located within the overlapping region of all four lensing kernels. The red solid lines denote line-of-sight regions that do not overlap with all four lensing kernels.  
    }
    \label{fig: SNR schematic diagram}
\end{figure}

Although the projected trispectrum formally involves line-of-sight integrations over a broad range of comoving distances, the dominant contribution to the trispectrum arises from configurations in which the four projected matter fluctuations are located at nearly the same comoving distance from the observer. \Cref{fig: SNR schematic diagram} schematically illustrates these two projection geometries. In the auto-correlation case, all source galaxies share the same source redshift $z_s$, leading to a fully overlapping line-of-sight integration region. In the cross-correlation case, the source galaxies are distributed across multiple redshifts $(z_1<z_2, z_3, z_4)$, resulting in a reduced overlap of the corresponding lensing kernels.  

\subsection{Weak Lensing Convergence Trispectrum}
Analogous to the angular power spectrum, the weak lensing convergence angular trispectrum is defined as the connected four-point correlation function of the harmonic coefficients $\kappa _{\ell m}$. By substituting  the projection expression from Eq.~\eqref{eq: weak lensing convergence projection for one point} into the expectation value $\left<\kappa_{\ell_1 m_1} \kappa_{\ell_2 m_2} \kappa_{\ell_3 m_3} \kappa_{\ell_4 m_4}\right>$, we obtain the full expression for the angular trispectrum: 
\begin{align}\label{eq: general full expression of trispectrum}
\left\langle \prod_{i=1}^4 \kappa _{\ell_i m_i} \right\rangle &= \prod_{i=1}^4 \left[ 4\pi i^{\ell_i} \int_{\boldsymbol{k}_i} \int_0^{\chi_{H_i}} \mathrm{d}\chi^{\prime}_i \, q_i(\chi^{\prime}_i)  j_{\ell_i}(k_i \chi^{\prime}_i) Y^*_{\ell_i m_i}(\boldsymbol{\hat{k}}_i) \right]
\left<\prod_{i=1}^4 \tilde{\delta}_m(\boldsymbol{k}_i, \chi^{\prime}_i) \right>
\end{align}


The full angular trispectrum depends on the multipoles $\ell$, which characterize the angular scales, as well as the magnetic quantum numbers $m$, which encode the phase information. We define the reduced angular trispectrum, $Q^{\ell_1 \ell_2}_{\ell_3 \ell_4} (L)$, as the rotationally invariant amplitude of the four-point correlation. Using the orthogonality properties of the Wigner 3-$j$ symbols, we can extract this quantity from the full trispectrum by contracting it with the appropriate geometric weights~\cite{Hu:2001fa}:
\begin{align}\label{eq: reduced trispectrum and full trispectrum}
        Q_{\ell _3\ell _4}^{\ell _1\ell _2}(L) = (2L+1) \sum_{M} (-1)^M \sum_{m_i} \begin{pmatrix} \ell_1 & \ell_2 & L \\ m_1 & m_2 & M \end{pmatrix} \begin{pmatrix} \ell_3 & \ell_4 & L \\ m_3 & m_4 & -M \end{pmatrix}  \langle \kappa_{\ell_1 m_1} \kappa_{\ell_2 m_2} \kappa_{\ell_3 m_3} \kappa_{\ell_4 m_4} \rangle .
\end{align}
Here, $L$ represents the diagonal angular momentum of the quadrilateral configuration coupling the two triangles $(\ell_1, \ell_2, L)$ and $(\ell_3, \ell_4, L)$. The reduced trispectrum $Q^{\ell_1 \ell_2}_{\ell_3 \ell_4}(L)$ isolates the rotationally invariant component of the angular trispectrum and removes the explicit dependence on the coordinate-dependent $m$ indices.

The three-dimensional matter density trispectrum is related to the primordial curvature trispectrum $T_{\mathcal{R}}$ via transfer function and the growth factor:
\begin{align}\label{eq: matter trispectrum general expression}
    \begin{split}
        \left< \prod_{i=1}^4{\tilde{\delta}_m\left( \boldsymbol{k}_i,\chi _{i}^{\prime} \right)} \right> =\left( 2\pi \right) ^3\delta _{D}^{\left( 3 \right)}\left( \boldsymbol{k}_{1234} \right) \prod_{i=1}^4{\left[ \frac{2D\left( \chi _{i}^{\prime} \right) k_{i}^{2}}{5\Omega _mH_{0}^{2}}\mathcal{T} \left( k_i \right) \right] T_{\mathcal{R}}\left( \boldsymbol{k}_1,\boldsymbol{k}_2,\boldsymbol{k}_3,\boldsymbol{k}_4 \right)}
    \end{split}
\end{align}

In the following derivation and discussion, we focus exclusively on the signature of parity violation. Consequently, we retain only the parity-odd component of the primordial trispectrum and set the parity-even real part to zero. In harmonic space, this implies that only configurations satisfying
\begin{equation}
    \ell_1 + \ell_2 + \ell_3 + \ell_4 = \text{odd}
\end{equation}
contribute to the signal, while the configurations with even multipoles sums vanish identically. 

\begin{figure}[htbp]
    \centering
    \begin{subfigure}[b]{0.495\textwidth}
        \centering
        \includegraphics[width=0.6\textwidth]{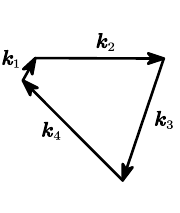}
        \caption{Squeezed configuration
        }
        \label{fig: geometric contact diagram}
    \end{subfigure}
    \hfill
    \begin{subfigure}[b]{0.495\textwidth}
        \centering
        \includegraphics[width=0.6\textwidth]{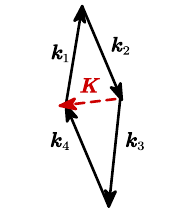}
        \caption{Collapsed configuration
        }
        \label{fig: geometric exchange diagram}
    \end{subfigure}

    \caption{ Momentum-space representation of the trispectrum configurations. (\subref{fig: geometric contact diagram}): The squeezed-type template depends only on the four side lengths and is most sensitive to configurations with one small side length. (\subref{fig: geometric exchange diagram}): The collapsed-type template additionally depends explicitly on the internal diagonal momentum $\boldsymbol{K}$ and is most sensitive to configurations with a small diagonal length. }
    \label{fig: geometric diagram}
\end{figure}

To connect our statistical templates with underlying inflationary physics, we consider two distinct models: the squeezed-type template and the collapsed-type template, as illustrated in \Cref{fig: geometric diagram}. The squeezed-type and collapsed-type templates are based on the momentum space configurations shown in \Cref{fig: geometric contact diagram,fig: geometric exchange diagram}, respectively, and are most sensitive to configurations with one small side length and one small diagonal length, respectively. 
The squeezed-type template provides the simplest realization, corresponding to a local four-point interaction and serving as a useful baseline for testing our projection formalism. Such a structure can arise, for example, in effective descriptions of inflationary dynamics with broken scale invariance~\cite{Cabass:2022rhr}. 
While this template is not tied to a specific microphysical model, its simple geometric structure makes it a convenient benchmark for isolating parity-violating signatures.

In contrast, collapsed-type trispectra naturally arise in scenarios where additional fields are present, such as in the cosmological collider framework~\cite{Liu:2019fag,Arkani-Hamed:2015bza,Noumi:2012vr}, where an intermediate particle mediates correlations between the external scalar modes. For our numerical predictions, we adopt a parity-odd template motivated by Axion-$U(1)$ models and proposed in Ref.~\cite{Shiraishi:2016mok}, in which gauge-field fluctuations induce characteristic trispectrum signatures. 
We note that the model in Ref.~\cite{Shiraishi:2016mok} involves a non-vanishing vacuum expectation value of the gauge field, leading to a mild breaking of statistical isotropy. In this work, we treat this setup as a convenient and well-defined phenomenological template to study parity-violating signals, in particular for configurations close to the collapsed limit. 


To streamline the notation and subsequent derivations, we explicitly derive the trispectrum expression for a single representative permutation of the external multipoles, labeled by $\pi$, where $\pi$ denotes one of the $24$ possible permutations. The remaining permutations follow analogously through relabeling of the external legs and are therefore not written out explicitly. 

For a given permutation, the reduced trispectrum can be factorized into an angular part and a radial part, schematically as:
\begin{equation}\label{eq: reduced trispectrum into angular and radial part}
\mathcal{Q}_{\pi} = \mathcal{W}_{\pi} \times \mathcal{I}_{\pi} ,
\end{equation}
where $\mathcal{W}_{\pi}$ encodes the angular dependence and $\mathcal{I}_{\pi}$ contains the line-of-sight integrals.

The full physical reduced trispectrum is then obtained by summing over all permutations of the external legs
\begin{equation}\label{eq: full reduced trispectrum and sum of single reduced trispectrum}
    Q = \sum_{\pi \in S_4} \mathcal{Q}_{\pi}. 
\end{equation}
where $S_4$ denotes the set of all $24$ permutations of the four external multipoles.

\subsubsection{Squeezed-Type Trispectrum}

To investigate the observational signatures of parity violation, we adopt the phenomenological model for primordial non-Gaussianity introduced by~\cite{Coulton:2023oug}. In this framework, the curvature perturbation is defined in real space as a Gaussian field modified by a quadratic parity-breaking correction:
\begin{equation}\label{eq: coulton template field}
    \xi \left( \boldsymbol{x} \right) =\xi _G\left( \boldsymbol{x} \right) +g_-\nabla \xi _{G}^{\left[ \alpha \right]}\left( \boldsymbol{x} \right) \cdot \left[ \nabla \xi _{G}^{\left[ \beta \right]}\left( \boldsymbol{x} \right) \times \nabla \xi _{G}^{\left[ \gamma \right]}\left( \boldsymbol{x} \right) \right] .
\end{equation}
Here $g_-$ parametrizes the strength of the parity violation. Following Refs.~\cite{Greco:2025xtt,Hou:2024udn}, we adopt the fiducial value $\left|g_-\right|=2 \times 10^7$, corresponding to the largest coupling for which the perturbative treatment remains valid and the non-Gaussian contribution remains subdominant to the Gaussian contribution. Since the trispectrum is linear in $g_-$, the fiducial choice serves only as the reference amplitude for the forecast. The dependence of our forecasts on the adopted fiducial value of $g_-$ is discussed in \Cref{app:parameter_constraints}. 
The filtered fields are defined in Fourier space as $\xi^{\left[\alpha\right]}_{G} \left(\boldsymbol{k}\right)= k^{\alpha}\xi _{G}\left(\boldsymbol{k}\right)$. By following the approximation of $n_s \simeq 1$ in~\cite{Jamieson:2024mau, Greco:2025xtt}, this interaction generates a leading-order imaginary trispectrum of the form:
\begin{equation}\label{eq: coulton template in fourier space}
    T_-\left( \boldsymbol{k}_1,\boldsymbol{k}_2,\boldsymbol{k}_3,\boldsymbol{k}_4 \right) =g_-\left( 2\pi ^2A_s \right) ^3\left[ \frac{k_{1}^{\alpha}k_{2}^{\beta}k_{3}^{\gamma}}{k_{1}^{3}k_{2}^{3}k_{3}^{3}}\boldsymbol{k}_1\cdot \left( \boldsymbol{k}_2\times \boldsymbol{k}_3 \right) +23\,\,\mathrm{perm} \right]. 
\end{equation}
The term in brackets involves a sum over the 24 permutations of the wavenumbers ${k_1, k_2, k_3, k_4}$, where the wavenumber arguments are permuted while the power indices remain fixed (see Ref.~\cite{Jamieson:2024mau} for a detailed description). Scale invariance requires only that the exponents satisfy $\alpha + \beta + \gamma = -3$. For concreteness, we adopt the representative choice $\alpha=-2, \beta = -1$ and  $\gamma = 0$, which is used in Ref.~\cite{Coulton:2023oug} and yields the characteristic $k^{-9}$ scaling of the trispectrum.  

The derivation involves the harmonic expansion of the parity-odd triple product, the projection of the primordial trispectrum onto the weak lensing convergence field, and the application of the reduced trispectrum formalism. The full parity-odd shape function introduced in Eq.~\eqref{eq: trispectrum into triple product and tau} is obtained by summing over all $24$ permutations of the representative contribution. For brevity, we defer the intermediate steps to \Cref{app:derivation contact diagram} and present only one representative permutation contribution~\footnote{The factor of $\left(\frac{5}{3}\right)^{4}$ arises because we are using primordial trispectrum of $\tilde{\mathcal{R}}$~\cite{Dodelson:2020bqr}.}:
\begin{align}
   \tau _{\pi}^{(s)}\left( k_1,k_2,k_3,k_4 \right) = \left( \frac{5}{3} \right) ^4\frac{\sqrt{2}}{3}\left( 4\pi \right) ^{3/2}g_-\left[ 2\pi ^2A_s \right] ^3 \left(\frac{k_{1}^{\alpha}k_{2}^{\beta}k_{3}^{\gamma}k_{4}^{0}}{k_{1}^{2}k_{2}^{2}k_{3}^{2}k_{4}^{0}} \right)_{\pi}.
\end{align}
where the superscript $(s)$ denotes the squeezed-limit template, and the subscript $\pi$ labels a single permutation of the external momenta among the full set of 24 permutations.

Combining the projection formalism with the reduced trispectrum definition of Eq.~\eqref{eq: reduced trispectrum and full trispectrum} and performing angular integration, we obtain the projected reduced trispectrum for a representative permutation:
\begin{align}
    \begin{split}
        {\mathcal{Q} _{\ell _3\ell _4}^{\ell _1\ell _2}}_{\pi}^{(s)}\left( L \right) &=\left( 2L+1 \right) \, i^{\ell _1+\ell _2+\ell _3}\, \left[ \frac{4}{5\pi \Omega _{m,0}H_{0}^{2}} \right] ^4
\\
&\times \sum_{L_1L_2L_3}{}\sum_{L^{\prime}}{}i^{L_1+L_2+L_3}\mathcal{F} _{L_1L_2L^{\prime}}\mathcal{F} _{L_3\ell _4L^{\prime}}\mathcal{F} _{L_11\ell _1}\mathcal{F} _{L_21\ell _2}\mathcal{F} _{L_31\ell _3}
\\
&\times \left\{ \begin{matrix}
	L&		L^{\prime}&		1\\
	\ell _2&		L_2&		1\\
	\ell _1&		L_1&		1\\
\end{matrix} \right\} \left\{ \begin{matrix}
	\ell _3&		\ell _4&		L\\
	L^{\prime}&		1&		L_3\\
\end{matrix} \right\} 
\\
&\times \int{\mathrm{d}x\,\,x^2\prod_{i=1}^4{\left[ \int{\mathrm{d}k_i\int_0^{\chi _H}{\mathrm{d}\chi _{i}^{\prime}\,\,q\left( \chi _{i}^{\prime} \right) D\left( \chi _{i}^{\prime} \right) k_{i}^{4}\mathcal{T} _{\delta}\left( k_i \right) j_{\ell _i}\left( k_i\chi _{i}^{\prime} \right) j_{L_i}(k_ix)}} \right]}}
\\
&\times \tau _{\pi}^{(s)}\left( k_1,k_2,k_3,k_4 \right) 
    \end{split}
\end{align}
where the superscript $(s)$ and subscript $\pi$ again denote the squeezed-limit model and a single permutation among all the $24$ permutations. For a different representative permutation $\pi$, the same expression is obtained after consistently permuting all external-leg labels throughout the projection formula. In the product, we set $L_4=\ell_4$ so that the fourth leg is written in the same form as the first three legs. The angular integrations introduce geometric coupling coefficients involving Wigner 3-$j$, 6-$j$ and 9-$j$ symbols. 
Here, the coefficient $\mathcal{F}$ is constructed from a Wigner 3-$j$ symbol:
\begin{align}
    \mathcal{F}_{\ell_1 \ell_2 \ell_3} \equiv \sqrt{\frac{\left(2\ell_1 + 1\right) \left(2 \ell_2 + 1\right) \left( 2 \ell_3 + 1\right)}{4\pi}} \left(\begin{matrix}
        \ell_1 & \ell_2 & \ell_3 \\
        0 & 0 & 0
    \end{matrix}\right) .
\end{align}
Following the schematic decomposition introduced in Eq.~\eqref{eq: reduced trispectrum into angular and radial part}, we now write the single-permutation reduced trispectrum explicitly in terms of a geometric coupling term, $\mathcal{W}^{(s)}_{\pi}$, and a radial projection integral, $\mathcal{I}^{(s)}_{\pi}$, for the squeezed-limit template. The geometric coupling term is given by:
\begin{align}\label{eq: contact geometric part}
    \begin{split}
        {\mathcal{W}_{\ell _3\ell _4}^{\ell _1\ell _2}}^{(s)}_{\pi}\left( L \right) &=\left( 2L+1 \right) \times i^{\ell _1+\ell _2+\ell _3}\times \sum_{L_1L_2L_3}{}\sum_{L^{\prime}}{} i^{L_1+L_2+L_3} 
\\
&\times \mathcal{F} _{L_1L_2L^{\prime}}\mathcal{F} _{L_3\ell _4L^{\prime}} \mathcal{F} _{L_11\ell _1}\mathcal{F} _{L_21\ell _2}\mathcal{F} _{L_31\ell _3}  \left\{ \begin{matrix}
	L&		L^{\prime}&		1\\
	\ell _2&		L_2&		1\\
	\ell _1&		L_1&		1\\
\end{matrix} \right\} \left\{ \begin{matrix}
	\ell _3&		\ell _4&		L\\
	L^{\prime}&		1&		L_3\\
\end{matrix} \right\} ,
    \end{split}
\end{align}
Then the line-of-sight integration takes the form:
\begin{align}\label{eq: coulton template full equation}
        {\mathcal I}_{\ell_1,\ell_2,\ell_3,\ell_4,\pi}^{(s)}=&\left[\frac{4}{5\pi\Omega_{m,0}H_0^2}\right]^4\int \mathrm{d} x\,x^2 \prod_{i=1}^{4}\Bigg[\int \mathrm{d}k_i\int_0^{\chi_H} \mathrm{d}\chi_i' \notag
        \\
& \times \Bigg. q(\chi_i')D(\chi_i') k_i^4 \mathcal T_\delta(k_i) j_{\ell_i}(k_i\chi_i') j_{L_i}(k_i x) \Bigg] \, \tau_\pi^{(s)}(k_1,k_2,k_3,k_4).
\end{align}
The numerical evaluation of this high-dimensional projection integral is described in \Cref{app:limber approximation}. 

For the squeezed-type template, the radial projection integral $\mathcal{I}^{(s)}$ is independent of the diagonal multipole $L$, reflecting the absence of explicit diagonal-momentum dependence in the primordial template. The remaining $L$-dependence of the reduced trispectrum is entirely encoded in the geometric coupling term $\mathcal{W}^{(s)}$, which arises from the harmonic coupling used to construct the reduced trispectrum in Eq.~\eqref{eq: reduced trispectrum and full trispectrum}.

\begin{figure}[htbp]
    \centering
    \includegraphics[width=0.9\linewidth]{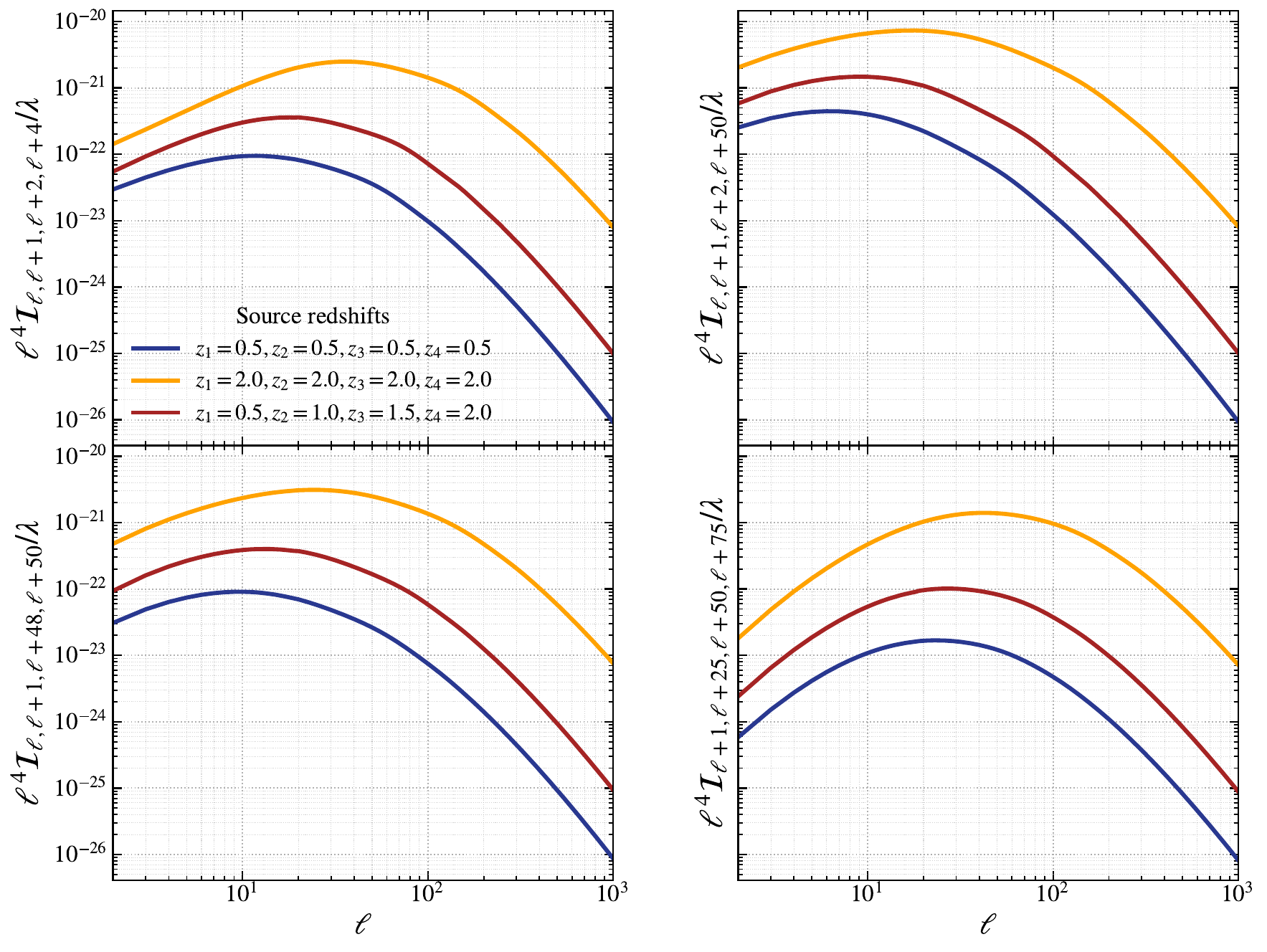}
    \caption{
    Magnitude of the line-of-sight projection term $\mathcal{I}$ from Eq.~\eqref{eq: coulton template full equation}. The four panels correspond to different external multipole configurations indicated on the vertical axes. Colors denote the tomographic source combinations shown in the legend: shallow auto-correlations (blue), deep auto-correlations (orange), and mixed-redshift cross-correlations (red). All curves are evaluated using the exact numerical integration. A factor of $\ell^4$ is included on the vertical axis to improve visual clarity. The template-dependent prefactor, $\lambda = (\frac{5}{3})^4\frac{\sqrt{2}}{3}(4\pi)^{3/2}g_-$,has been divided out to facilitate a direct comparison with the collapsed-type template. Since the figure shows the projection term alone, only the parity odd condition is imposed, and no quadrilateral closure condition is enforced. 
    }
    \label{fig: coulton template projection full integration}
\end{figure}

\begin{figure}[htbp]
    \centering
    \includegraphics[width=1.0\linewidth]{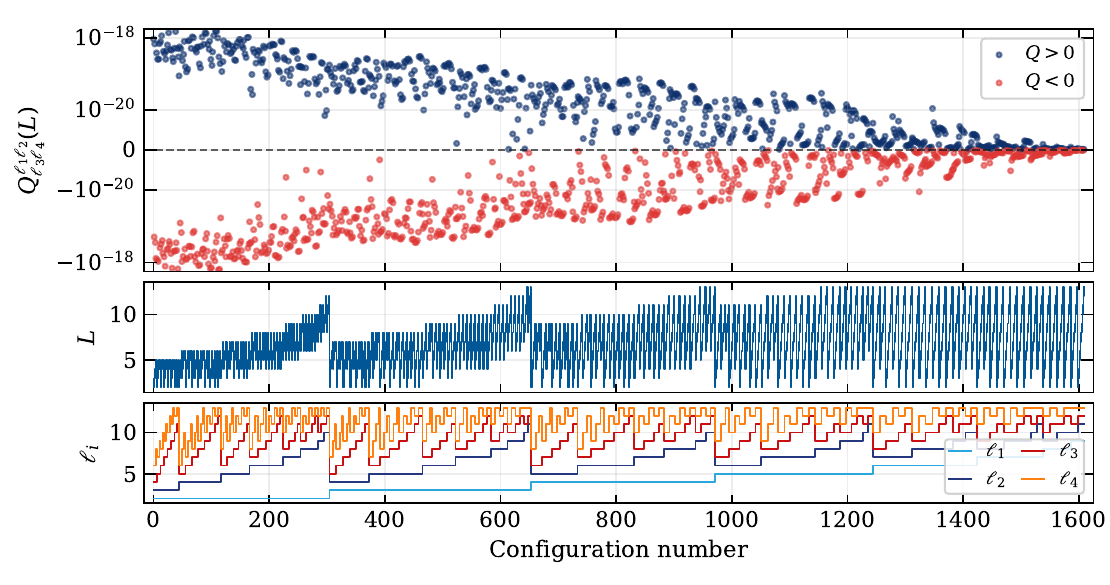}
    \caption{Reduced trispectrum for the squeezed-limit template, obtained by summing over all permutations as defined in Eq.~\eqref{eq: full reduced trispectrum and sum of single reduced trispectrum}. A cutoff $\ell_{\max}=13$ is applied, and we impose $\ell_1 < \ell_2 < \ell_3 < \ell_4$ and also vary the diagonal $L$ for fixed external multipoles. Top panel: the reduced trispectrum $Q$, with positive (blue) and negative (red) configurations highlighted. The alternating signs reflect the oscillatory angular dependence of the parity-odd trispectrum. Middle panel: the corresponding diagonal $L$. Bottom panel: the external multipoles $\left(\ell_1, \ell_2, \ell_3, \ell_4 \right)$ for each configuration. }
    \label{fig: reduced trispectrum for contact diagram}
\end{figure}

\Cref{fig: coulton template projection full integration} presents the radial projection term defined in Eq.~\eqref{eq: coulton template full equation}. The amplitude of the spectra is primarily governed by the lensing kernels: as the source galaxies are located in higher redshift bins, the kernels assign greater weight to the line-of-sight integration, resulting a stronger signal. To further illustrate how the signal is assembled across configurations, we also examine the reduced trispectrum obtained by summing over all permutations, as defined in Eq.~\eqref{eq: full reduced trispectrum and sum of single reduced trispectrum}. \Cref{fig: reduced trispectrum for contact diagram} shows this quantity as a function of all the multipoles. We observe that the parity-odd signal alternates between positive and negative values across different configurations, reflecting its geometric dependence from the angular part of the trispectrum. Interestingly, the signal remains significant when $\ell_1$ is small, even if the other three multipoles become large. This behavior already hints at the enhanced sensitivity of the parity-odd trispectrum to squeezed configurations. By contrast, when all multipoles increase simultaneously, the overall amplitude gradually decreases. This behavior suggests that, in the large multipole limit where all side lengths become large together, the geometric configuration becomes compressed toward a coplanar structure, such that the genuinely three-dimensional geometric information becomes less pronounced. As a result, the parity-odd reduced trispectrum becomes suppressed. The detailed configuration dependence will be discussed in \Cref{sec: numerical result geometric configuration sensitivity}. 

\subsubsection{Collapsed-Type Trispectrum}

While the template in the previous section was parameterized to probe the squeezed limit (where one external mode is soft, e.g., $k_1 \ll k_2, k_3, k_4$~\cite{Bao:2025onc}), it does not capture the physical dependence on the diagonal momentum. To explore regimes where this intermediate scale plays a central role, we employ 
a distinct trispectrum template characterized by an explicit dependence on the diagonal mode, particularly in the collapsed limit where $\left|\boldsymbol{K}\right| \equiv \left|\boldsymbol{k}_{1}+\boldsymbol{k}_2\right| \rightarrow 0$. Unlike the previous model in Eq.~\eqref{eq: coulton template field}, this template is constructed such that the primary parity information is encoded directly in the collapsed configurations. 

To model the parity-violating signal in this regime, we adopt the linear parameterization proposed by Ref.~\cite{Shiraishi:2016mok}. This ansatz expands the trispectrum using Legendre polynomials, $\mathcal{L}_n( \hat{\boldsymbol{k}}\cdot \hat{\boldsymbol{k}}^{\prime} ) $. We define the primordial curvature trispectrum in terms of the interaction kernel $t^{\boldsymbol{k}_1 \boldsymbol{k}_2}_{\boldsymbol{k}_3 \boldsymbol{k}_4}(\boldsymbol{K})$ as:
\begin{align}\label{eq: shiraishi full split}
    \begin{split}
        \left\langle\prod_{i=1}^4 \zeta \left( \boldsymbol{k}_i \right)  \right\rangle &=\left( 2\pi \right) ^6\int_{\vec K}{\,\delta _{D}^{\left( 3 \right)}\left( \boldsymbol{k}_{12}+\boldsymbol{K} \right)} \delta _{D}^{\left( 3 \right)}\left( \boldsymbol{k}_{34}-\boldsymbol{K} \right) t_{\boldsymbol{k}_3\boldsymbol{k}_4}^{\boldsymbol{k}_1\boldsymbol{k}_2}\left( \boldsymbol{K} \right)+23 \ \mathrm{perms}.
    \end{split}
\end{align}
Eq.~\eqref{eq: shiraishi full split} involves a diagonal wavevector $\boldsymbol{K}$ associated with the pairing of the external momenta. The trispectrum can be organized into contributions from different pairing structures, classified by the choice of diagonal.

While the full trispectrum involves all $24$ permutations of the four wavevectors, these permutations organize into three pairing channels, characterized by the diagonals
\begin{equation}
    \boldsymbol{K}_{ij} = -\left(\boldsymbol{k}_i + \boldsymbol{k}_j \right).
\end{equation}
These correspond to the $(12)$-$(34)$, $(13)$-$(24)$, and $(14)$-$(23)$ channels, respectively. 

\begin{figure}[htbp]
    \centering
    \begin{subfigure}[b]{0.32\textwidth}
        \centering
        \includegraphics[width=1.0\textwidth]{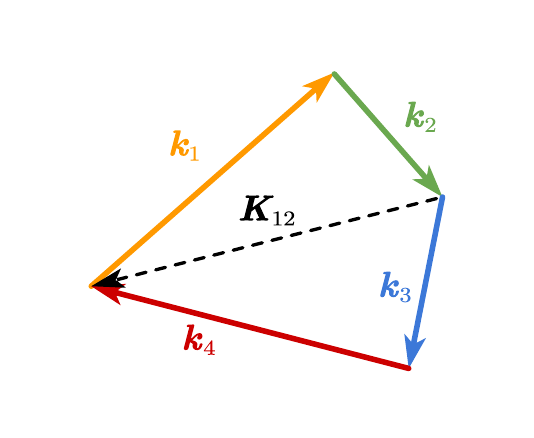}
        \caption{$\boldsymbol{K}_{12} = -\left(\boldsymbol{k}_1 + \boldsymbol{k}_2 \right)$}
        \label{fig: exchange K12 channel}
    \end{subfigure}
    \hfill
    \begin{subfigure}[b]{0.32\textwidth}
        \centering
        \includegraphics[width=1.0\textwidth]{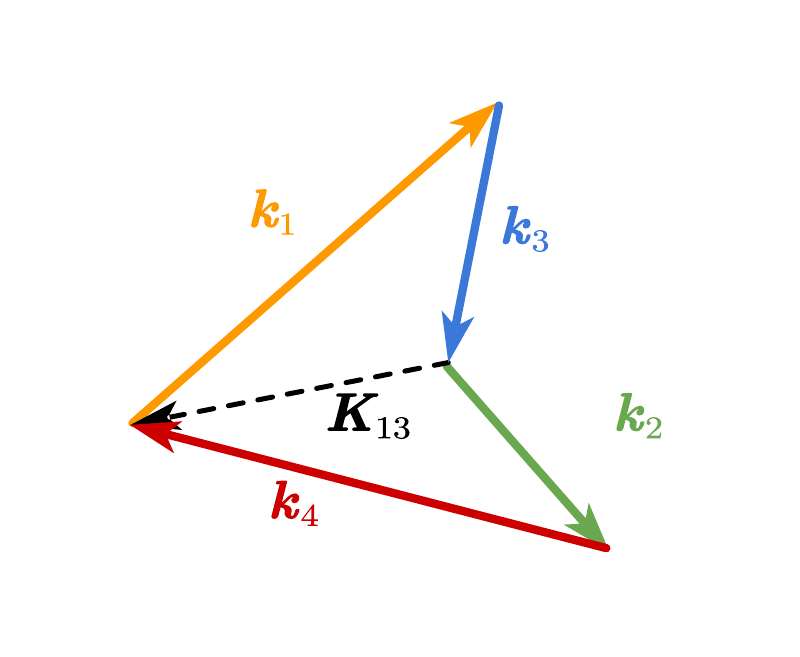}
        \caption{$\boldsymbol{K}_{13} = -\left( \boldsymbol{k}_1 + \boldsymbol{k}_3 \right)$}
        \label{fig: exchange K13 channel}
    \end{subfigure}
    \hfill
    \begin{subfigure}[b]{0.32\textwidth}
        \centering
        \includegraphics[width=1.0\textwidth]{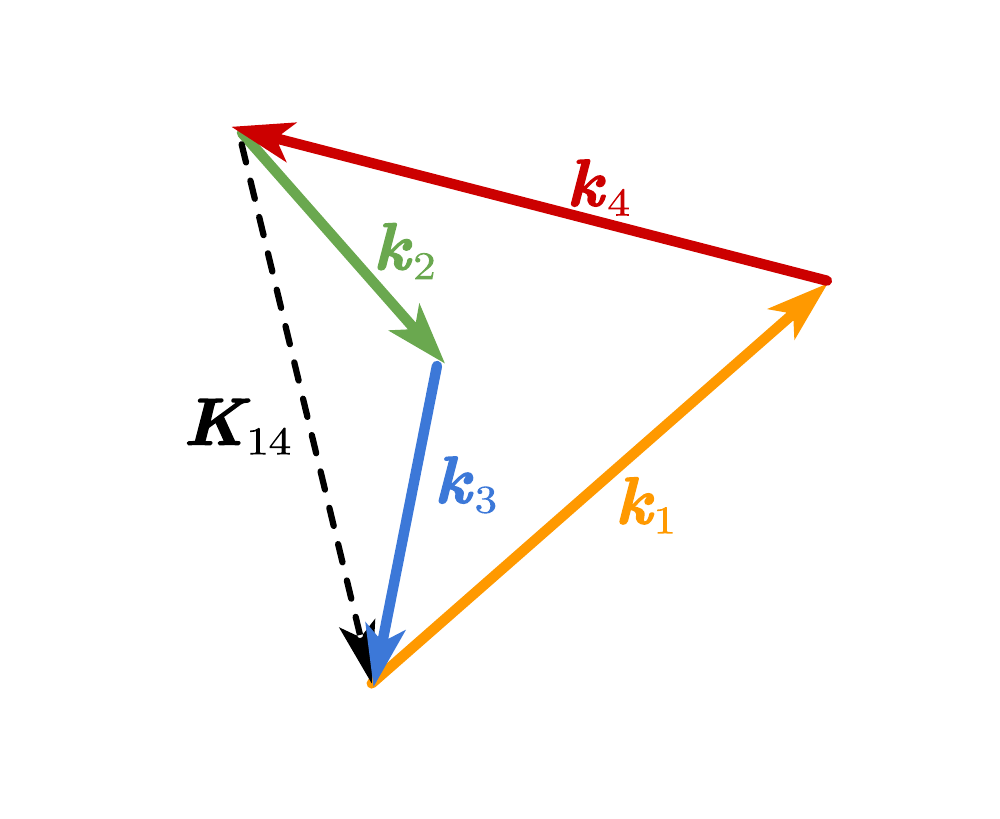}
        \caption{$\boldsymbol{K}_{14} = -\left( \boldsymbol{k}_1 + \boldsymbol{k}_4 \right)$}
        \label{fig: exchange K14 channel}
    \end{subfigure}

    \caption{Three independent diagonal channels of the trispectrum. Each panel shows a different pairing of the four wavevectors, defining a diagonal $\boldsymbol{K}_{ij} = -\left( \boldsymbol{k}_i + \boldsymbol{k}_j \right)$. From left to right, the panels correspond to the $(12)$-$(34)$, $(13)$-$(24)$ and $(14)$-$(23)$ pairings, respectively. These represent three inequivalent internal momentum configurations entering the trispectrum. }
    \label{fig: exchange diagram three channels}
\end{figure}

This structure is illustrated in \Cref{fig: exchange diagram three channels}, where each pairing defines a distinct internal momentum $\boldsymbol{K}_{ij}$ flowing through the configuration. The trispectrum template depends explicitly on this diagonal momentum, and the three channels therefore represent physically inequivalent contributions, analogous to the $s$-, $t$-, and $u$-channels in scattering theory~\cite{Mandelstam:1958xc}.

Crucially, although $\boldsymbol{K}$ appears as an integration variable associated with the Dirac delta functions, it enters explicitly in the interaction kernel $t^{\boldsymbol{k}_1 \boldsymbol{k}_2}_{\boldsymbol{k}_3 \boldsymbol{k}_4 }\left(\boldsymbol{K}\right)$. As a result, the integral does not collapse trivially, and different choices of diagonal correspond to physically distinct contributions.

We parameterize the interaction kernel $t_{\boldsymbol{k}_3\boldsymbol{k}_4}^{\boldsymbol{k}_1\boldsymbol{k}_2}( \boldsymbol{K} )$ using a phenomenological template that respects statistical homogeneity, rotational invariance, and symmetry under the exchange of external momenta. The resulting expression is given by~\cite{Shiraishi:2016mok}:
\begin{equation}\label{eq: shiraishi template 3D original version}
    \begin{split}
        t_{\boldsymbol{k}_3\boldsymbol{k}_4}^{\boldsymbol{k}_1\boldsymbol{k}_2}\left( \boldsymbol{K} \right) &\equiv i\sum_n{d_{n}^{\mathrm{odd}}\left[ \mathcal{L}_n\left( \hat{\boldsymbol{k}}_1\cdot \hat{\boldsymbol{k}}_3 \right) +\mathcal{L}_n\left( \hat{\boldsymbol{k}}_1\cdot \hat{\boldsymbol{K}} \right) +\left( -1 \right) ^n \mathcal{L}_n\left( \hat{\boldsymbol{k}}_3\cdot \hat{\boldsymbol{K}} \right) \right]}
\\
&\times \left[ \hat{\boldsymbol{K}}\cdot \left( \hat{\boldsymbol{k}}_1\times \hat{\boldsymbol{k}}_3 \right) \right] \times P_{\mathcal{R}}\left( k_1 \right) P_{\mathcal{R}}\left( k_3 \right) P_{\mathcal{R}}\left( K \right) ,
    \end{split}
\end{equation}
where $\mathcal{L}_{n}$ denotes the $n$-th order Legendre polynomial, $d_n^{\rm odd}$ parametrizes the amplitude of the parity-odd signal, $d_{0}^{\text{odd}} = -d_{1}^{\text{odd}}/3$ and $d_{n\ge 2}^{\text{odd}}=0$, and we set $\left|d_1^{\text{odd}}\right|=9\times 10^5$, in accordance with the parameter choice used in Ref.~\cite{Kurita:2025hmp}. For a discussion of different choices of $d_{1}^{\rm odd}$ and their impact on the SNR, we refer reader to \Cref{app:parameter_constraints}.

To facilitate the angular projection, we expand both the parity-odd triple product and the Legendre polynomials into spherical harmonics. This yields a fully harmonic representation of the interaction kernsl,
\begin{align}\label{eq: shiraishi template 3D version}
    \begin{split}
        t_{\boldsymbol{k}_3\boldsymbol{k}_4}^{\boldsymbol{k}_1\boldsymbol{k}_2}\left( \boldsymbol{K} \right) &=P_{\mathcal{R}}\left( k_1 \right) P_{\mathcal{R}}\left( k_3 \right) P_{\mathcal{R}}\left( K \right) \sum_n{}d_{n}^{\mathrm{odd}}\sum_{L_{1}^{\prime}L_{3}^{\prime}L_K}{}G_{L_{1}^{\prime}L_{3}^{\prime}L_K}^{n}\sum_{M_{1}^{\prime}M_{3}^{\prime}M_K}{}
\\
&\times Y_{L_{1}^{\prime}M_{1}^{\prime}}^{*}\left( \hat{\boldsymbol{k}}_1 \right) Y_{L_{3}^{\prime}M_{3}^{\prime}}^{*}\left( \hat{\boldsymbol{k}}_3 \right) Y_{L_KM_K}^{*}\left( \hat{\boldsymbol{K}} \right) \left( \begin{matrix}
	L_{1}^{\prime}&		L_{3}^{\prime}&		L_K\\
	M_{1}^{\prime}&		M_{3}^{\prime}&		M_K\\
\end{matrix} \right) ,
    \end{split}
\end{align}
where the geometric coupling coefficient $G_{L_{1}^{\prime}L_{3}^{\prime}L_K}^{n}$ arises from the angular-momentum coupling of the spherical harmonic expansion and can be expressed in terms of Wigner symbols as:
\begin{align}
  \begin{split}
      G_{L_{1}^{\prime}L_{3}^{\prime}L_K}^{n}&\equiv \frac{4\pi}{2n+1}\sqrt{6}\left( \frac{4\pi}{3} \right) ^{3/2}\left[ \left( -1 \right) ^n\mathcal{F} _{n1L_{1}^{\prime}}\mathcal{F} _{n1L_{3}^{\prime}}\delta _{L_K,1}^{K}\left\{ \begin{matrix}
	L_{1}^{\prime}&		L_{3}^{\prime}&		L_K\\
	1&		1&		n\\
\end{matrix} \right\} \right. 
\\
&\left. +\left( -1 \right) ^n\mathcal{F} _{n1L_{1}^{\prime}}\delta _{L_{3}^{\prime},1}^{K}\mathcal{F} _{n1L_K}\left\{ \begin{matrix}
	L_{1}^{\prime}&		L_K&		L_{3}^{\prime}\\
	1&		1&		n\\
\end{matrix} \right\} +\delta _{L_{1}^{\prime},1}^{K}\mathcal{F} _{n1L_{3}^{\prime}}\mathcal{F} _{n1L_K}\left\{ \begin{matrix}
	L_{3}^{\prime}&		L_K&		L_{1}^{\prime}\\
	1&		1&		n\\
\end{matrix} \right\} \right] .
  \end{split}
\end{align}


The detailed derivation is presented in \Cref{app:derivation exchange diagram}. Here we only note that the momentum-conserving Dirac delta function introduces an internal diagonal momentum $\boldsymbol{K}$, which naturally decomposes the quadrilateral  configuration into two coupled triangles.

Combining the harmonic representation of the interaction kernel with the projection formalism, the plane-wave expansion of the Dirac delta functions, and reduced trispectrum definition, we obtain the projected reduced trispectrum:
\begin{align}\label{eq: shiraishi reduced trispectrum}
    \begin{split}
        {\mathcal{Q} _{\ell _3\ell _4}^{\ell _1\ell _2}}_{\pi}^{\left( c \right)}\left( L \right)  &=\left( 2L+1 \right) \left( \frac{2}{\pi} \right) ^5\times i^{\ell _1+\ell _2+\ell _3+\ell _4}\left[ \frac{2}{5}\frac{1}{\Omega _{m,0}H_{0}^{2}} \right] ^4
\\
&\times \sum_{L_1L_{K_+}}{}\sum_{L_3L_{K_-}}{}\sum_{L_{1}^{\prime}L_{3}^{\prime}L_K}{}\sum_n{}d_{n}^{\text{odd}}G_{L_{1}^{\prime}L_{3}^{\prime}L_K}^{n}
\\
& \times \mathcal{F} _{L_1\ell _2L_{K_+}}\mathcal{F} _{L_3\ell _4L_{K_-}}\mathcal{F} _{L_1\ell _1L_{1}^{\prime}}\mathcal{F} _{L_3\ell _3L_{3}^{\prime}}\mathcal{F} _{L_{K_+}L_{K_-}L_K} 
\\
&\times \left\{ \begin{matrix}
	L_K&		L_{K_+}&		L_{K_-}\\
	L&		L_{3}^{\prime}&		L_{1}^{\prime}\\
\end{matrix} \right\} \left\{ \begin{matrix}
	\ell _1&		\ell _2&		L\\
	L_{K_+}&		L_{1}^{\prime}&		L_1\\
\end{matrix} \right\} \left\{ \begin{matrix}
	\ell _4&		L_3&		L_{K_-}\\
	L_{3}^{\prime}&		L&		\ell _3\\
\end{matrix} \right\} 
\\
&\times \left( -1 \right) ^{L_{K_-}+L_{3}^{\prime}+L_{K_-}+\ell _1+\ell _3}\times i^{L_1+\ell _2+L_3+\ell _4+L_{K_+}+L_{K_-}}
\\
&\times \int{}\mathrm{d}K\int{}\mathrm{d}x\,\,x^2\int{}\mathrm{d}y\,\,y^2j_{L_1}\left( k_1x \right) j_{\ell _2}\left( k_2x \right) j_{L_{K_+}}\left( Kx \right) j_{L_3}\left( k_3y \right) j_{\ell _4}\left( k_4y \right) j_{L_{K_-}}\left( Ky \right) 
\\
&\times \prod_{n=1}^4{}\left[ \int{}\mathrm{d}k_n \ k_{n}^{4}\mathcal{T} _{\delta}\left( k_n \right) \int_0^{\chi _{H_n}}{}\mathrm{d}\chi _{n}^{\prime} \ q\left( \chi _{n}^{\prime} \right) j_{\ell _n}\left( k_n\chi _{n}^{\prime} \right) D\left( \chi _n \right) \right] K^2
\\
&\times P_{\mathcal{R}}\left( k_1 \right) P_{\mathcal{R}}\left( k_3 \right) P_{\mathcal{R}}\left( K \right) .
    \end{split}
\end{align}
where the superscript $(c)$ denotes the collapsed-type template, and the subscript $\pi$ labels a representative permutation of the external momenta. $L_{K_+}$ and $L_{K_-}$ are the angular momentum indices introduced by the spherical harmonic expansions of the first and second Dirac delta functions, respectively, and are both associated with the same internal diagonal momentum $\boldsymbol{K}$.  

Eq.~\eqref{eq: shiraishi reduced trispectrum} represents the base reduced trispectrum evaluated specifically for the $(12)$-$(34)$ coupling. Because the collapsed-type template depends explicitly on the intermediate diagonal momentum $\boldsymbol{K}$, the corresponding diagonal momentum and external leg labels must be relabeled for different pairings. The complete physical signal is therefore obtained by summing this base configuration over all distinct permutations of the external indices. 

Following the same strategy adopted for the squeezed-limit model, we separate the reduced trispectrum into a purely geometric coupling term and a radial projection term. The geometric term $\mathcal{W}^{(c)}_{\pi}$ encapsulates the angular momentum selection rules for the collapsed model 
\begin{align}\label{eq: shiraishi geometry before simplification}
    \begin{split}
        {\mathcal{W} _{\ell _3\ell _4}^{\ell _1\ell _2}}_{\pi}^{\left( c \right)}\left( L \right) &=\left( 2L+1 \right) \, i^{\ell _1+\ell _3}\sum_{L_1L_{K_+}}{}\sum_{L_3L_{K_-}}{}\mathcal{F} _{L_1\ell _2L_{K_+}}\mathcal{F} _{L_3\ell _4L_{K_-}}
\\
&\times \sum_{L_{1}^{\prime}L_{3}^{\prime}L_K}{}\sum_n{}d_{n}^{\text{odd}}G_{L_{1}^{\prime}L_{3}^{\prime}L_K}^{n}\mathcal{F} _{L_1\ell _1L_{1}^{\prime}}\mathcal{F} _{L_3\ell _3L_{3}^{\prime}}\mathcal{F} _{L_{K_+}L_{K_-}L_K}
\\
&\times \left\{ \begin{matrix}
	L_K&		L_{K_+}&		L_{K_-}\\
	L&		L_{3}^{\prime}&		L_{1}^{\prime}\\
\end{matrix} \right\} \left\{ \begin{matrix}
	\ell _1&		\ell _2&		L\\
	L_{K_+}&		L_{1}^{\prime}&		L_1\\
\end{matrix} \right\} \left\{ \begin{matrix}
	\ell _4&		L_3&		L_{K_-}\\
	L_{3}^{\prime}&		L&		\ell _3\\
\end{matrix} \right\} 
\\
&\times \left( -1 \right) ^{L_{K_-}+L_{3}^{\prime}+L_{K_-}+\ell _1+\ell _2+\ell _3+\ell _4}\, i^{L_1+L_3+L_{K_+}+L_{K_-}} .
    \end{split}
\end{align}

And the radial projection term, ${\mathcal{I} _{\ell _3\ell _4}^{\ell _1\ell _2}}_{\pi}^{\left( c \right)}\left( L \right) $, encodes the signal magnitude through the line-of-sight projection integrals:
\begin{align}\label{eq: shiraishi full trispectrum before Limber}
    \begin{split}
&{\mathcal{I} _{\ell _3\ell _4}^{\ell _1\ell _2}}_{\pi}^{\left( c \right)}\left( L \right) =\left[ \frac{2}{5\Omega _{m,0}H_{0}^{2}} \right] ^4\left( \frac{2}{\pi} \right) ^5\int{\mathrm{d}K\,K^2\int{\mathrm{d}x\,x^2j_{L_{K_+}}(Kx)\int{\mathrm{d}y\,y^2j_{L_{K_-}}(Ky)}}}
\\
&\times \prod_{n=1}^4{\left[ \int{\mathrm{d}k_n \ k_{n}^{4}\mathcal{T} (k_n)j_{L_n}(k_nr_n)\int{d\chi _{n}^{\prime}\,q(\chi _{n}^{\prime})D(\chi _{n}^{\prime})j_{\ell _n}(k_n\chi _{n}^{\prime})}} \right] P_{\mathcal{R}}(K)P_{\mathcal{R}}(k_1)P_{\mathcal{R}}(k_3)} .
    \end{split}
\end{align}
where $r_n$ is $x$ when $n=1, 2$ and is $y$ when $n=3, 4$. The numerical implementation of this high-dimensional projection integral is described in \Cref{app:limber approximation}.

\begin{figure}[htbp]
    \centering
    \includegraphics[width=1.0\linewidth]{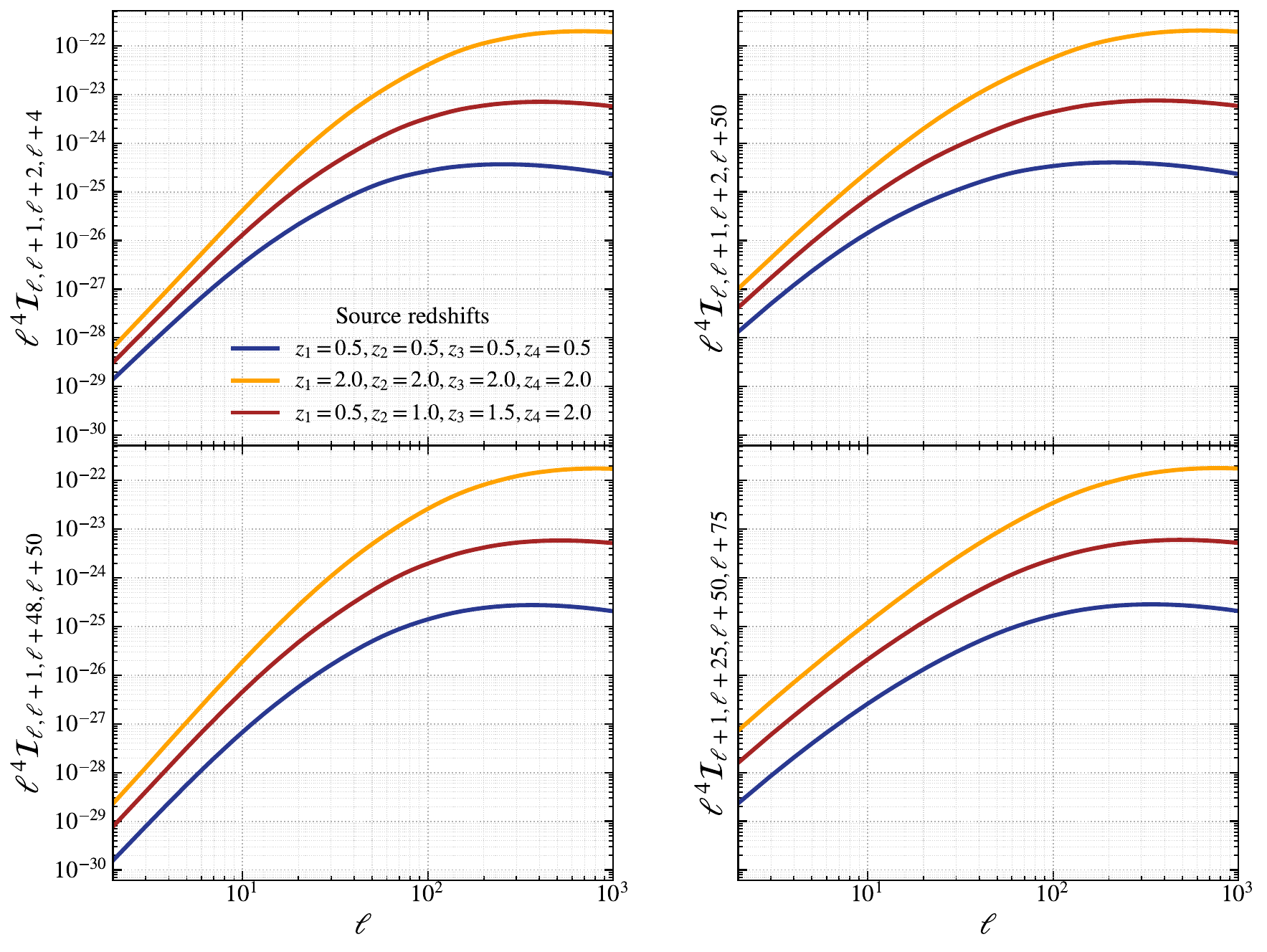}
    \caption{
    Magnitude of the line-of-sight projection term $\mathcal{I}$ from Eq.~\eqref{eq: shiraishi full trispectrum before Limber}. The internal diagonal multipole is fixed to $L=2$ in all panels. The four panels correspond to different external multipole configurations indicated on the vertical axes. Colors denote the tomographic source combinations shown in the legend: shallow auto-correlations (blue), deep auto-correlations (orange), and mixed-redshift cross-correlations (red). All curves are evaluated using the exact numerical integration. A factor of $\ell^4$ is included on the vertical axis to improve visual clarity. Since the figure shows the projection term alone, only the parity odd condition is imposed, and no quadrilateral closure condition is enforced.  
    }
    \label{fig: shiraishi template projection full integration}
\end{figure}

\begin{figure}[htbp]
    \centering
    \includegraphics[width=1.0\linewidth]{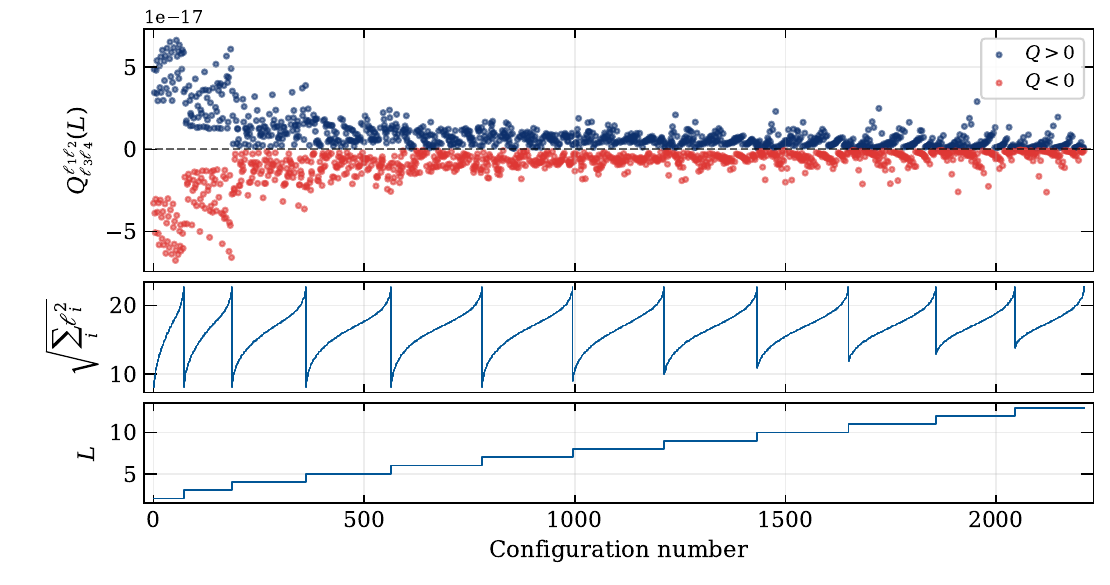}
    \caption{
    Reduced trispectrum for the collapsed-limit template, obtained by summing over all permutations as defined in Eq.~\eqref{eq: full reduced trispectrum and sum of single reduced trispectrum}. A cutoff of $\ell_{\max}=13$ is applied, with configurations ordered by the diagonal multipole $L$ and, within each fixed-$L$ block, by increasing $\sqrt{\ell_1^2+\ell_2^2+\ell_3^2+\ell_4^2}$. Top panel: reduced trispectrum $Q$, where positive and negative values are shown in blue and red, respectively. The alternating signs reflect the oscillatory angular structure of the parity-odd trispectrum. Middle panel: the corresponding external multipole scale, characterized by $\sqrt{\ell_1^2+\ell_2^2+\ell_3^2+\ell_4^2}$. Bottom panel: the diagonal multipole $L$. The number of non-vanishing configurations differs from that of the squeezed-limit template because the geometric selection rules remove different configurations in the two cases.
    }
    \label{fig: reduced trispectrum for exchange diagram}
\end{figure}

\Cref{fig: shiraishi template projection full integration} displays the projection part of the trispectrum for this diagram. To isolate the dependence on the external side lengths, the internal diagonal multipole is fixed to $L=2$ across all panels. The overall redshift dependence follows a similar hierarchy to the squeezed-limit template, while the scale dependence differs from the squeezed model case. 
We also examine the reduced trispectrum obtained by summing over all permutations for the collapsed template, shown in \Cref{fig: reduced trispectrum for exchange diagram}. Similar to the squeezed-limit template, the parity-odd signal alternates between positive and negative values across different configurations, reflecting the oscillatory angular dependence of the trispectrum. As the diagonal multipole $L$ increases, the overall signal amplitude exhibits a gradual decrease, indicating that the collapsed-type template is primarily sensitive to small diagonal lengths. Moreover, for fixed small diagonal multipoles $L$, the signal generally increases as the characteristic external multipole scale $\sqrt{\ell_1^2 + \ell_2^2 + \ell_3^2 + \ell_4^2}$ increases. Consequently, the ratio $L/\sqrt{\ell_1^2 + \ell_2^2 + \ell_3^2 + \ell_4^2}$ becomes smaller, corresponding to increasingly collapsed configurations, which exhibit stronger parity-odd signal. This trend becomes less pronounced for larger $L$, since the finite multipole cutoff $\ell_{\max}$ limits how collapsed the configurations can become. A more detailed comparison of the dependence on the diagonal and external multipoles is presented in \Cref{sec: numerical result geometric configuration sensitivity}.


\section{Numerical Results}
\label{sec: numerical result}

In this section, we present the numerical evaluation of the parity-violating trispectrum and the resulting signal-to-noise ratio (SNR) for the two templates derived in \Cref{sec: weak lensing convergence trispectrum}.  We first describe the SNR formalism, and then investigate the dependence of the signal on the trispectrum template, source redshift distribution, and geometric configuration. Finally, we present results based on realistic DES Y3 and LSST-like Y10 source distributions, both with and without galaxy shape noise. Throughout this section, the cumulative SNR is evaluated for the fiducial template amplitudes introduced in \Cref{sec: weak lensing convergence trispectrum}. The corresponding Fisher forecasts for the amplitude parameters are presented in \Cref{app:parameter_constraints}.

\subsection{Signal-to-Noise Estimation}
\label{sec: signal-to-noise estimation}

To quantify the theoretical detectability of the parity-violating signatures derived in \Cref{sec: weak lensing convergence trispectrum}, we adopt an unbiased estimator for the parity-odd component of the harmonics four-point correlation function. Following the formalism of Refs.~\cite{Hu:2001fa, Greco:2025xtt}, we define the estimator for the reduced trispectrum, $\hat{Q}^{\ell_1 \ell_2}_{\ell_3 \ell_4} (L)$, as 
\begin{align}
    \begin{split}
        \hat{Q}^{\ell_1 \ell_2}_{\ell_3 \ell_4} \left(L\right) \equiv \left(2L+1 \right) \sum_{m_i} \sum_M \left(-1\right)^M \left(\begin{matrix}
    \ell_1 & \ell_2 & L \\
    m_1 & m_2 & M
\end{matrix}\right)
\left(\begin{matrix}
    \ell_3 & \ell_4 & L \\
    m_3 & m_4 & -M
\end{matrix}\right) \kappa_{\ell_1 m_1} \kappa_{\ell_2 m_2} \kappa_{\ell_3 m_3} \kappa_{\ell_4 m_4} ,
    \end{split}
\end{align}
for $i=1, 2, 3, 4$, where $\ell_1 + \ell_2 + \ell_3 + \ell_4$ is odd to ensure the parity-odd signal, guaranteeing $\langle\hat{Q}^{\ell_1 \ell_2}_{\ell_3 \ell_4}(L)\rangle = Q^{\ell_1 \ell_2}_{\ell_3 \ell_4}(L)$. Due to the complexity of accurately modeling the non-Gaussian covariance of the weak lensing trispectrum, we adopt a simplified Gaussian covariance approximation in this work. Under this assumption, the covariance is dominated by the disconnected Gaussian component and is approximated as diagonal, depending only on the products of the angular power spectra. A more realistic covariance treatment will be left for future work.

To estimate the cumulative SNR, we sum the signal over all unique geometric configurations up to a maximum multipole $\ell_{\max}$, defined such that all angular multipoles entering the estimator satisfy $\ell_i \leq \ell_{\max}$ and $L \leq \ell_{\max}$.
For a tomographic source-bin combination $\boldsymbol{i}=(i_1,i_2,i_3,i_4)$, where the multipole $\ell_a$ is associated with the source bin $i_a$, we restrict the summation to the ordered domain $\ell_1 < \ell_2 < \ell_3 < \ell_4$ to avoid overcounting redundant permutations of the four multipole labels:
\begin{equation}\label{eq: SNR calculation equation}
    \mathrm{SNR}_{\boldsymbol{i}}\simeq \sqrt{\sum_{L=2}^{\ell _{\max}}{\left( 2L+1 \right) ^{-1}\sum_{\ell _1<\ell _2<\ell _3<\ell _4}^{\ell _{\max}}{\frac{\left| Q_{\ell _3\ell _4}^{\ell _1\ell _2}\left( L \right) \right|^2}{C_{\ell _1}^{i_1i_1}C_{\ell _2}^{i_2i_2}C_{\ell _3}^{i_3i_3}C_{\ell _4}^{i_4i_4}}}}}.
\end{equation}
The summation over $L$ starts from $2$, since configurations with $L=0$ or $L=1$ require $\ell_1+\ell_2+\ell_3+\ell_4$ to be even by the Wigner 3-$j$ selection rules, and therefore cannot contribute to the parity-odd trispectrum. Here $C_{\ell}^{ii}$ denotes the auto angular convergence power spectrum of the $i$-th tomographic source bin. The reduced trispectrum in the numerator is evaluated for the same tomographic source-bin combination $\boldsymbol{i}$. Each tomographic combination is evaluated independently throughout this work. For compactness, we suppress tomographic indices in the following unless they are needed explicitly.


\subsection{Template sensitivity}

\subsubsection{Squeezed-Type Template}

\begin{figure}[htbp]
    \centering
    \includegraphics[width=1.0\linewidth]{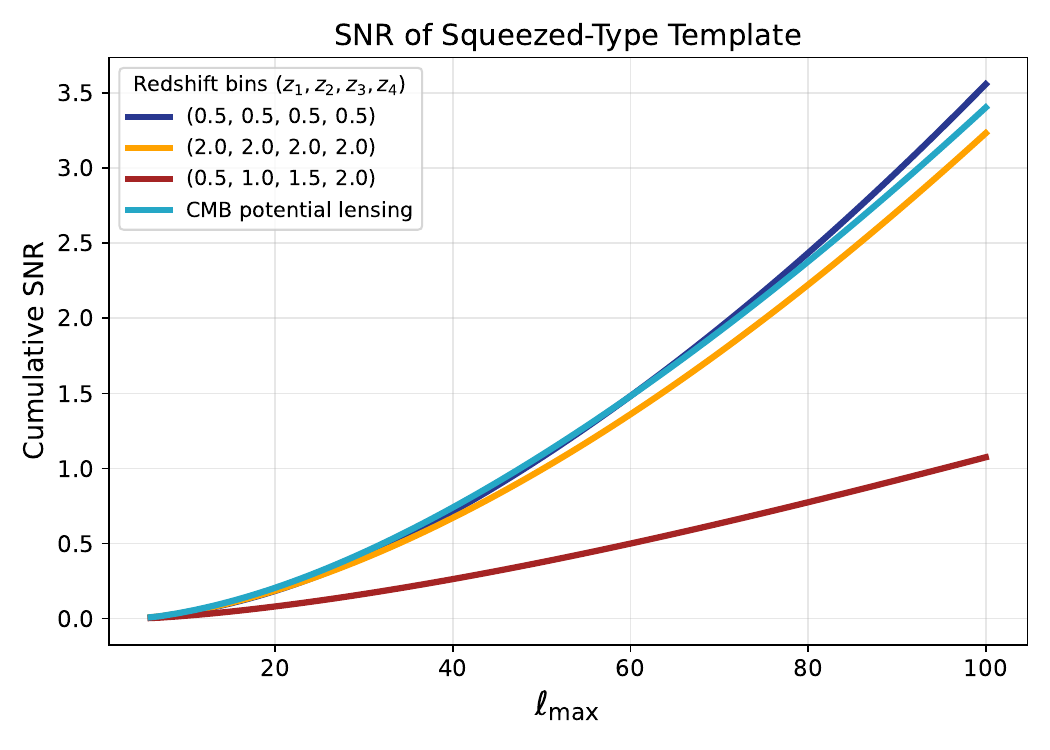}
    \caption{
    Cumulative SNR of the parity-odd squeezed trispectrum as a function of the maximum multipole $\ell_{\max}$, evaluated using the exact FFTLog integration. Different colors correspond to the source redshift combinations indicated in the legend, assuming Dirac delta source redshift distributions. The cyan curves denote the CMB potential lensing case. The results are evaluated assuming $\left| g_- \right|=2\times 10^7$. No galaxy shape noise is included. 
    }
    \label{fig: coulton template SNR fftlog}
\end{figure}


\Cref{fig: coulton template SNR fftlog} presents the cumulative SNR for the squeezed-type template as a function of the maximum multipole $\ell_{\max}$.  Interestingly, the cumulative SNR demonstrates that lower-redshift source distributions yield a higher theoretical detectability. The configuration with all source galaxies at $z=0.5$ produces the strongest cumulative signal, whereas the deeper $z=2.0$ configuration yields a correspondingly lower total SNR. The theoretical expectation for CMB potential lensing is included as an extreme high-redshift reference, which yields a lower SNR than the uniform galaxy lensing cases. Furthermore, the mixed-redshift cross-correlation bin $(0.5, 1.0, 1.5, 2.0)$ produces the lowest overall detectability, highlighting a strong geometric suppression when correlating widely separated source planes. Within the accessible multipole range, the cumulative SNR does not exhibit a clear saturation behavior. Extending the calculation to higher $\ell_{\max}$ becomes computationally expensive due to the large summations over Wigner symbols entering the angular coupling terms, and the asymptotic saturation behavior is therefore investigated separately in \Cref{app:saturation noiseless} through an analysis of the line-of-sight projection integrals.





\subsubsection{Collapsed-Type Template}

\begin{figure}[htbp]
    \centering
    \includegraphics[width=1.0\linewidth]{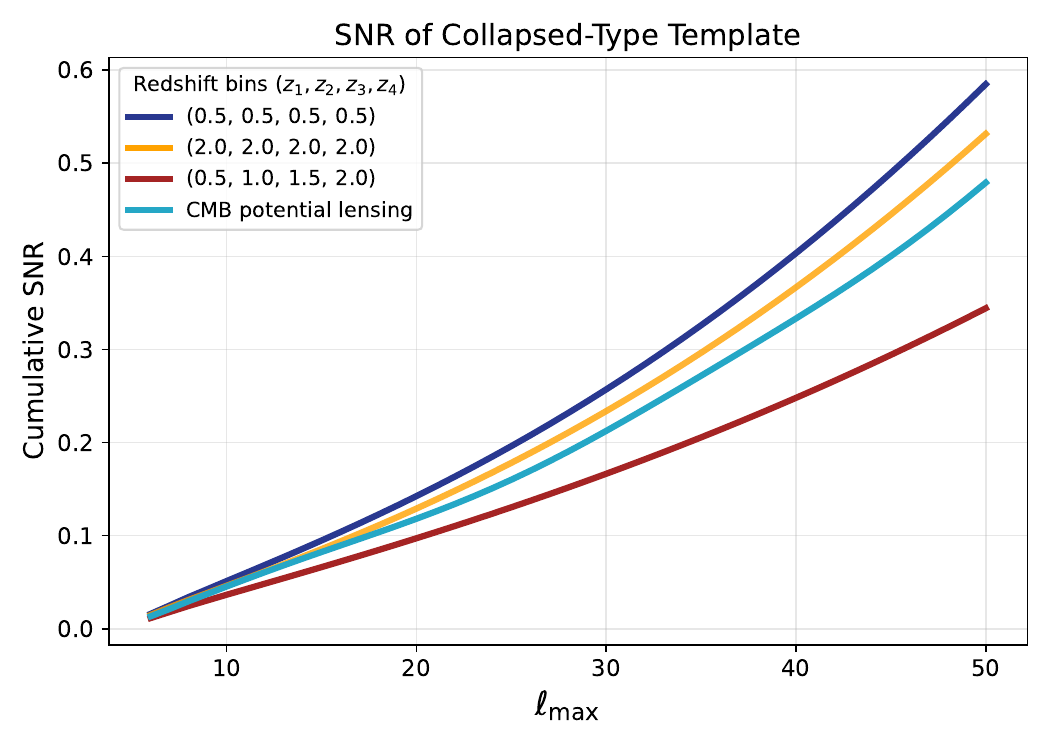}
    \caption{
    Cumulative SNR of the parity-odd collapsed-type trispectrum as a function of the maximum multipole $\ell_{\max}$, evaluated using FFTLog integration. Different colors correspond to the source redshift combinations indicated in the legend, assuming Dirac delta source redshift distributions. The cyan curves denote the CMB potential lensing case. The results are evaluated assuming $\left| d^{\rm odd}_{1} \right|=9\times 10^5$. No galaxy shape noise is included. 
    }
    \label{fig: shiraishi template SNR fftlog}
\end{figure}

The corresponding cumulative SNR for the collapsed-type template is shown in \Cref{fig: shiraishi template SNR fftlog}. We observe the same tomographic trend as the squeezed model: the shallowest redshift bin ($ z=0.5 $) yields the highest SNR, while the deeper tomographic bins and CMB lensing yield lower detectability. The mixed-redshift combination yields the lowest cumulative SNR among all cases considered. The physical origin of this suppression will be discussed in \Cref{sec: physical interpretation of redshift dependence}. Due to the significantly higher computational cost of the collapsed-type model, primarily arising from the additional summations over angular coupling terms, the numerical evaluation is restricted to $\ell_{\max}=50$ in this work.

\subsection{Physical Interpretation of the Redshift Dependence}
\label{sec: physical interpretation of redshift dependence}

As shown in \Cref{fig: coulton template SNR fftlog,fig: shiraishi template SNR fftlog}, the cumulative SNR exhibits a counter-intuitive dependence on the source redshift: shallower tomographic bins (e.g.\ $z=0.5$) yield a higher SNR than deeper bins ($z=2.0$), and mixed-redshift configurations strongly suppress the detectability. One might naively assume that extending the integration to higher redshifts would accumulate more parity-odd trispectrum signal, and that the cross-correlation would capture more 3D information compared to configurations restricted to a single spherical shell. To understand this behavior, we must examine the geometric projection of the 3D parity-violating signal onto the 2D sky. 

To gain physical intuition for these trends, we make use of the geometric picture provided by the Limber approximation~\cite{Limber:1954zz, Lemos:2017arq}, which provides a simplified description of the line-of-sight projection. Examining the SNR estimator in Eq.~\eqref{eq: SNR calculation equation} alongside the trispectrum expressions, we establish a fundamental baseline: for a fixed geometric configuration defined by the multipoles $\{\ell_1, \ell_2, \ell_3, \ell_4, L \}$, the geometric term is constant. Consequently, the sole variable governing the observed amplitude is the line-of-sight integration domain, which is strictly determined by the source galaxy redshift distributions.


Building on this, we first address why deeper auto-correlations yield a lower total SNR than shallower ones. Within the Limber approximation, the angular power spectrum and trispectrum exhibit markedly different radial dependence. The angular power spectrum integrand scales as $\chi^{-6}$, whereas the trispectrum integrand scales significantly more steeply, as $\chi^{-14}$ (see \Cref{app:limber_derivation}). Physically, this reflects the fundamental nature of projecting a 3D field onto a 2D sky: higher-order statistics are more sensitive to line-of-sight projection effects than two-point statistics. Because the trispectrum requires four spatial points to be simultaneously correlated within the same local 3D volume, extending the line-of-sight integration over large cosmological distances causes the trispectrum contribution to the signal to grow more slowly with redshift than the covariance contribution from the angular power spectra. Consequently, although extending the integration to higher redshift bins accumulates a larger absolute parity-odd signal in the numerator, the much steeper radial scaling of the trispectrum relative to the angular power spectra suppresses the growth of the cumulative SNR.

The second question is why the SNR is strongly suppressed when source galaxies are distributed across different tomographic bins. A useful way to visualize this effect is through the Limber approximation, where the dominant contribution to the projected correlation function arises from structures located on the same spherical shell along the line of sight. Consequently, when all four points are drawn from the same tomographic source bin, as shown in \Cref{fig: schematic diagram same tomo}, the parity-odd trispectrum signal accumulates continuously from the source redshift $z_s$ to the observer. Crucially, the auto power spectra, $C_{\ell}$, which constitute the variance in the denominator of the SNR estimator, also accumulate over this exact same integration volume. However, this overlap structure changes when the source galaxies are distributed across different tomographic bins as seen in \Cref{fig: schematic diagram diff tomo}. Because the lensing kernel essentially drops to zero beyond the distance of the source galaxy, the overlap of the four kernels, and thus the accumulation of the trispectrum signal, is abruptly truncated at the redshift of the closest galaxy ($z_1$). In contrast, the variance (noise) continues to grow: the individual auto power spectra $C_{\ell_i}$ for the deeper background galaxies ($z_2, z_3, z_4$) continue to accumulate along the line-of-sight up to their respective redshifts. The red segments in \Cref{fig: schematic diagram diff tomo} visually represent this discrepancy. These line-of-sight volumes act as pure noise penalties; they contribute heavily to the sample variance in the SNR denominator but provide zero structural overlap for parity-odd signal in the numerator. This geometric ``noise-only'' accumulation accounts for the severe SNR suppression observed in mixed source galaxy distributions.

\subsection{Geometric Configuration Sensitivity}
\label{sec: numerical result geometric configuration sensitivity}

Having established the cumulative detectability of the two templates, we now investigate which specific geometric configurations drive the signal and how these features can be used to distinguish the underlying microphysics. We first perform a sensitivity test on the absolute SNR by introducing lower-bound cutoffs in the multipole summations. 

\begin{figure}[htbp]
    \centering

    \begin{subfigure}[b]{0.495\textwidth}
        \centering
        \includegraphics[width=\textwidth]{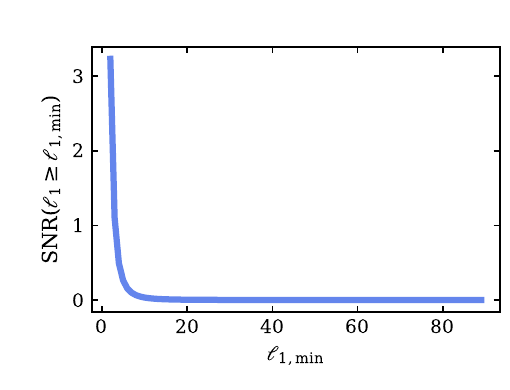}
        \caption{External-side cutoff.}
        \label{fig: coulton SNR as a function of ell1}
    \end{subfigure}
    \hfill
    \begin{subfigure}[b]{0.495\textwidth}
        \centering
        \includegraphics[width=\textwidth]{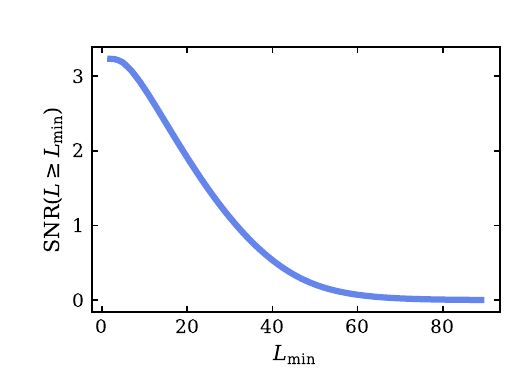}
        \caption{Diagonal cutoff.}
        \label{fig: coulton SNR as a function of L}
    \end{subfigure}

    \caption{
    Cumulative SNR of the squeezed-type template as a function of the minimum external multipoles cutoff $\ell_{1, \min}$ (left) and the minimum diagonal multipole cutoff $L_{\min}$ (right) cutoff. 
    }
    \label{fig: coulton cumulative SNR sensitivity}
\end{figure}

\Cref{fig: coulton cumulative SNR sensitivity} shows the cumulative SNR as a function of the minimum allowed external side length, $\ell_{1, \min}$, and the minimum internal diagonal, $L_{\min}$, for the squeezed-type template. In both cases, the SNR drops as the minimum multipole cutoff increases, indicating that the signal is dominated by configurations containing small multipoles, which correspond to large separations in real space. The signal is, however, substantially more sensitive to the external side length cutoff $\ell_{1,\min}$ than to the internal diagonal cutoff $L_{\min}$. This indicates that, for the squeezed-type template, small external side lengths carry more of the dominant parity-odd signal than the internal diagonal configurations. This is also consistent with the behavior observed in \Cref{fig: reduced trispectrum for contact diagram}, where the strongest signal contributions are associated with configurations containing small external multipoles. 

\begin{figure}[htbp]
    \centering
    \begin{subfigure}[b]{0.495\textwidth}
        \centering
        \includegraphics[width=\textwidth]{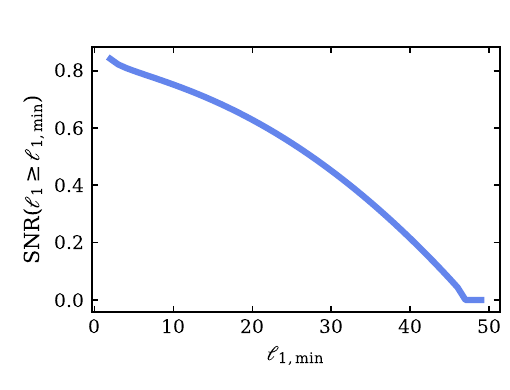}
        \caption{External-side cutoff}
        \label{fig: shiraishi SNR as a function of ell1}
    \end{subfigure}
    \hfill
    \begin{subfigure}[b]{0.495\textwidth}
        \centering
        \includegraphics[width=\textwidth]{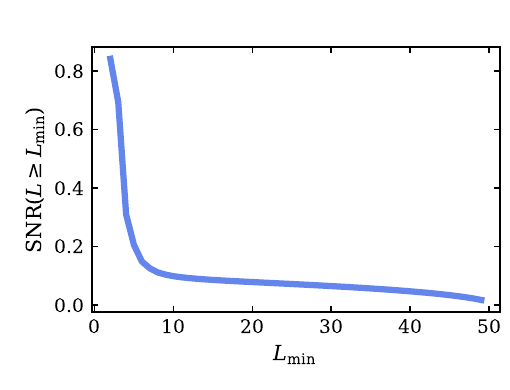}
        \caption{Diagonal cutoff}
        \label{fig: shiraishi SNR as a function of L}
    \end{subfigure}
    \caption{
     Cumulative SNR of the collapsed-limit template as a function of the minimum external multipoles cutoff $\ell_{1, \min}$ (left) and the minimum diagonal multipole cutoff $L_{\min}$ (right) cutoff. 
    }
    \label{fig: shiraishi cumulative SNR sensitivity}
\end{figure}

\Cref{fig: shiraishi cumulative SNR sensitivity} also displays the cumulative SNR as a function of the minimum allowed external side length $\ell_{1,\min}$ and the minimum internal diagonal $L_{\min}$ for the collapsed-type template. In contrast to the behavior of the squeezed-type template, the collapsed-type model exhibits a distinct response to the two geometric cutoffs. As the minimum external side length increases, the cumulative SNR decreases gradually. However, the SNR drops precipitously as the diagonal cutoff $L_{\min}$ increases, with the vast majority of the signal concentrated at the smallest diagonal configurations. This confirms that the observable collapsed-type signal is overwhelmingly dominated by configurations in the collapsed geometric limit ($L \rightarrow 0$), whereas the signal strength remains broadly distributed across the external multipoles. This behavior also explains the strong configuration-dependent fluctuations observed in \Cref{fig: reduced trispectrum for exchange diagram}, where the signal amplitude remains highly sensitive to the choice of diagonal length even at large external multipoles. 

\begin{figure}[htbp]
    \centering

    \begin{subfigure}[b]{0.495\textwidth}
        \centering
        \includegraphics[width=\textwidth]{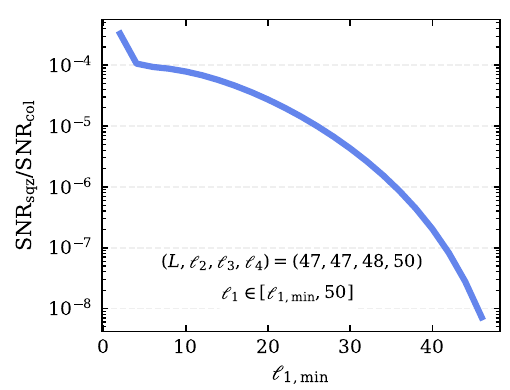}
        \caption{Squeezed limit}
        \label{fig: squeezed limit}
    \end{subfigure}
    \hfill
    \begin{subfigure}[b]{0.495\textwidth}
        \centering
        \includegraphics[width=\textwidth]{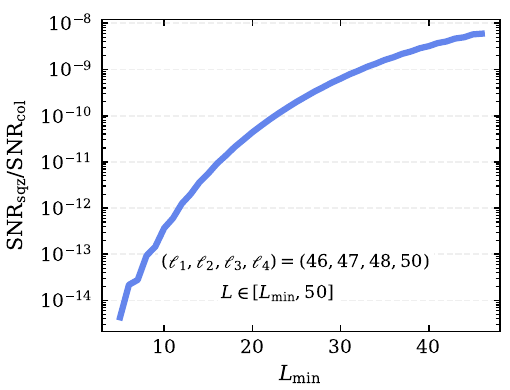}
        \caption{Collapsed limit}
        \label{fig: collapsed limit}
    \end{subfigure}

    \caption{
    Ratio of cumulative SNRs, $\text{SNR}_{\text{sqz}}/\text{SNR}_{\text{col}}$, highlighting the distinct geometric sensitivities of the squeezed and collapsed models. Panel (\subref{fig: squeezed limit}) probes the squeezed limit through the external multipole $\ell_{1,\min}$, while panel (\subref{fig: collapsed limit}) probes the collapsed limit through the diagonal multipole $L_{\min}$.
    }
    \label{fig: configuration sensitivity test}
\end{figure}

The underlying physical mechanism driving this sensitivity to small multipoles stems from the intrinsic 3D nature of parity violation. In harmonic space, small multipoles (whether external sides $\ell$ or internal diagonals $L$) correspond to large angular separations on the sky. When the four points are widely separated, their distribution across the spherical shell naturally traces a distinct 3D volume. Due to the intrinsic curvature of the sky, these widely spaced points cannot be approximated as lying on a flat 2D plane. Conversely, configurations entirely composed of large multipoles correspond to very small angular separations, meaning the four points are closely packed and effectively coplanar. Because parity violation requires a defined 3D handedness or chirality, its observable signature identically vanishes when the configuration collapses to a 2D plane. Consequently, the 3D parity information is heavily suppressed in the large-multipole regime, explaining why the bulk of the detectability relies strictly on configurations containing small sides or diagonals, where the 3D structural geometry is preserved. 

To explicitly quantify these distinct geometric sensitivities, we evaluate the cumulative SNR ratio of the squeezed-type to collapsed-type templates under fixed configuration in \Cref{fig: configuration sensitivity test}. The left panel isolates the squeezed limit, where the rapid decrease of the ratio as $\ell_{1, \min}$ increases confirms that the squeezed configuration uniquely dominates the signal at the lowest external multipoles ($\ell_1 \rightarrow 0$). The right panel isolates the collapsed limit, showing that the SNR ratio increases rapidly as the lowest internal multipoles are excluded. This demonstrates that the collapsed-type model receives a dominant contribution from collapsed configurations with small diagonal multipoles $L$. Together, these distinct geometric dependencies provide a clear observational strategy to separate the two parity-violating templates based on their 3D shape dependence. 


\subsection{Realistic Source Distributions and Shape Noise}


While the preceding sections utilized idealized Dirac delta source distributions to isolate geometric effects, realistic weak lensing observations involve extended source redshift distributions and additional shape noise contributions. We first replace the idealized source planes with tomographic redshift distributions representative of the DES Y3 and LSST-like Y10 analyses, and then include the effect of shape noise. The corresponding tomographic bin selections, redshift ranges, effective galaxy number densities, and intrinsic ellipticity dispersions are summarized in \Cref{tab:survey_spec}.

\begin{table}[htbp]
\centering
\caption{
Survey specifications adopted in this work for the DES Y3 and LSST-like Y10 weak lensing analyses. For each tomographic bin, we list the corresponding redshift range, effective galaxy number density $n_{\text{eff}}$, and the intrinsic ellipticity dispersion $\sigma_{e}$. The DES Y3 and LSST-like Y10 specifications are adopted from Refs.~\cite{DES:2020ebm,LSSTDarkEnergyScience:2018jkl} .
}
\begin{tabular}{ccccc}
\hline
Survey & Bin & Redshift range & $n_{\rm eff}$ [arcmin$^{-2}$] & $\sigma_e$ \\
\hline
DES Y3 & $1$ & $0.000<z\le0.358$ & $1.476$ & $0.243$ \\
DES Y3 & $2$ & $0.358<z\le0.631$ & $1.479$ & $0.262$ \\
DES Y3 & $3$ & $0.631<z\le0.872$ & $1.484$ & $0.259$ \\
DES Y3 & $4$ & $0.872<z\le2.000$ & $1.461$ & $0.301$ \\
\hline
LSST Y10 & $1$ & $0.200<z\le0.300$ & $2.600$ & $0.260$ \\
LSST Y10 & $2$ & $0.500<z\le0.600$ & $2.600$ & $0.260$ \\
LSST Y10 & $3$ & $0.800<z\le0.900$ & $2.600$ & $0.260$ \\
LSST Y10 & $4$ & $1.100<z\le1.200$ & $2.600$ & $0.260$ \\
\hline
\end{tabular}
\label{tab:survey_spec}
\end{table}

\begin{figure}[htbp]
\centering

\begin{subfigure}{0.495\textwidth}
    \centering
    \includegraphics[width=\linewidth]{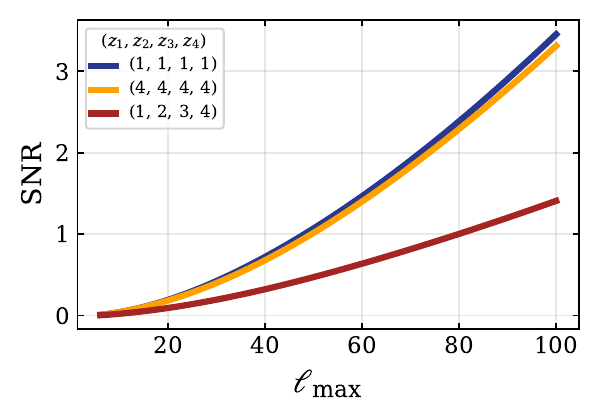}
    \caption{DES Y3 -- Squeezed-Type}
    \label{fig: DES coulton SNR}
\end{subfigure}
\hfill
\begin{subfigure}{0.495\textwidth}
    \centering
    \includegraphics[width=\linewidth]{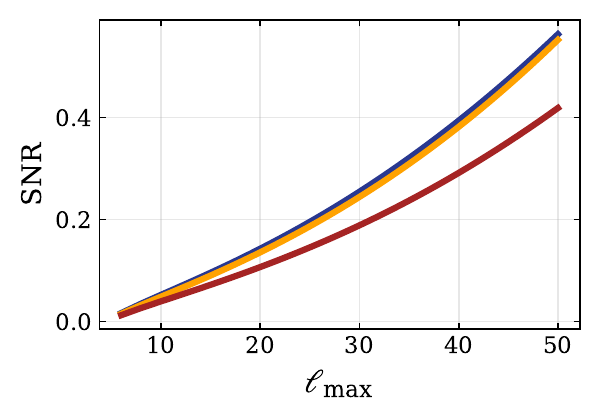}
    \caption{DES Y3 -- Collapsed-Type}
    \label{fig: DES shiraishi SNR}
\end{subfigure}

\vspace{0.5cm}

\begin{subfigure}{0.495\textwidth}
    \centering
    \includegraphics[width=\linewidth]{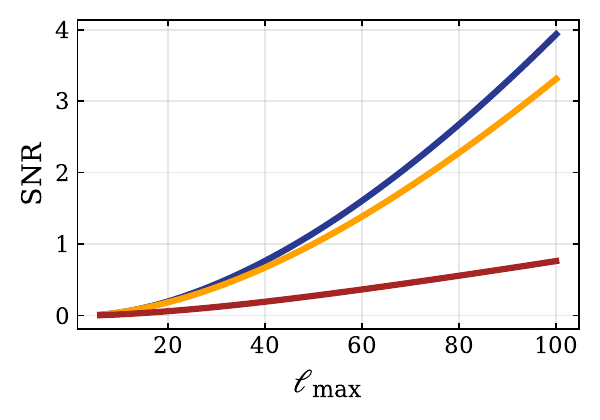}
    \caption{LSST-like Y10 -- Squeezed-Type}
    \label{fig: LSST contact SNR}
\end{subfigure}
\hfill
\begin{subfigure}{0.495\textwidth}
    \centering
    \includegraphics[width=\linewidth]{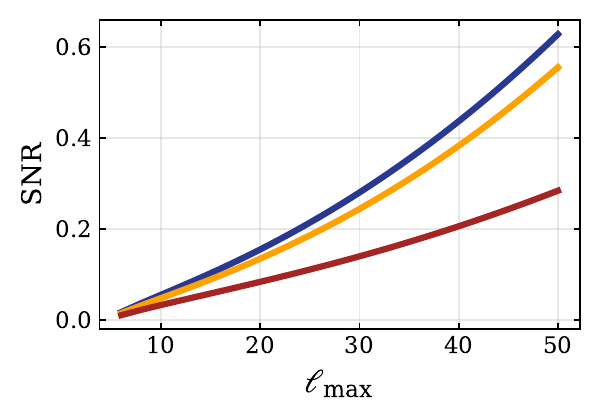}
    \caption{LSST-like Y10 -- Collapsed-Type}
    \label{fig: LSST exchange SNR}
\end{subfigure}

\caption{
Cumulative SNR of the parity-odd trispectrum as a function of the maximum multipole $\ell_{\max}$ for DES Y3 and LSST-like Y10 lensing kernels, evaluated with FFTLog integration. Panels (\subref{fig: DES coulton SNR}, \subref{fig: LSST contact SNR}) show the squeezed-type template, while panels (\subref{fig: DES shiraishi SNR}, \subref{fig: LSST exchange SNR}) show the collapsed-type model. Different colors correspond to different tomographic source bin combinations. Galaxy shape noise is not included. 
}
\label{fig: SNR_all}
\end{figure}

\subsubsection{Realistic Redshift Distributions}

\Cref{fig: DES coulton SNR,fig: DES shiraishi SNR} present the cumulative SNR for the parity-odd squeezed and collapsed templates, respectively, evaluated using these realistic DES Y3 lensing kernels via the exact FFTLog numerical integration method. We also include the corresponding results with LSST-like Y10 kernels in \Cref{fig: LSST contact SNR,fig: LSST exchange SNR}. For the squeezed-type template in \Cref{fig: DES coulton SNR,fig: LSST contact SNR}, the solid curves demonstrate that the phenomenological trends observed with idealized kernels remain robust. The single-bin configuration utilizing the shallower "Bin $1$" kernels yields a higher SNR than the deeper "Bin $4$" configuration. Crucially, the cross-correlation configuration encompassing four distinct tomographic bins clearly demonstrates the expected geometric SNR suppression associated with mixed-redshift source distributions. 

\Cref{fig: DES shiraishi SNR,fig: LSST exchange SNR} demonstrates that the collapsed template follows the identical tomographic hierarchy. Consistent with the idealized delta-kernel analysis, the mixed-bin configuration yields the lowest cumulative detectability. These results confirm that the physical suppression mechanisms governing the 3D-to-2D projection of higher-order-statistics, specifically the line-of-sight signal truncation and consequent "noise-only" accumulation, profoundly impact the detectability of parity-violation signatures in realistic photometric surveys. 

The difference between single-bin SNR and the suppression of cross-correlations is more pronounced in the LSST-like case due to the reduced overlap of the lensing kernels in redshift, which enhances the contrast between different tomographic configurations compared to DES Y3.

\subsubsection{Shape Noise}


\begin{figure}[htbp]
\centering

\begin{subfigure}{0.495\textwidth}
    \centering
    \includegraphics[width=\linewidth]{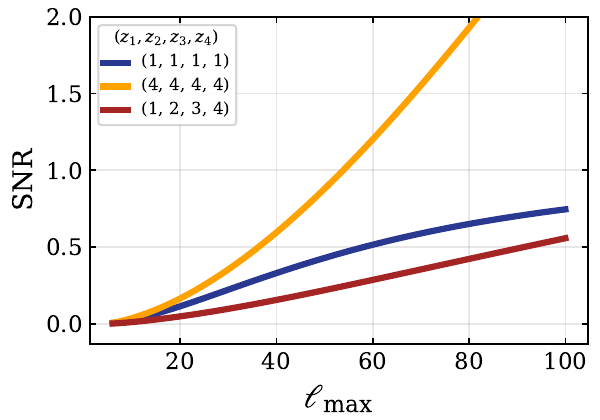}
    \caption{DES Y3 -- Squeezed-Type}
    \label{fig: DES coulton noise SNR}
\end{subfigure}
\hfill
\begin{subfigure}{0.495\textwidth}
    \centering
    \includegraphics[width=\linewidth]{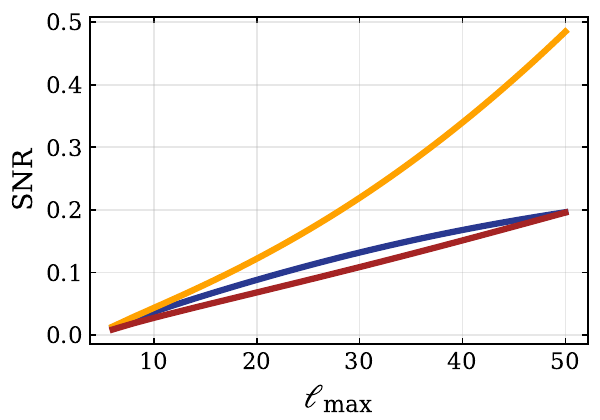}
    \caption{DES Y3 -- Collapsed-Type}
    \label{fig: DES shiraishi noise SNR}
\end{subfigure}

\vspace{0.5cm}

\begin{subfigure}{0.495\textwidth}
    \centering
    \includegraphics[width=\linewidth]{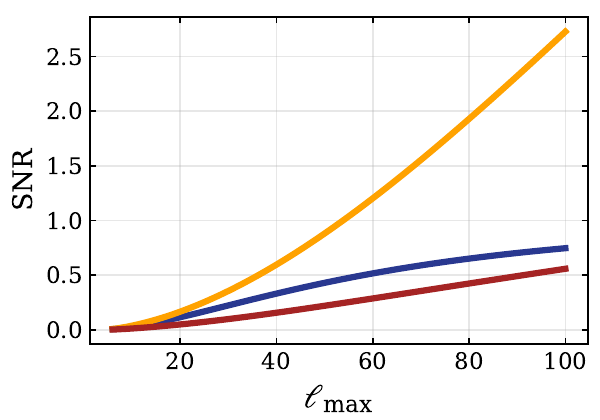}
    \caption{LSST-like Y10 -- Squeezed-Type}
    \label{fig: LSST contact noise SNR}
\end{subfigure}
\hfill
\begin{subfigure}{0.495\textwidth}
    \centering
    \includegraphics[width=\linewidth]{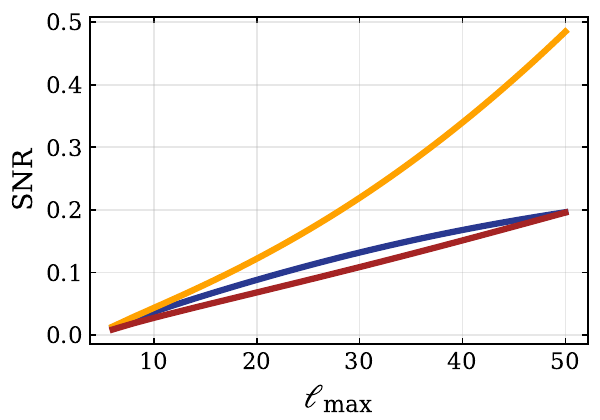}
    \caption{LSST-like Y10 -- Collapsed-Type}
    \label{fig: LSST exchange noise SNR}
\end{subfigure}

\caption{
Cumulative SNR of the parity-odd trispectrum as a function of the maximum multipole $\ell_{\max}$ for DES Y3 and LSST-like Y10 lensing kernels, evaluated with FFTLog integration, including tomographic galaxy shape noise. Panels (\subref{fig: DES coulton noise SNR}, \subref{fig: LSST contact noise SNR}) and (\subref{fig: DES shiraishi noise SNR}, \subref{fig: LSST exchange noise SNR}) correspond to the squeezed and collapsed templates, respectively. 
}
\label{fig: SNR_all noise}
\end{figure}

The cumulative SNR estimates presented thus far incorporate realistic DES Y3 and LSST-like Y10 redshift distributions but neglect galaxy shape noise. In realistic weak lensing observations, the measured shear field is contaminated by the intrinsic ellipticity dispersion of galaxies, which introduces an additional shape noise contribution to the angular power spectra entering the covariance. To account for this observational effect, we include the standard tomographic shape noise term \cite{Deshpande:2020jjs}:
\begin{equation}\label{eq: shape noise equation}
    N_{\ell}^{(i)} = \frac{\sigma_{e}^2}{n_{\text{eff}}^{(i)}},
\end{equation}
where $\sigma_{e}$ denotes the intrinsic ellipticity dispersion and $n_{\text{eff}}^{(i)}$ is the effective galaxy number density of the $i$-the tomographic bin. The values of $\sigma_e$ and $n_{\text{eff}}$ adopted for the DES Y3 and LSST-like Y10 analyses are summarized in \Cref{tab:survey_spec}. The total angular power spectrum entering the trispectrum covariance is therefore modified according to \cite{Deshpande:2020jjs}
\begin{equation}
    C_{\ell}^{(i)} \rightarrow C_{\ell}^{(i)} + N_{\ell}^{(i)}, 
\end{equation}

\Cref{fig: SNR_all noise} shows the cumulative SNR after including galaxy shape noise. In contrast to the noise free case, the relative hierarchy between different tomographic bins changes significantly once the additional noise contribution is included. Since the shape noise term primarily enters through the denominator of Eq.~\eqref{eq: SNR calculation equation}, it effectively increases the total angular power spectrum appearing in the covariance. Moreover, the shape noise contribution, $N_{\ell}^{(i)}$, behaves approximately as a white noise component and has similar amplitudes across different tomographic bins. As a result, high-redshift bins, which typically possess larger angular power spectra, are less affected by the additional noise contribution, whereas low-redshift bins with smaller power spectra are more strongly suppressed by the same noise level. Consequently, the cumulative trispectrum SNR becomes comparatively larger for higher-redshift tomographic bins, while the low-redshift bins experience a much stronger degradation after shape noise is included. A mild saturation trend can already be observed for some tomographic configurations, particularly in the low-redshift bins. The asymptotic saturation behavior at large $\ell_{\max}$ is discussed separately in \Cref{app:saturation noise}.

\section{Conclusion}
\label{sec: conclusion}


The search for parity-violating signatures has traditionally relied on the early-universe cosmic microwave background (CMB) and more recently, three-dimensional galaxy clustering. In this study, we established the weak gravitational lensing convergence trispectrum as a late-time geometric probe of these fundamental symmetries, which is complementary to CMB lensing probe \cite{Greco:2025xtt}. By adapting and extending the theoretical framework of parity-odd correlators to the LSS, we demonstrated that this observable provides an independent avenue for testing parity-breaking inflationary models.  

We first verified the consistency of our theoretical pipeline by applying it to the standard convergence angular power spectrum, utilizing both idealized Dirac-delta redshift distributions and realistic tomographic profiles from the DES Y3 \cite{DES:2020ebm} and LSST-like Y10 \cite{LSSTDarkEnergyScience:2018jkl} surveys. We then constructed the full projection formalism for the parity-odd angular trispectrum. By projecting established primordial curvature perturbation templates,specifically evaluating both squeezed and collapsed templates \cite{Coulton:2023oug,Shiraishi:2016mok}, we successfully mapped three-dimensional primordial non-Gaussianities into two-dimensional observable statistics.

A central insight from our SNR forecasts is the strong geometric dependence of the parity-odd observable. For both primordial templates considered in this work, the projected signal is largely dominated by angular configurations containing at least one low multipole mode. The detailed origin of this behavior is template dependent: the squeezed-type template is enhanced when one external multipole is small, whereas the collapsed-type template is particularly sensitive to configurations with a small diagonal multipole. Therefore, although both templates favor configurations involving long-wavelength modes, they do so through different shape dependences. Furthermore, the SNR is highly sensitive to the source galaxy distribution: auto-correlations within a single tomographic bin yield a substantially stronger signal than cross-tomographic combinations. While different primordial templates naturally favor different detailed geometries, our results demonstrate that the observable parity-odd signal exhibits a strong dependence on the underlying trispectrum configuration.


We emphasize that these detectability forecasts should still be regarded as idealized theoretical predictions, even though the galaxy shape noise is included. A comprehensive observational search must ultimately account for realistic survey systematics, such as masking effects and galaxy intrinsic alignments. Bridging the gap between these primitive theoretical limits and raw observational data remains a critical next step for future studies.

Ultimately, this work positions the weak lensing trispectrum as an independent link in the broader search for cosmological parity violation. By directly tracing the dark matter distribution, this approach circumvents galaxy bias and naturally complements existing constraints from both
CMB observables and 3D galaxy clustering. With upcoming Stage-IV observatories mapping the late-time universe with unprecedented volume and resolution, the theoretical framework developed here provides a robust, unbiased pathway for testing the fundamental symmetries of the early universe.

\appendix
\section{Fisher Forecast for the Template Amplitude}
\label[appendix]{app:parameter_constraints}

In this appendix, we present the Fisher forecast~\cite{Komatsu:2001rj} for the overall amplitude parameters for both parity-odd templates. Since the reduced trispectrum depends linearly on the template amplitudes, we parameterize it as
\begin{equation}
    Q = A Q_{\rm temp},
\end{equation}
where $Q_{\rm temp}$ denotes the fixed template shape and $A$ is the only free parameter. Following the Gaussian covariance approximation adopted in \Cref{sec: signal-to-noise estimation},  the Fisher matrix for the template amplitude $A$ is:
\begin{equation}
    F_{AA}=\sum_L{\sum_{\ell _1<\ell _2<\ell _3<\ell _4}{\left( 2L+1 \right) ^{-1}\frac{\left| \partial Q_{\ell _3\ell _4}^{\ell _1\ell _2}\left( L \right) /\partial A \right|^2}{C_{\ell _1}C_{\ell _2}C_{\ell _3}C_{\ell _4}}}},
\end{equation}
where the summation is performed over all unique configurations up to the chosen multipole cutoff, following the same ordering convention as \Cref{sec: signal-to-noise estimation}. The corresponding forecasted $1 \sigma$ uncertainty is 
\begin{equation}
    \sigma\left(A\right) = \frac{1}{\sqrt{F_{AA}}}. 
\end{equation}

\begin{figure}[htbp]
    \centering
    \begin{subfigure}[b]{0.495\textwidth}
        \centering
        \includegraphics[width=\textwidth]{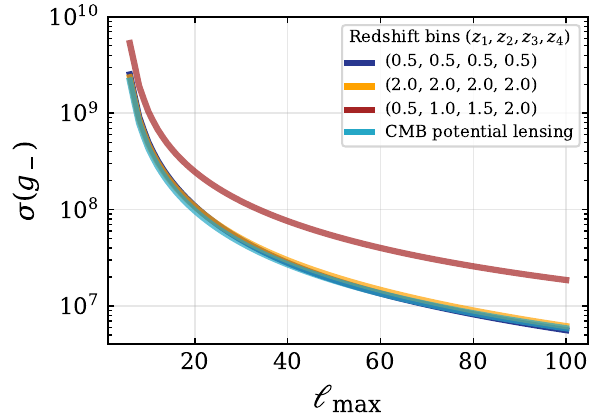}
        \caption{Squeezed-type template}
        \label{fig: app parameter constraints coulton template}
    \end{subfigure}
    \hfill
    \begin{subfigure}[b]{0.495\textwidth}
        \centering
        \includegraphics[width=\textwidth]{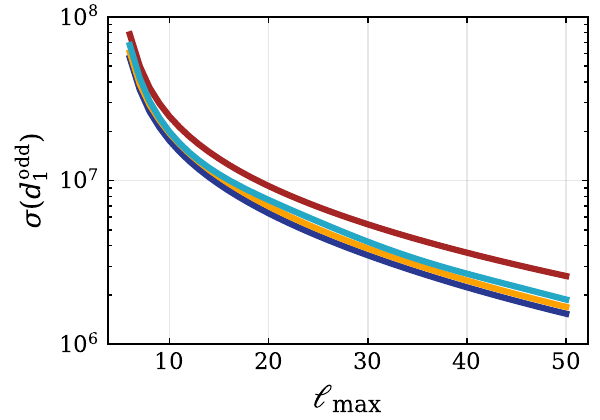}
        \caption{Collapsed-type template}
        \label{fig: app parameter constraints shiraishi template}
    \end{subfigure}
    \caption{
     Forecasted $1 \sigma$ uncertainty on the amplitude parameters of the squeezed-type (left) and collapsed-type (right) parity-odd templates as functions of the maximum multipole $\ell_{\max}$ using idealized Dirac delta galaxy distribution. Different curves correspond to different source-redshift configurations. 
    }
    \label{fig: app parameter constraints}
\end{figure}

\Cref{fig: app parameter constraints} presents the resulting forecasted $1\sigma$ uncertainties for the overall amplitude parameters of the squeezed- and collapsed-type templates. 


\section{Theoretical Calculation of the Template}
\label[appendix]{app:derivation}


In this appendix, we summarize the main mathematical steps leading from the primordial parity-odd trispectrum templates to the reduced angular convergence trispectrum. The derivation proceeds through harmonic space decomposition of both the momentum-conserving Dirac delta function and the angular dependent kernels. 

\subsection{Squeezed-Type Template Formalism}
\label[appendix]{app:derivation contact diagram}
We first introduce the plane-wave expansion, which provides the connection between Fourier and harmonic domains~\cite{Mehrem:2009ip}:
\begin{equation}\label{eq: plane-wave expansion}
    e^{i\boldsymbol{k}\cdot \chi ^{\prime}\boldsymbol{\hat{\theta} }}=4\pi \sum_{\ell ^{\prime}=0}^{\infty}{}\sum_{m^{\prime}=-\ell ^{\prime}}^{\ell ^{\prime}}{}i^{\ell ^{\prime}}j_{\ell ^{\prime}}\left( k\chi ^{\prime} \right) Y_{\ell ^{\prime}m^{\prime}}^{*}\left( \hat{\boldsymbol{k}} \right) Y_{\ell ^{\prime}m^{\prime}}\left( \boldsymbol{\hat{\theta} } \right),
\end{equation}
We then rewrite the Dirac delta function using its integral representation:
\begin{equation}
    \begin{split}
\delta _{D}^{\left( 3 \right)}\left( \boldsymbol{k}_{1234} \right) &=\int_{\boldsymbol{x}}{\exp \left[ i\boldsymbol{k}_{1234}\cdot \boldsymbol{x} \right]}
\\
&=32\pi \sum_{L_aM_a}^{}{\int_0^{\infty}{\mathrm{d}x\,x^2\prod_{n=1}^4{\biggl[ i^{L_n}j_{L_n}(k_nx)Y_{L_nM_n}(\hat{\boldsymbol{k}}_n) \biggr]}}}
\\
&\times \int{\mathrm{d}^2\hat{\boldsymbol{x}}\,\,Y_{L_1M_1}^{*}\left( \hat{\boldsymbol{x}} \right) Y_{L_2M_2}^{*}\left( \hat{\boldsymbol{x}} \right) Y_{L_3M_3}^{*}\left( \hat{\boldsymbol{x}} \right) Y_{L_4M_4}^{*}\left( \hat{\boldsymbol{x}} \right) ,}
    \end{split}
\end{equation}
where $a=1, 2, 3, 4$. To perform the integration of four spherical harmonics over $\hat{\boldsymbol{x}}$, we make use of the definition of the Gaunt integral and the addition of angular momenta theorem~\cite{Varshalovich:1988ifq}:
\begin{equation}\label{eq: left two spherical harmonics right gaunt and spherical harmonics}
    Y_{L_1M_1}^{*}\left( \hat{\boldsymbol{x}} \right) Y_{L_2M_2}^{*}\left( \hat{\boldsymbol{x}} \right) =\sum_{L^{\prime}M^{\prime}}{\mathcal{G} _{L_1L_2L^{\prime}}^{M_1M_2M^{\prime}}Y_{L^{\prime}M^{\prime}}}\left( \hat{\boldsymbol{x}} \right) ,
\end{equation}
where $L^{\prime}$ and $M^{\prime}$ are total angular momentum and magnetic number respectively. And the Gaunt coefficient $\mathcal{G}$ is defined as~\cite{Varshalovich:1988ifq}:
\begin{equation}\label{eq: gaunt integral}
    \begin{split}
        \mathcal{G} _{L_1L_2L_3}^{M_1M_2M_3}&\equiv \int{\mathrm{d}^2\hat{\boldsymbol{x}}\,\,Y_{L_1M_1}\left( \hat{\boldsymbol{x}} \right) Y_{L_2M_3}\left( \hat{\boldsymbol{x}} \right) Y_{L_3M_3}\left( \hat{\boldsymbol{x}} \right)}
\\
&=\sqrt{\frac{\left( 2L_1+1 \right) \left( 2L_2+1 \right) \left( 2L_3+1 \right)}{4\pi}}\left( \begin{matrix}
	L_1&		L_2&		L_3\\
	0&		0&		0\\
\end{matrix} \right) \left( \begin{matrix}
	L_1&		L_2&		L_3\\
	M_1&		M_2&		M_3\\
\end{matrix} \right) .
    \end{split}
\end{equation}

The integration over the four spherical harmonics decomposes the quadrilateral configuration into two coupled triangles characterized by the angular momenta $\left\{L_1, L_2, L^{\prime}\right\}$ and $\left\{L_3, L_4, L^{\prime}\right\}$ respectively, with $L^{\prime}$ acting as the diagonal length (see Ref.~\cite{Cahn:2020axu} for a detailed derivation): 
\begin{equation}
    \begin{split}
        \mathcal{G} _{L_1L_2L_3L_4}^{M_1M_2M_3M_4}&=\int{\mathrm{d}^2\hat{\boldsymbol{x}}\,\,Y_{L_1M_1}^{*}\left( \hat{\boldsymbol{x}} \right) Y_{L_2M_2}^{*}\left( \hat{\boldsymbol{x}} \right) Y_{L_3M_3}^{*}\left( \hat{\boldsymbol{x}} \right) Y_{L_4M_4}^{*}\left( \hat{\boldsymbol{x}} \right)}
\\
&=\sum_{L^{\prime}M^{\prime}}{\left( -1 \right) ^{M^{\prime}}\mathcal{G} _{L_1L_2L^{\prime}}^{M_1M_2-M^{\prime}}\mathcal{G} _{L_3L_4L^{\prime}}^{M_3M_4M^{\prime}}} .
    \end{split}
\end{equation}

Since direct integration of the triple product is computationally expensive, we adopt an expansion into an isotropic basis function~\cite{Cahn:2020axu}. This formalism exploits the rotational invariance of the field by decomposing the angular dependence into spherical harmonics~\cite{Newman:1966ub} coupled through Wigner $3$-$j$ symbols~\cite{1927ZPhy...43..624W, Eckart:1930utp}:
\begin{equation}\label{eq: triple product and isotropic expansion}
    \begin{split}
        \hat{\boldsymbol{k}}_1\cdot \left( \hat{\boldsymbol{k}}_2\times \hat{\boldsymbol{k}}_3 \right) =-i\sqrt{6}\left( \frac{4\pi}{3} \right) ^{3/2}\sum_{m_{1}^{\prime}m_{2}^{\prime}m_{2}^{\prime}}{Y_{1m_{1}^{\prime}}\left( \hat{\boldsymbol{k}}_1 \right)}Y_{1m_{2}^{\prime}}\left( \hat{\boldsymbol{k}}_2 \right) Y_{1m_{3}^{\prime}}\left( \hat{\boldsymbol{k}}_3 \right) \left( \begin{matrix}
	1&		1&		1\\
	m_{1}^{\prime}&		m_{2}^{\prime}&		m_{3}^{\prime}\\
\end{matrix} \right) .
    \end{split}
\end{equation}

Substituting the harmonic expansions into the trispectrum expression, performing the angular integrations, and applying the reduced trispectrum definition introduced in Eq.~\eqref{eq: reduced trispectrum and full trispectrum}, we obtain:
\begin{equation}
    \begin{split}
        {\mathcal{Q} _{\ell _3\ell _4}^{\ell _1\ell _2}}_{\pi}^{(s)}\left( L \right) &=\left( 2L+1 \right) \times i^{\ell _1+\ell _2+\ell _3}(-1)^{\ell _4}\times \left[ \frac{4}{5\pi \Omega _{m,0}H_{0}^{2}} \right] ^4
\\
&\times \sum_{L_1L_2L_3}{}\sum_{L^{\prime}}{}\mathcal{F} _{L_1L_2L^{\prime}}\mathcal{F} _{L_3\ell _4L^{\prime}}\mathcal{F} _{L_11\ell _1}\mathcal{F} _{L_21\ell _2}\mathcal{F} _{L_31\ell _3}\times i^{L_1+L_2+L_3}
\\
&\times \sum_{M_1M_2M_3}{}\sum_{M^{\prime}}{}\sum_{m_{1}^{\prime}m_{2}^{\prime}m_{3}^{\prime}}{}\sum_{m_1m_2m_3m_4}{}\sum_M{\left( -1 \right) ^{M+M^{\prime}+m_1+m_2+m_3}}
\\
&\times \left( \begin{matrix}
	\ell _1&		\ell _2&		L\\
	m_1&		m_2&		M\\
\end{matrix} \right) \left( \begin{matrix}
	\ell _3&		\ell _4&		L\\
	m_3&		m_4&		-M\\
\end{matrix} \right) \left( \begin{matrix}
	L_1&		L_2&		L^{\prime}\\
	M_1&		M_2&		-M^{\prime}\\
\end{matrix} \right) \left( \begin{matrix}
	L_3&		\ell _4&		L^{\prime}\\
	M_3&		m_4&		M^{\prime}\\
\end{matrix} \right) 
\\
&\times \left( \begin{matrix}
	L_1&		1&		\ell _1\\
	M_1&		m_{1}^{\prime}&		-m_1\\
\end{matrix} \right) \left( \begin{matrix}
	L_2&		1&		\ell _2\\
	M_2&		m_{2}^{\prime}&		-m_2\\
\end{matrix} \right) \left( \begin{matrix}
	L_3&		1&		\ell _3\\
	M_3&		m_{3}^{\prime}&		-m_3\\
\end{matrix} \right) \left( \begin{matrix}
	1&		1&		1\\
	m_{1}^{\prime}&		m_{2}^{\prime}&		m_{3}^{\prime}\\
\end{matrix} \right) 
\\
&\times \int{\mathrm{d}x\,\,x^2}\prod_{n=1}^4{\left[ \int{\mathrm{d}k_n}\int_0^{\chi _H}{\mathrm{d}\chi _{n}^{\prime}}\,\,q\left( \chi _{n}^{\prime} \right) D\left( \chi _{n}^{\prime} \right) k_{n}^{4}\mathcal{T} _{\delta}\left( k_n \right) j_{\ell _n}\left( k_n\chi _{n}^{\prime} \right) j_{L_n}(k_nx) \right]}
\\
&\times \tau _{\pi}^{(s)}\left( k_1,k_2,k_3,k_4 \right) 
    \end{split}
\end{equation}
where the superscript $(s)$ and subscript $\pi$ denote the squeezed model and a single permutation among all the $24$ permutations. In the product, we set $L_4=\ell_4$ so that the fourth leg is written in the same form as the first three legs. 

To perform the summation over the magnetic quantum numbers $m$, we employ the \textsc{Python} package \texttt{Reduce3j}~\cite{Xiang:2021mzd}, which reduces products of Wigner $3$-$j$ symbols summed over magnetic indices into expressions involving Wigner $6$-$j$ and $9$-$j$ symbols. This procedure yields the compact expression presented in Eq.~\eqref{eq: contact geometric part}.

\subsection{Collapsed-Type Template Formalism}
\label[appendix]{app:derivation exchange diagram}
Unlike the squeezed template, the collapsed template contains an additional internal diagonal mode $\boldsymbol{K}$. To facilitate the angular projection, we first decompose the Legendre polynomial into spherical harmonics using the addition theorem, 
\begin{equation}\label{eq: Legendre polynomial}
    \mathcal{L}_L\left( \hat{\boldsymbol{k}}\cdot \hat{\boldsymbol{k}}^{\prime} \right) =\frac{4\pi}{2L+1}\sum_M{Y_{LM}^{*}\left( \hat{\boldsymbol{k}} \right) Y_{LM}\left( \hat{\boldsymbol{k}}^{\prime} \right)}.
\end{equation}

As indicated in Eq.~\eqref{eq: shiraishi full split}, the Dirac delta function effectively split the quadrilateral configuration into two triangles sharing the common diagonal $\boldsymbol{K}$ in Fourier space. To evaluate the angular dependence explicitly, we apply the same plane-wave expansion introduced in Eq.~\eqref{eq: plane-wave expansion} to both delta functions. 

The projection introduces two independent sets of expansion indices for the spatial integrals over $\boldsymbol{x}$ and $\boldsymbol{y}$. We distinguish the angular momenta associated with the diagonal mode $\boldsymbol{K}$ as $L_{K_+}$ (arising from the first delta function) and $L_{K_-}$ (from the second). Consequently, the full Dirac delta term can be recast in the following compact harmonic form: 
\begin{align}\label{eq: shiraishi template dirac delta expansion}
    \begin{split}
	\delta _{D}^{\left( 3 \right)}\left( \boldsymbol{k}_{1234} \right) &=\int{}\mathrm{d}^3\boldsymbol{K}\,\,\delta _{D}^{\left( 3 \right)}\left( \boldsymbol{k}_{12}+\boldsymbol{K} \right) \delta _{D}^{\left( 3 \right)}\left( \boldsymbol{k}_{34}-\boldsymbol{K} \right) 
\\
&=64\int{}\mathrm{d}^3 \boldsymbol{K}\int{}\mathrm{d}x\,\,x^2\int{}\mathrm{d}y\,\,y^2
\\
&\times \sum_{L_1L_2L_{K_+}}{}\sum_{L_3L_4L_{K_-}}{}\sum_{M_1M_2M_{K_+}}{}\sum_{M_3M_4M_{K_-}}{}
\\
&\times \left( -1 \right) ^{L_{K_-}}\mathcal{G} _{L_1L_2L_{K_+}}^{M_1M_2M_{K_+}}\mathcal{G} _{L_3L_4L_{K_-}}^{M_3M_4M_{K_-}}
\\
&\times \prod_{p=1,2,K_+}{}\left[ i^{L_p}j_{L_p}\left( k_px \right) Y_{L_pM_p}^{*}\left( \hat{\boldsymbol{k}}_p \right) \right] 
\\
&\times \prod_{q=3,4,K_-}{}\left[ i^{L_q}j_{L_q}\left( k_qy \right) Y_{L_qM_q}^{*}\left( \hat{\boldsymbol{k}}_q \right) \right] ,
    \end{split}
\end{align}
where the indices $p$ and $q$ run over the sets $\{1, 2, K_+ \}$ and $\{3, 4, K_-\}$ respectively.

With all the ingredients prepared, we can express the reduced angular trispectrun for collapsed model as:
\begin{equation}
    \begin{split}
        {\mathcal{Q} _{\ell _3\ell _4}^{\ell _1\ell _2}}_{\pi}^{\left( c \right)}\left( L \right) &=\left( 2L+1 \right) \times i^{\ell _1+\ell _2+\ell _3+\ell _4}\left[ \frac{2}{\pi} \right] ^2\left[ \frac{2}{5}\frac{1}{\Omega _{m,0}H_{0}^{2}} \right] ^4
\\
&\times \sum_M{}\sum_{m_1m_2m_3m_4}{}\sum_{L_1L_{K_+}}{}\sum_{L_3L_{K_-}}{}\sum_{M_1M_{K_+}}{}\sum_{M_3M_{K_-}}{}\sum_{L_{1}^{\prime}L_{3}^{\prime}L_K}{}\sum_{M_{1}^{\prime}M_{3}^{\prime}M_K}{}\sum_n{}d_{n}^{odd}
\\
&\times G_{L_{1}^{\prime}L_{3}^{\prime}L_K}^{n}\mathcal{F} _{L_1\ell _2L_{K_+}}\mathcal{F} _{L_3\ell _4L_{K_-}}\mathcal{F} _{L_1\ell _1L_{1}^{\prime}}\mathcal{F} _{L_3\ell _3L_{3}^{\prime}}\mathcal{F} _{L_{K_+}L_{K_-}L_K}
\\
&\times \left( \begin{matrix}
	\ell _1&		\ell _2&		L\\
	m_1&		m_2&		M\\
\end{matrix} \right) \left( \begin{matrix}
	\ell _3&		\ell _4&		L\\
	m_3&		m_4&		-M\\
\end{matrix} \right) \left( \begin{matrix}
	L_1&		\ell _2&		L_{K_+}\\
	M_1&		-m_2&		M_{K_+}\\
\end{matrix} \right) \left( \begin{matrix}
	L_3&		\ell _4&		L_{K_-}\\
	M_3&		-m_4&		M_{K_-}\\
\end{matrix} \right) 
\\
&\times \left( \begin{matrix}
	L_{1}^{\prime}&		L_{3}^{\prime}&		L_K\\
	M_{1}^{\prime}&		M_{3}^{\prime}&		M_K\\
\end{matrix} \right) \left( \begin{matrix}
	L_1&		\ell _1&		L_{1}^{\prime}\\
	-M_1&		-m_1&		-M_{1}^{\prime}\\
\end{matrix} \right) \left( \begin{matrix}
	L_3&		\ell _3&		L_{3}^{\prime}\\
	-M_3&		-m_3&		-M_{3}^{\prime}\\
\end{matrix} \right) \left( \begin{matrix}
	L_{K_+}&		L_{K_-}&		L_K\\
	-M_{K_+}&		-M_{K_-}&		-M_K\\
\end{matrix} \right) 
\\
&\times \left( -1 \right) ^{m_1+m_2+m_3+m_4+M_1+M_{1}^{\prime}+M_3+M_{3}^{\prime}+M_{K_+}+M_{K_-}+M_K+M+L_{K_-}}\times i^{L_1+\ell _2+L_3+\ell _4+L_{K_+}+L_{K_-}}
\\
&\times \int{}\mathrm{d}KK^2\int{}\mathrm{d}x\,\,x^2\int{}\mathrm{d}y\,\,y^2\prod_{n=1}^4{}\left[ \int{\mathrm{d}k_n}\int_0^{\chi _H}{\mathrm{d}\chi _{n}^{\prime}\,\,}q\left( \chi _{n}^{\prime} \right) k_{n}^{4}j_{\ell _n}\left( k_n\chi _{n}^{\prime} \right) D\left( \chi _n \right) \mathcal{T} _{\delta}\left( k_n \right) \right] 
\\
&\times j_{L_1}\left( k_1x \right) j_{\ell _2}\left( k_2x \right) j_{L_{K_+}}\left( Kx \right) j_{L_3}\left( k_3y \right) j_{\ell _4}\left( k_4y \right) j_{L_{K_-}}\left( Ky \right) 
\\
&\times P_{\xi}\left( k_1 \right) P_{\xi}\left( k_3 \right) P_{\xi}\left( K \right) 
    \end{split}
\end{equation}
where the superscript $(c)$ denotes the collapsed template, and the subscript $\pi$ labels a representative permutation of the external momenta. Again, \texttt{Reduce3j} is used to calculation the summation over $m$ and simplify the angular part into Eq.~\eqref{eq: shiraishi geometry before simplification}.

\section{Limber Approximation}
\label[appendix]{app:limber_derivation}

In this appendix, we derive the expressions for both the squeezed and collapsed templates under Limber approximation. The approximation is implemented through
\begin{equation}\label{eq: limber approximation}
    j_{\ell} \left(k\chi\right) \rightarrow \sqrt{\frac{\pi }{2\ell}} \frac{1}{\chi}\delta _{D}^{\left(1\right)} \left(k - \frac{\ell}{\chi}\right). 
\end{equation}
Under Limber approximation, angular power spectrum is written as:
\begin{equation}\label{eq: angular power spectrum under limber approximation}
    C_{\ell}\approx \frac{4\ell ^4}{25\Omega _{m,0}^{2}H_{0}^{4}}\int_0^{\chi _H}{}\,\,\frac{\mathrm{d}\chi}{\chi ^6}q^2\left( \chi \right) D^2\left( \chi \right) \mathcal{T} ^2\left( \frac{\ell}{\chi} \right) P_{\mathcal{R}}\left( \frac{\ell}{\chi} \right) .
\end{equation}

\subsection{Squeezed-Type Template Limber Expression}
Applying Eq.~\eqref{eq: limber approximation} to the squeezed-type template expression in Eq.~\eqref{eq: coulton template full equation}, the correspondence between multipoles and wavenumbers becomes:
\begin{align}
    \begin{cases}
	k_i\simeq \frac{L_i}{x}\simeq \frac{\ell _i}{\chi _{i}^{\prime}}&		\mathrm{for} \,\, i=1,2,3,\\
	k_4\simeq \frac{\ell _4}{x}\simeq \frac{\ell _4}{\chi _{4}^{\prime}}.&		\\
\end{cases}
\end{align}
The Gaunt coefficients in Eq.~\eqref{eq: contact geometric part} impose the triangle conditions
\begin{equation}
    \left| L_i - \ell_i \right| \le 1 \le L_i + \ell_i \quad \quad \text{for} \quad i=1,2,3 .
\end{equation}
Since $L_n$ and $\ell_n$ differ by at most unity, we can validly approximate $L_n \simeq \ell_n$. Under the Limber approximation, the relationship between the comoving wavenumbers and radial distances simplifies to: 
\begin{equation}
    k_i\simeq \frac{\ell _i}{x}, \; \; x\simeq \chi _{i}^{\prime}. 
\end{equation}
Then the line-of-sight integration takes the form:
\begin{align}\label{eq: coulton template projection part under limber approximation}
    \begin{split}
         {\mathcal{I}_{\ell_1, \ell_2, \ell_3, \ell_4}}^{(c)}_{\pi} = \left[ \frac{2}{5}\frac{1}{\Omega _{m,0}H_{0}^{2}} \right] ^4\int_0^{\infty}{\frac{\mathrm{d}x}{x^{14}}}\prod_{i=1}^4{\left[ q_i\left( x \right) D\left( x \right) \mathcal{T} \left( \frac{\ell _i}{x} \right) \ell _{i}^{2} \right] \tau^{(c)} _{\pi}\left( \frac{\ell _1}{x},\frac{\ell _2}{x},\frac{\ell _3}{x},\frac{\ell _4}{x} \right)} .
    \end{split}
\end{align}

\subsection{Collapsed-Type Template Limber Expression}
Applying Eq.~\eqref{eq: limber approximation} to the collapsed template expression shown in Eq.~\eqref{eq: shiraishi full trispectrum before Limber} yields
\begin{align}
    \begin{cases}
	k_i\simeq \frac{\ell _i}{\chi _{n}^{\prime}}&		\mathrm{for}\,\,i=1,2,3,4\\
	k_i\simeq \frac{L_i}{x}&		\mathrm{for} \,\, i=1, 2, K_+\\
	k_i\simeq \frac{L_i}{y}&		\mathrm{for} \,\, i=3, 4, K_- .\\
\end{cases}
\end{align}
Furthermore, inspecting the triangle inequalities in the Wigner 3-$j$ symbols of Eq.~\eqref{eq: shiraishi geometry before simplification}, we note that the auxiliary momenta $L_1^{\prime}, L_3^{\prime}, L_K$ are restricted to values near unity ($\left|n\pm1\right|$ or $1$). Since the non-vanishing geometric terms require $n \in \{0, 1\} $, we are justified in making the identifications:
\begin{equation}
    L_i \approx \ell_i \quad (i=1,3), \qquad L \approx L_{K_+} \approx L_{K_-}, \qquad x \approx y.
\end{equation}
Then the radial integration simplifies into:
\begin{align}\label{eq: exchange diagram projection limber}
    \begin{split}
        {\mathcal{I} _{\ell _3\ell _4}^{\ell _1\ell _2}}_{\pi}^{\left( e \right)}\left( L \right) &=\left[ \frac{2}{5\Omega _{m,0}H_{0}^{2}} \right] ^4\times \int_0^{\chi _H}{}\,\,\frac{\mathrm{d}x}{x^{14}}\prod_{i=1}^4{}\left[ q_i\left( x \right) D\left( x \right) \mathcal{T} _{\delta}\left( \frac{\ell _i}{x} \right) \ell _{i}^{2} \right] 
        \\
        &\times P_{\mathcal{R}}\left( \frac{\ell _1}{x} \right) P_{\mathcal{R}}\left( \frac{\ell _3}{x} \right) P_{\mathcal{R}}\left( \frac{L}{x} \right) .
    \end{split}
\end{align}

\section{Factorized Projection and Limber Validation}
\label[appendix]{app:limber approximation}

In this appendix, we present the numerical pipeline used to evaluate the high-dimensional projection integrals entering the weak lensing trispectrum. We also assess the accuracy of the Limber approximation, which is widely employed in two-point statistics due to its ability to reduce multi-dimensional projection integrals to one-dimensional integrations, as derived in \Cref{app:limber_derivation}. 

The main conclusion of this appendix is that, while the standard Limber approximation performs well for the angular trispectrum when \textit{all} constituent multipoles are large, it can break down when at least one multipole is small. This is particularly important for the parity-odd signal studied in this work. As shown in \Cref{sec: numerical result}, the SNR is dominated by geometric configurations involving low-$\ell$ modes, such as squeezed- and collapsed-limit configurations, where the projected observable retains sensitivity to the three-dimensional parity-violating information. Therefore, although the Limber approximation greatly simplifies the computation, it is not sufficiently accurate for our theoretical SNR forecasts, and the exact FFTLog numerical integration pipeline developed in this work is required.

\begin{figure}[htbp]
    \centering

    \begin{subfigure}[b]{0.495\textwidth}
        \centering
        \includegraphics[width=\textwidth]{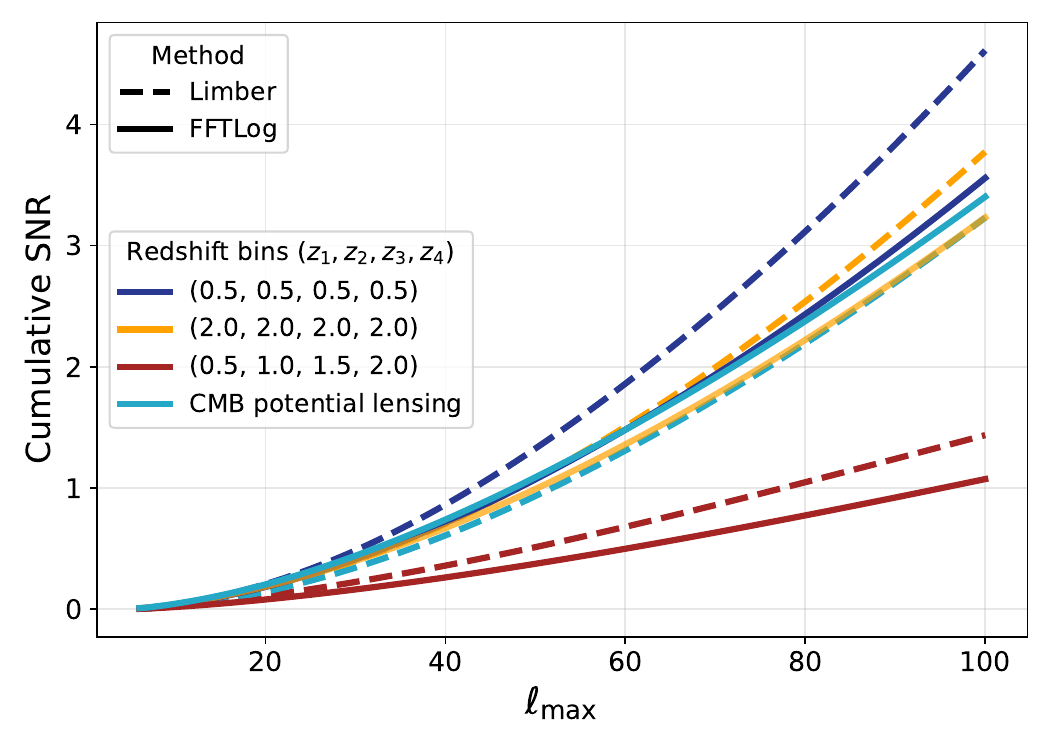}
        \caption{Squeezed-Type Template}
        \label{fig: coulton template SNR}
    \end{subfigure}
    \hfill
    \begin{subfigure}[b]{0.495\textwidth}
        \centering
        \includegraphics[width=\textwidth]{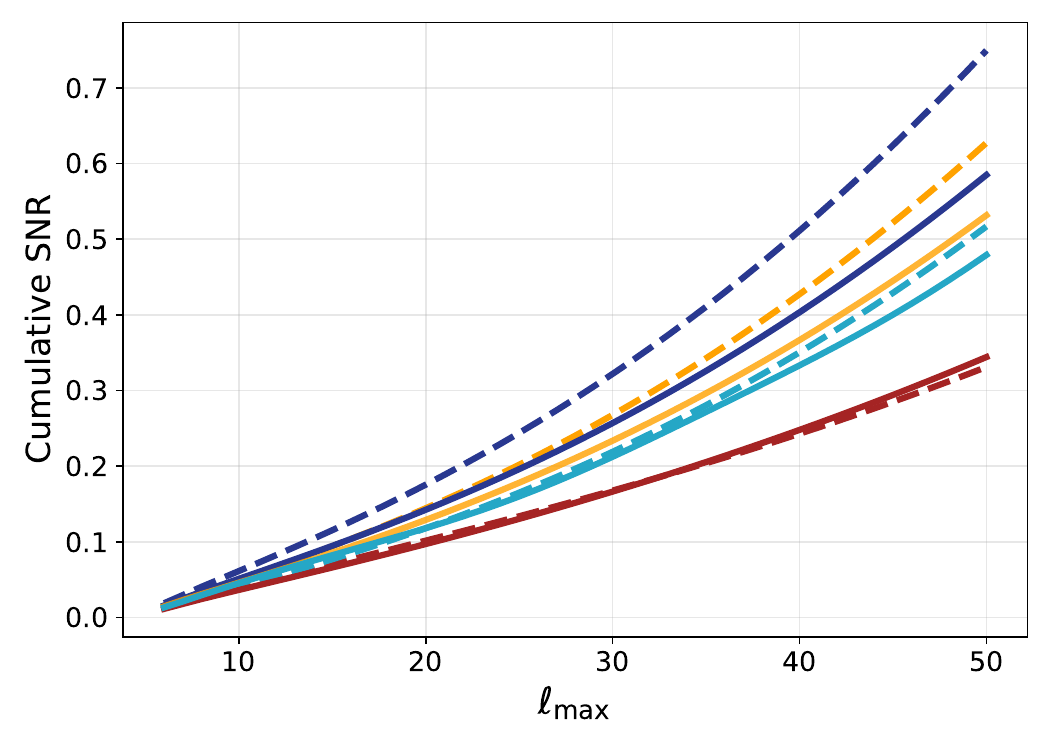}
        \caption{Collapsed-Type Template}
        \label{fig: shiraishi template SNR}
    \end{subfigure}

    \caption{Cumulative SNR of both squeezed and collapsed templates as a function of the maximum multipole $\ell_{\max}$ using Dirac delta redshift distribution. Different colors correspond to the source redshift combinations indicated in the legend. Dashed and solid curves show the Limber approximation and exact FFTLog evaluation, respectively. The cyan curves denote the CMB potential lensing case. No galaxy shape noise is included.}
    \label{fig: contact and exchange SNR fftlog and limber comparison}
\end{figure}

To validate our numerical implementation, we compare the FFTLog-based integration against independent calculations performed with \texttt{Mathematica 14.0}~\cite{Mathematica}. \Cref{fig: contact and exchange SNR fftlog and limber comparison} presents the cumulative SNR obtained using both the Limber approximation and the exact FFTLog integration, providing an overview of the discrepancies that motivate the analysis presented in this appendix.

\subsection{Factorized Integration Formalism}
The trispectrum projection involves a product of multiple Bessel functions coupled by the auxiliary variable $x$. To isolate the integration challenges, we utilize a factorizable method \cite{Smith:2015uia} to perform the integration, decomposing the high-dimensional radial integrations into a series of one-dimensional projections. The integration over the auxiliary variable couples the four fields, while the integrals over $\chi_i ^{\prime}$ and $k_i$ can be treated independently. We define the foundational single-point projection integral as:
\begin{equation}\label{eq:app single kernel}
    \mathcal{S} _{\ell ,n}\left( x \right) \equiv \int{\mathrm{d}k \ k^{4+n}\mathcal{T} \left( k \right) j_{\ell}\left( kx \right)}\int{\mathrm{d}\chi ^{\prime} \ q\left( \chi ^{\prime} \right) D\left( \chi ^{\prime} \right) j_{\ell}\left( k\chi ^{\prime} \right)} ,
\end{equation}
where $n$ represents the power-law index derived from our spectrum templates. 

Usually, the convention for integrating with spherical Bessel functions is to convert them into cylindrical Bessel functions using the relation:
\begin{equation}
    j_{\ell}(x) = \sqrt{\frac{\pi}{2x}} J_{\nu} (x),
\end{equation}
where strictly $\nu = \ell+0.5$. However, when performing the actual computation, we set $\nu = \ell$ to maintain a fair and direct comparison with the standard Limber approximation. We first perform the integration over $\chi^{\prime}$ to eliminate one spherical Bessel function, defining the intermediate radial window function:
\begin{equation}
    W_{\ell} (k) \equiv \sqrt{\frac{\pi}{2k}} \int \mathrm{d}\chi^{\prime} \ \frac{q(\chi^\prime) D(\chi^{\prime})}{\sqrt{\chi^{\prime}}} J_{\ell} (k \chi^{\prime}).
\end{equation}
We evaluate this integral utilizing the Python package \texttt{hankel}  \cite{2019JOSS....4.1397M} to efficiently perform the Hankel transformation. The subsequent integrand over $k$ then becomes:
\begin{equation}
    \mathcal{S}_{\ell, n}(x) = \sqrt{\frac{\pi}{2x}} \int \mathrm{d}k \ k^{3.5+n} \mathcal{T}(k) W_{\ell}(k) J_{\ell}(kx).
\end{equation}
We test and quantify the accuracy of the Limber approximation compared to the numerical integration result. 

\subsection{Factorized Angular Power Spectrum}

Before proceeding to the full trispectrum integration, we first examine the angular power spectrum case to establish a theoretical baseline. Expanding the Dirac delta function allows us to express the power spectrum projection as the direct product of two single-point kernels. Factoring out the constants from the primordial power spectrum, we obtain
\begin{align}\label{eq: appendix factorization angular power spectrum}
    \left<\kappa_{\ell_1 m_1}^{*} \kappa_{\ell_2 m_2} \right> \propto \delta^{K}_{\ell_1 \ell_2} \delta^{K}_{m_1 m_2} \int \mathrm{d} x \ x^2 \mathcal{S}_{\ell_1, n_1}(x) \mathcal{S}_{\ell_2, n_2}(x). 
\end{align}
where we have the product of two single-point projections for the angular power spectrum. Here, we have $n_1 = -3$ derived from the primordial power spectrum and $n_2 = 0$. These two indices are completely interchangeable, as the primordial power spectrum can be placed symmetrically in either the $k_1$ or $k_2$ integration. 

\begin{figure}[htbp]
    \centering
    \includegraphics[width=1.0\linewidth]{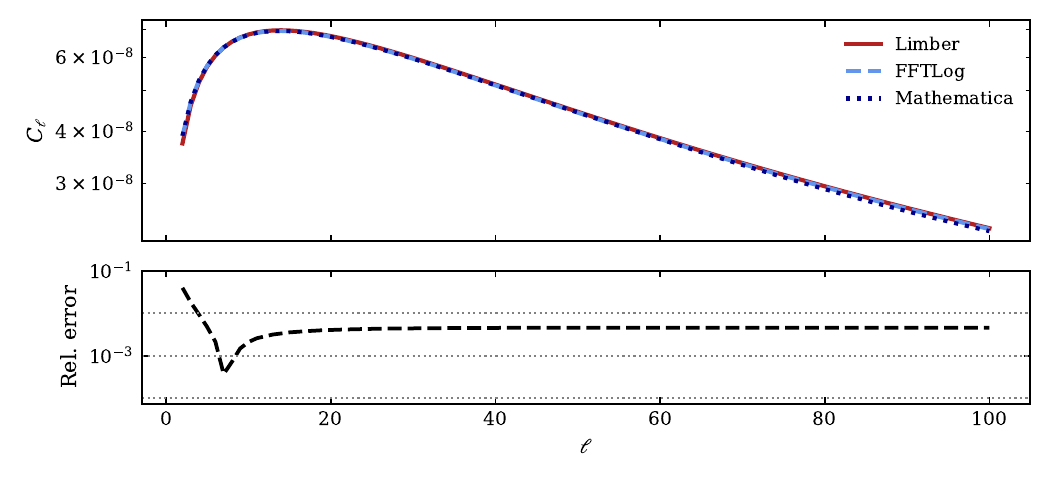}
    \caption{
    Validation of the FFTLog integration pipeline using the angular power spectrum. The Limber approximation is compared with the exact FFTLog result, and the Mathematica is included as an independent cross-check. The lower panel shows the relative Limber error. 
    }
    \label{fig: angular power spectrum limber accuracy comparison}
\end{figure}

We define the theoretical systematic error introduced by the Limber approximation as the relative difference against the exact FFTLog result, shown in \Cref{fig: angular power spectrum limber accuracy comparison}. As expected from the known behavior of the Limber approximation for two-point statistics, the approximation performs very well for the configurations with large multipole numbers. The relative error smoothly decays to $1\%$ at $\ell \ge 10$, and reaches the baseline numerical convergence level at higher multipoles. This clean, converging profile acts as a sanity check, confirming the numerical stability of our exact FFTLog pipeline.

\subsection{Limber Approximation on Squeezed-Type Template}
\label[appendix]{app:limber approximation contact diagram}

Similarly, we can write down the projection part of the weak lensing trispectrum for the squeezed-type template in a factorized format:
\begin{equation}\label{eq: appendix factorization contact trispectrum}
    {\mathcal{I}^{(s)}_{\ell_1, \ell_2, \ell_3, \ell_4}}_{\pi} \propto \int{\mathrm{d}x\,x^2\,}\mathcal{S} _{\ell _1,n_1}\left( x \right) \mathcal{S} _{\ell _2,n_2}\left( x \right) \mathcal{S} _{\ell _3,n_3}\left( x \right) \mathcal{S} _{\ell _4,n_4}\left( x \right) ,
\end{equation} 
where the multipoles $\{\ell_1, \ell_2, \ell_3, \ell_4\}$ must form a closed geometric configuration to ensure rotational invariance. Crucially, the power indices $\{n_1, n_2, n_3, n_4 \}$ must form a strict permutation of the set $\{-4, -3, -2, 0\}$. Each branch uniquely takes on one of these distinct values; we cannot assign the same power index to multipole branches within the squeezed-type template. 

What makes the trispectrum fundamentally different from the power spectrum is that we do not require the four multipoles to be equal. This flexibility means that some multipoles within a given configuration can be very small while the rest are very large. There are many configurations that feature one or two small side lengths, and these specific geometric configurations lead directly to the breakdown of the Limber approximation.

Because the Limber approximation performs poorly when $\ell$ is very small, these short side lengths yield poorly approximated $\mathcal{S}_{\ell, n}$ kernels. In the context of the full trispectrum integration, when these poorly evaluated kernels are multiplied with large-multipole $\mathcal{S}_{\ell, n}$ kernels, which are well approximated by Limber, the bad approximation systematically infects the final line-of-sight integration, causing large systematic errors.

\begin{figure}[htbp]
    \centering
    \begin{subfigure}[b]{0.495\textwidth}
        \centering
        \includegraphics[width=\textwidth]{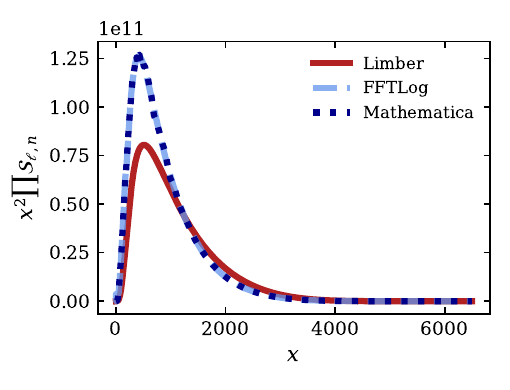}
        \caption{All short legs: $\ell = \left[2, 3, 4, 6\right]$}
        \label{fig: appendix limber mathematic different configuration four small ell one}
    \end{subfigure}
    \hfill
    \begin{subfigure}[b]{0.495\textwidth}
        \centering
        \includegraphics[width=\textwidth]{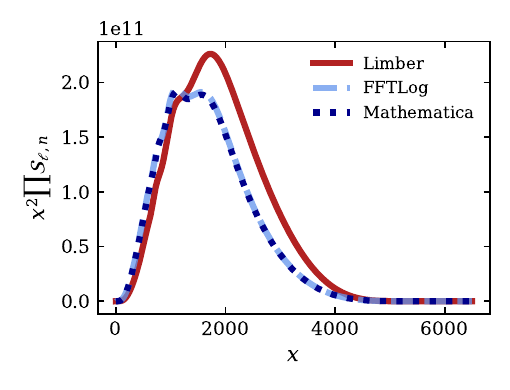}
        \caption{Two short legs: $\ell=\left[2, 3, 96, 98\right]$}
        \label{fig: appendix limber mathematic different configuration four small ell two}
    \end{subfigure}
    
    \vspace{0.5cm}
    
    \begin{subfigure}[b]{0.495\textwidth}
         \centering
        \includegraphics[width=\textwidth]{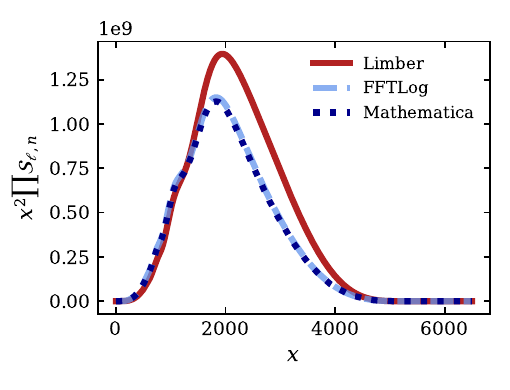}
        \caption{One short leg: $\ell = \left[2, 95, 96, 98\right]$}
        \label{fig: appendix limber mathematic different configuration squeezed limit}
    \end{subfigure}
    \hfill
    \begin{subfigure}[b]{0.495\textwidth}
        \centering
        \includegraphics[width=\textwidth]{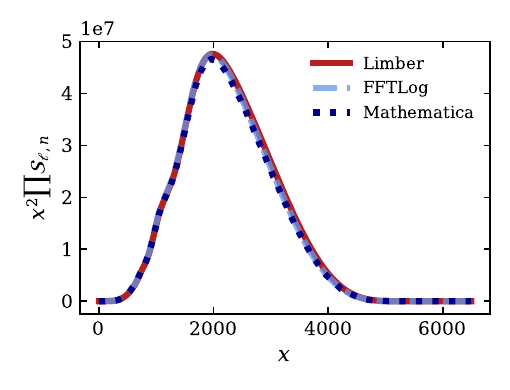}
        \caption{All long legs: $\ell = \left[10, 95, 96, 98\right]$}
        \label{fig: appendix limber mathematic different configuration four large ell one}
    \end{subfigure}
    \caption{
    Comparison of the radial projection integrand for the squeezed template evaluated using the Limber approximation and exact FFTLog method. All source galaxies are placed at redshift of $2.0$, with power indices assigned as $n=[-4, -3, -2, 0]$ following the multipole ordering shown in each panel. Panels (a-d) illustrate representative geometric configurations from low-$\ell$ to high-$\ell$. Significant discrepancies appear whenever at least one multipole lies in the low-$\ell$ regime, while excellent agreement is recovered when all multipoles are large. 
    } 
    \label{fig: appendix limber mathematic different configuration redshift all 2.0}
\end{figure}

\begin{figure}[htbp]
    \centering
    \begin{subfigure}[b]{0.75\textwidth}
        \centering
        \includegraphics[width=\textwidth]{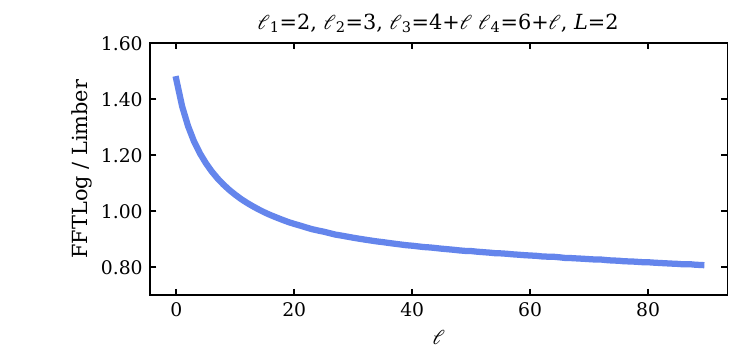}
        \caption{Fix two short legs}
        \label{fig: appendix limber FFTLog different configuration two small ell}
    \end{subfigure}
    \vfill
    \begin{subfigure}[b]{0.75\textwidth}
        \centering
        \includegraphics[width=\textwidth]{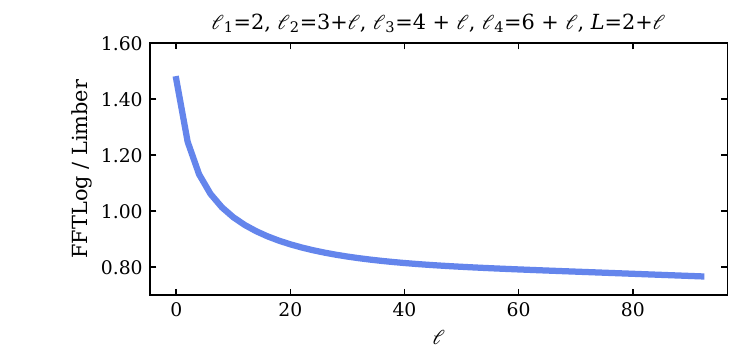}
        \caption{Fix one short leg}
        \label{fig: appendix limber FFTLog different configuration one small ell}
    \end{subfigure}
    \vfill
    \begin{subfigure}[b]{0.75\textwidth}
        \centering
        \includegraphics[width=\textwidth]{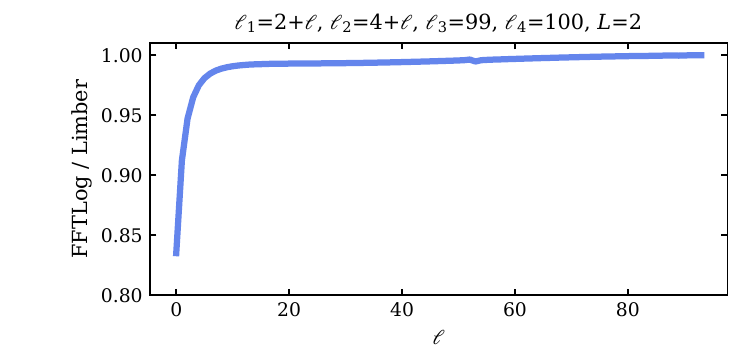}
        \caption{Fix two long legs}
        \label{fig: appendix limber FFTLog different configuration two large ell}
    \end{subfigure}
    \caption{
    Ratio of the reduced trispectrum evaluated using FFTLog and the Limber approximation for representative squeezed-type template configurations. Panels (a-c) correspond to configurations with two, one, and zero fixed low-$\ell$ external lengths, respectively. The discrepancy decreases as the configuration becomes increasingly dominated by large multipoles. The diagonal $L$ does not directly affect the Limber error in the squeezed model since it enters only through the angular coupling terms. 
    }
    \label{fig: appendix limber FFTLog different configuration}
\end{figure}

\Cref{fig: appendix limber mathematic different configuration redshift all 2.0} demonstrates this effect by comparing the full integrand evaluated using the Limber approximation against the exact FFTLog and Mathematica results. We isolate four distinct geometric cases:
\begin{itemize}
    \item \Cref{fig: appendix limber mathematic different configuration four small ell one}: When all four external legs are short, the exact numerical integration yields a higher amplitude than the Limber approximation.
    \item \Cref{fig: appendix limber mathematic different configuration four small ell two}: For configurations with two short external legs and two long legs, the Limber approximation overestimates the exact numerical integration.
    \item \Cref{fig: appendix limber mathematic different configuration squeezed limit}: When three external legs are long and one is short, the Limber approximation still overestimates the result.
    \item \Cref{fig: appendix limber mathematic different configuration four large ell one}: Finally, for configurations where all four external legs are relatively long ($\ell \ge 10$), the Limber approximation shows excellent agreement with the numerical integration. In this regime, the constituent $\mathcal{S}_{\ell, n}$ kernels begin to reside safely within the high-multipole limit, confirming that the approximation yields sufficient accuracy for configurations dominated by higher multipoles. 
\end{itemize}

To quantify how these integration errors propagate into the final physical observables, \Cref{fig: appendix limber FFTLog different configuration} demonstrates the ratio of the single-configuration trispectrum signal evaluated via the Limber approximation to the exact FFTLog result. This signal is constructed from the summation of all $24$ signed permutations of the geometric and projection terms. \Cref{fig: appendix limber FFTLog different configuration two small ell} isolates configurations that fix two short side lengths while allowing the other two to vary. When all four side lengths are small, the Limber approximation under-predicts the FFTLog integration. However, as the other two side lengths increase to large values, the ratio rapidly drops below $1$, indicating a severe over-prediction by the Limber approximation. 

This conclusion remains robust for configurations containing only a single short side length, as shown in \Cref{fig: appendix limber FFTLog different configuration one small ell}. Finally, \Cref{fig: appendix limber FFTLog different configuration two large ell} confirms that when all four side lengths become very large, the Limber approximation perfectly agrees with the exact numerical integration, entirely fitting our physical intuition that the approximation remains valid for higher-order statistics when all constituent multipoles are in the high-$\ell$ limit. 

When the maximum multipole cutoff $\ell_{\max}$ is very small, the total signal is dominated by configurations where all four side lengths are short. In this regime, the Limber approximation underestimates the integration, yielding a lower SNR. However, as $\ell_{\max}$ increases, the summation incorporates a vast, growing number of mixed configurations involving both short and long side lengths. The accumulated systematic overestimation of these mixed configurations by the Limber approximation causes its predicted SNR to overtake the exact FFTLog result, explaining why the exact numerical integration ultimately yields a lower total SNR at large cutoffs. This behavior is consistent with \Cref{fig: coulton template SNR}.

The comparison above shows that the Limber approximation is already accurate for sufficiently large multipoles, while noticeable deviations remains only at low multipoles. This therefore motivates a hybrid implementation, in which the low-$\ell$ kernels are evaluated exactly using FFTLog and the remaining high-$\ell$ kernels are computed with the Limber approximation. This significantly reduces the computational cost while preserving the accuracy of the projection integrals.

\subsection{Limber Approximation on Collapsed-Type Template}
\label[appendix]{app:limber approximation exchange diagram}

For the collapsed-type template, the application of the FFTLog method is slightly more complicated, since there is one more layer of integration with spherical Bessel functions. Eq.~\eqref{eq: shiraishi full trispectrum before Limber} shows the full integration, where we have the projection integration of each individual point together with the diagonal integration. But we could still follow the factorizable method to do the integration. Following the same procedure used for the angular power spectrum and squeezed-type trispectrum cases, we write down the whole integration as factorizable form:
\begin{align}
    \begin{split}
        {\mathcal{I} _{\ell_3 \ell_4}^{\ell_1 \ell_2}}^{(c)}_{\pi} (L) \propto &\times \int{}\mathrm{d}x\,x^2\mathcal{S} _{\ell _1,n_1}\left( x \right) \mathcal{S} _{\ell _2,n_2}\left( x \right) j_{L}(Kx)
\\
&\times \int{}\mathrm{d}y\,y^2\mathcal{S} _{\ell _3,n_3}\left( y \right) \mathcal{S} _{\ell _4,n_4}\left( y \right) j_{L}(Ky) 
\\
& \times\int{}\mathrm{d}K\,K^{-1}
,
    \end{split}
\end{align}
where $\{n_1, n_2\}$ and $\{n_3, n_4\}$,  are either $\{0, -3\}$. For the integration over the auxiliary variable, $x$ and $y$, this is just another layer of integration, and we write down it as:
\begin{align}\label{eq: exchange diagram second layer integrand}
    \begin{split}
        \mathcal{B} _{\ell \ell^{\prime} L,n n^{\prime}}\left( K \right) =\int{}\mathrm{d}r\,r^2\mathcal{S} _{\ell ,n}\left( r \right) \mathcal{S} _{\ell ^{\prime},n^{\prime}}\left( r \right) j_L(Kr).
    \end{split}
\end{align}
Thus, the final integration could be written as:
\begin{align}\label{eq: exchange final integrand}
    \begin{split}
        {\mathcal{I} _{\ell_3 \ell_4}^{\ell_1 \ell_2}}^{(c)}_{\pi}(L) \propto \int{}\mathrm{d}K\,K^{-1}\mathcal{B} _{\ell _1\ell _2L,n_1n_2}\left( K \right) \mathcal{B} _{\ell _3\ell _4L,n_3n_4}\left( K \right) .
    \end{split}
\end{align}

\begin{figure}[htbp]
    \centering
    \begin{subfigure}[b]{0.495\textwidth}
        \centering
        \includegraphics[width=\textwidth]{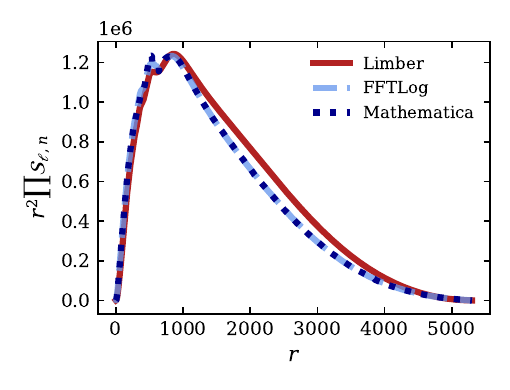}
        \caption{One short leg $\ell = [2, 47]$}
        \label{fig: exchange one small one large}
    \end{subfigure}
    \hfill
    \begin{subfigure}[b]{0.495\textwidth}
        \centering
        \includegraphics[width=\textwidth]{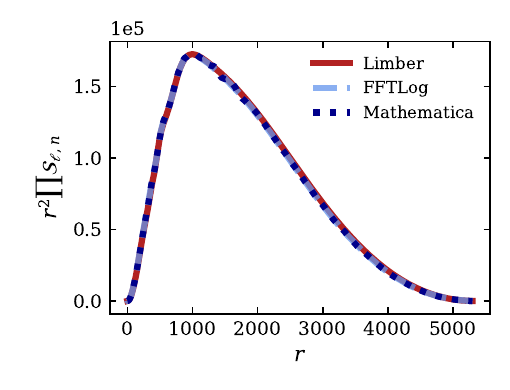}
        \caption{Two long legs $\ell = [10, 47]$}
        \label{fig: exchange two large}
    \end{subfigure}
    \caption{
    Comparison of the radial kernel $r^2S_{\ell, n}(r)S_{\ell^\prime, n^\prime}(r)$ entering Eq.~\eqref{eq: exchange diagram second layer integrand} for the collapsed-type template, evaluated using the Limber approximation and exact FFTLog. Both source galaxies are placed at $z=2.0$, with power indices $n=[-3, 0]$. Panels (\subref{fig: exchange one small one large}) and (\subref{fig: exchange two large}) show representative mixed-scale and high-$\ell$ configurations, respectively. Significant discrepancies appear when a low-$\ell$ multipole is present, while excellent agreement is recovered in the high-$\ell$ limit. 
    }
    \label{fig: exchange second layer configuration}
\end{figure}

As we can see, similar to the case of the power spectrum, but these are only two external lengths of a triangle, and the two $\ell$'s do not have to be equal. \Cref{fig: exchange second layer configuration} shows the integrand of Eq.~\eqref{eq: exchange diagram second layer integrand}. As we can see, when one $\ell$ is small while another is very large, Limber approximation overestimates the result, shown in \Cref{fig: exchange one small one large}, while \Cref{fig: exchange two large} demonstrates the integrand with two large multipoles, which indicates that Limber approximation agrees very well with the exact numerical integrations. 

\begin{figure}[htbp]
    \centering
    \begin{subfigure}[b]{0.495\textwidth}
        \centering
        \includegraphics[width=\textwidth]{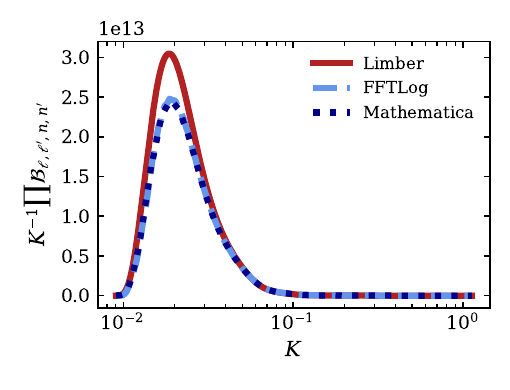}
        \caption{One short leg: $[2, 47, 48, 50, 48]$}
        \label{fig: appendix exchange limber mathematic different configuration one small side}
    \end{subfigure}
    \hfill
    \begin{subfigure}[b]{0.495\textwidth}
        \centering
        \includegraphics[width=\textwidth]{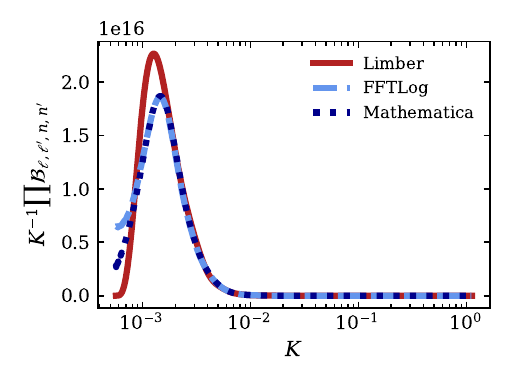}
        \caption{Three short legs: $[4, 5, 48, 50, 3]$}
        \label{fig: appendix exchange limber mathematic different configuration two small sides small diagonal}
    \end{subfigure}
    
    \vspace{0.5cm}
    
    \begin{subfigure}[b]{0.495\textwidth}
         \centering
        \includegraphics[width=\textwidth]{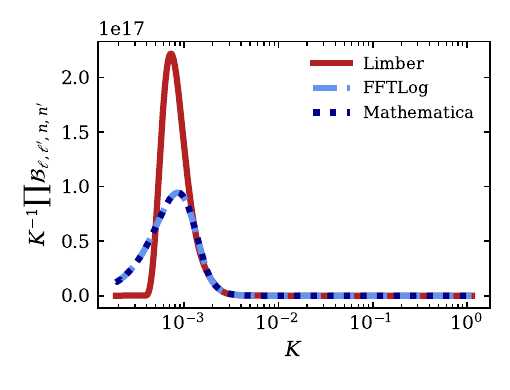}
        \caption{One short diagonal: $[46, 47, 48, 50, 2]$}
        \label{fig: appendix exchange limber mathematic different configuration small diagonal}
    \end{subfigure}
    \hfill
    \begin{subfigure}[b]{0.495\textwidth}
        \centering
        \includegraphics[width=\textwidth]{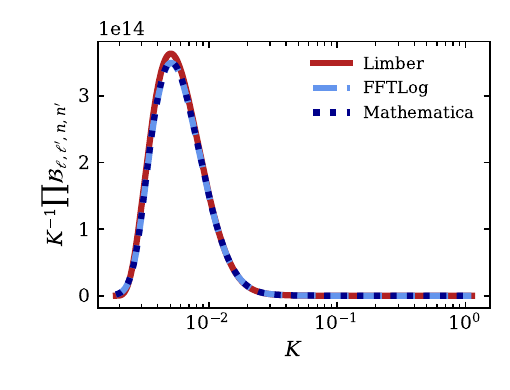}
        \caption{All long legs: $[10, 11, 12, 14, 10]$}
        \label{fig: appendix exchange limber mathematic different configuration four large sides}
    \end{subfigure}
    \caption{Comparison of the radial projection integrand $\mathcal{I}^{(c)}_{\pi}$ for the collapsed model as a function of the diagonal momentum $K$. All the four points are located in the redshift of $2.0$ and have the power indices $n=\left[-3, 0, -3, 0\right]$. The tuple shown in each subcaption follows the ordering $[\ell_1,\ell_2,\ell_3,\ell_4,L]$. The panels isolate four distinct geometric regimes: (\subref{fig: appendix exchange limber mathematic different configuration one small side}) the collapsed limit, (\subref{fig: appendix exchange limber mathematic different configuration two small sides small diagonal}) the squeezed limit, (\subref{fig: appendix exchange limber mathematic different configuration small diagonal}) a mixed-scale configuration, and (\subref{fig: appendix exchange limber mathematic different configuration four large sides}) an equilateral high-$\ell$ limit. The discrepancy between the Limber and FFTLog methods is highly sensitive to the exact momentum pairing.}
    \label{fig: appendix exchange limber mathematic different configuration redshift all 2.0}
\end{figure}

\begin{figure}[htbp]
    \centering
    \begin{subfigure}[b]{0.75\textwidth}
        \centering
        \includegraphics[width=\textwidth]{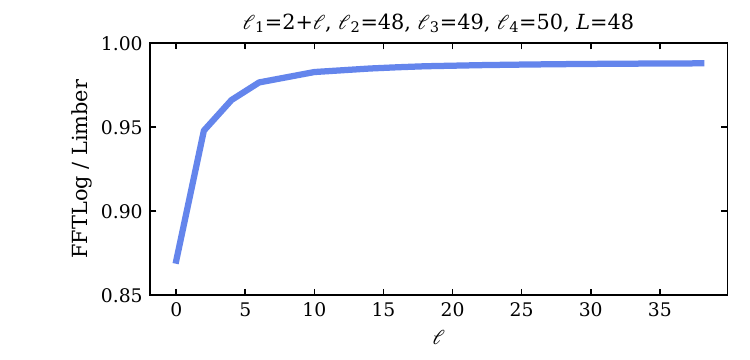}
        \caption{Vary one external leg}
        \label{fig: appendix exchange vary one short side length}
    \end{subfigure}
    \vfill
    \begin{subfigure}[b]{0.75\textwidth}
        \centering
        \includegraphics[width=\textwidth]{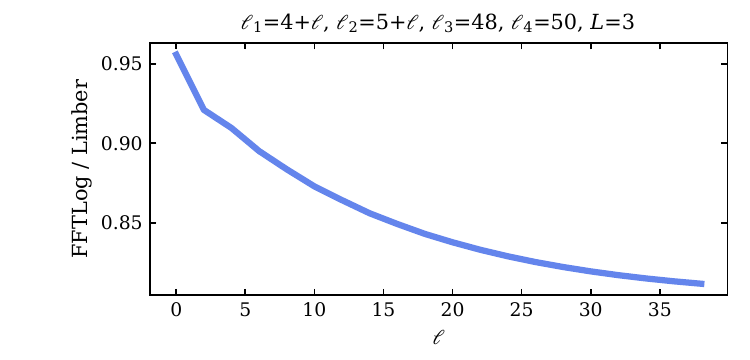}
        \caption{Vary two external legs}
        \label{fig: appendix exchange vary two side lengths}
    \end{subfigure}
    \vfill
    \begin{subfigure}[b]{0.75\textwidth}
        \centering
        \includegraphics[width=\textwidth]{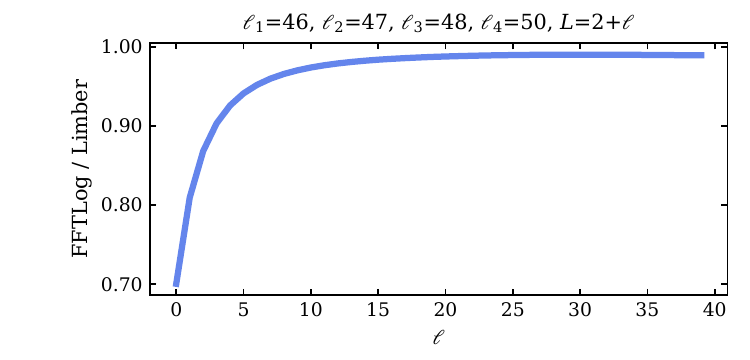}
        \caption{Vary diagonal leg}
        \label{fig: appendix exchange vary diagonal}
    \end{subfigure}
    \caption{The ratio of SNR of the reduced trispectrum signal evaluated via FFTLog over the Limber approximation for the given collapsed template. All the four points are located in the redshift of $2.0$. Similar to the squeezed-type template, mixed multipole configurations (panels \subref{fig: appendix exchange vary one short side length} and \subref{fig: appendix exchange vary two side lengths}) drive severe departures from unity, whereas configurations dominated by purely large multipoles (panel \subref{fig: appendix exchange vary diagonal}) rapidly converge to agreement.}
    \label{fig: exchange diagram Limber FFTLog ratio}
\end{figure}

\Cref{fig: appendix exchange limber mathematic different configuration redshift all 2.0} demonstrates the final integrand of Eq.~\eqref{eq: exchange final integrand} under different configurations evaluated using the Limber approximation against the exact FFTLog and Mathematica results. We isolate four distinct geometric cases:
\begin{itemize}
    \item \Cref{fig: appendix exchange limber mathematic different configuration one small side}: When one external leg is short and rest of the multipoles are very long, Limber approximation yields a higher integrand value than numerical integration. Since \Cref{fig: exchange one small one large} shows the higher value of integrand using Limber approximation, the error propagates into the final integrand even though the diagonal $L$ is already in the regime where Limber approximation works quite well.
    \item \Cref{fig: appendix exchange limber mathematic different configuration two small sides small diagonal} demonstrates the integrand which contains three short multipoles and two long external legs. Limber approximation again gives a higher amplitude of integrand than the numerical integration results.
    \item \Cref{fig: appendix exchange limber mathematic different configuration small diagonal} shows the collapsed limit where the diagonal leg is very short while the four external legs are long. 
    The integration result shows that Limber approximation is about $35\%$ larger than the numerical integration result.
    \item \Cref{fig: appendix exchange limber mathematic different configuration four large sides} shows the configuration where all the four external legs together with the diagonal leg are not that short, and then Limber approximation agrees well with the numerical integration results. 
\end{itemize}

\Cref{fig: exchange diagram Limber FFTLog ratio} shows the ratio of Limber approximation and FFTLog on different configurations. \Cref{fig: appendix exchange vary one short side length} demonstrates that when one side length is very short while other multipoles are very large, Limber overestimates the amplitude, which agrees with what \Cref{fig: appendix exchange limber mathematic different configuration one small side} shows, and the error goes down quickly when the short side length becomes a relatively large value. \Cref{fig: appendix exchange vary two side lengths} shows the ratio of the configurations which have one short diagonal length while varying the other short side lengths. When the configuration becomes more and more squeezed limit, the Limber approximation overestimates the signal more significantly. Finally, for the collapsed configuration shown in \Cref{fig: appendix exchange vary diagonal}, Limber approximation still overestimates the amplitude and when the diagonal length becomes large, the error goes down. The same conclusion still holds for the case where the four source galaxies are located in the redshift bin of $0.5$, but the value of ratio is different due to different lensing kernels. 

\begin{figure}[htbp]
    \centering
    \begin{subfigure}[b]{0.495\textwidth}
        \centering
        \includegraphics[width=\textwidth]{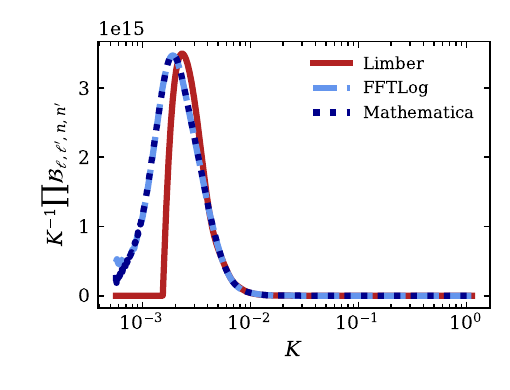}
        \caption{$\{\ell_1, \ell_2, \ell_3, \ell_4, L\} = [4, 5, 48, 50, 3]$}
        \label{fig: exchange three small diff tomo}
    \end{subfigure}
    \hfill
    \begin{subfigure}[b]{0.495\textwidth}
        \centering
        \includegraphics[width=\textwidth]{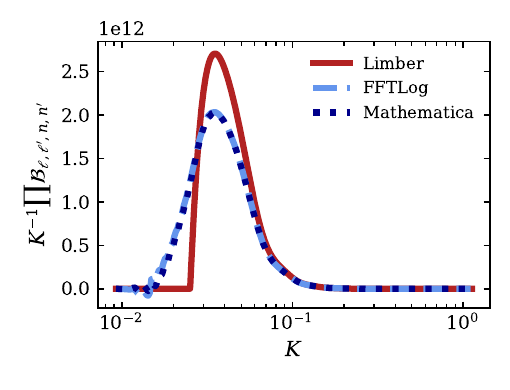}
        \caption{$\{\ell_1, \ell_2, \ell_3, \ell_4, L\} = [2, 48, 49, 50, 49]]$}
        \label{fig: exchange one small diff tomo}
    \end{subfigure}
    \caption{Comparison of the projection integrand for the collapsed template evaluated under mixed-tomography constraints (source galaxies distributed across $z \in \{0.5, 1.0, 1.5, 2.0\}$ and have the power indices $n=[-3, 0, -3, 0]$. Panel (\subref{fig: exchange three small diff tomo}) illustrates a configuration with three small multipoles causing a Limber underestimation, while Panel (\subref{fig: exchange one small diff tomo}) demonstrates a single small multipole leading to a Limber overestimation.}
    \label{fig: exchange second layer configuration diff tomo}
\end{figure}

\begin{figure}[htbp]
    \centering
    \begin{subfigure}[b]{0.8\textwidth}
        \centering
        \includegraphics[width=\textwidth]{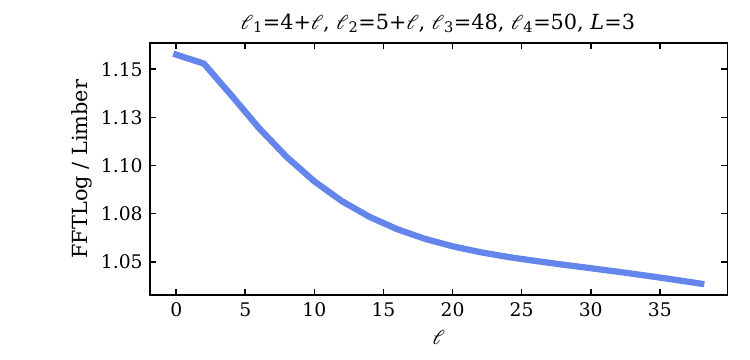}
        \caption{Fix two short length}
        \label{fig: appendix limber FFTLog different configuration two small ell diff tomo}
    \end{subfigure}
    \vfill
    \begin{subfigure}[b]{0.8\textwidth}
        \centering
        \includegraphics[width=\textwidth]{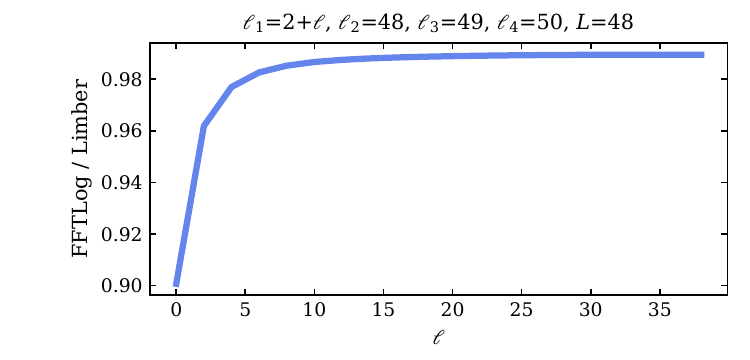}
        \caption{Fix one short length}
        \label{fig: appendix limber FFTLog different configuration one small ell diff tomo}
    \end{subfigure}
    \caption{The corresponding FFTLog-to-Limber signal ratio for the mixed-tomography configurations presented in \Cref{fig: exchange second layer configuration diff tomo}. The alternating behavior between under-prediction (ratio $> 1$, panel \subref{fig: appendix limber FFTLog different configuration two small ell diff tomo}) and 
    over-prediction (ratio $< 1$, panel \subref{fig: appendix limber FFTLog different configuration one small ell diff tomo}) ultimately drives the systematic cancellation of errors observed in the final cumulative SNR.}
    \label{fig: appendix limber FFTLog different configuration diff tomo}
\end{figure}

In \Cref{fig: shiraishi template SNR}, the SNR of the collapsed template evaluated with different tomographic source distributions reveals a complex relationship between the Limber approximation and exact FFTLog integration. This setup yields four projection kernels with distinct shapes, though the underlying analysis remains consistent. To understand this behavior, \Cref{fig: exchange second layer configuration diff tomo} illustrates the projection integrand for a single representative permutation. As shown in \Cref{fig: exchange three small diff tomo}, for configurations characterized by three small multipoles, which have the same geometric configuration as those in \Cref{fig: appendix exchange limber mathematic different configuration two small sides small diagonal}, the Limber approximation systematically underestimates the integral, consistent with the SNR ratios plotted in \Cref{fig: appendix limber FFTLog different configuration two small ell diff tomo}. Conversely, when only one side length falls into the low-$\ell$ regime where the approximation breaks down (\Cref{fig: exchange one small diff tomo}), the Limber approach overestimates the projection value (corresponding to \Cref{fig: appendix limber FFTLog different configuration one small ell diff tomo}). Ultimately, these competing effects partially cancel each other out. It is this systematic cancellation between overestimation and underestimation that explains the surprisingly small overall relative error between the Limber approximation and the exact numerical integration in \Cref{fig: shiraishi template SNR}.


These tests confirm that the exact FFTLog pipeline is necessary for accurately evaluating the parity-odd weak-lensing trispectrum in the low-$\ell$ regimes that dominate the signal.

\section{Asymptotic Scaling and Saturation Behavior}
\label[appendix]{app:saturation}

In the cumulative SNR calculation, neither the Limber approximation nor the FFTLog calculation shows a clear saturation behavior within the accessible multipole range. To gain a better physical understanding of this behavior, we perform a simple asymptotic scaling analysis based on the Limber approximation. The purpose of this analysis is not to derive an exact prediction for the cumulative SNR, but rather to identify the dominant scaling behavior of different configurations in the large-multipole limit. 

Under Limber approximation, the kernel in Eq.~\eqref{eq:app single kernel} can be written as:
\begin{equation}\label{eq: appendix saturation limber kernel}
    \mathcal{S} _{\ell ,n}\left( x \right) =\left( \frac{\ell}{x} \right) ^n\ell ^2\mathcal{T} \left( \frac{\ell}{x} \right) q\left( x \right) D\left( x \right) . 
\end{equation}

To isolate the asymptotic multipole dependence, we factorize the explicit $\ell$-dependent terms from the line-of-sight integral, leaving only the radial integration variable $x$ inside the integrand. The transfer function behaves approximately as \cite{Dodelson:2020bqr}:
\begin{equation}
\mathcal{T}(k)\sim 1,
\qquad k\ll 1,
\end{equation}
and 
\begin{equation}
\mathcal{T}(k)\sim \frac{\ln k}{k^2},
\qquad k\gg 1.
\end{equation}

For simplicity, we neglect the residual logarithmic correction in the large-$k$ regime and approximate $\mathcal{T}(k) \propto k^{-2}$ since power law dominate the changing trend. Since the integration range for $x$ is typically $x \in [1, 7000]$, the kernels associated with sufficiently large multipoles mainly probe the high $k$ regime, while kernels associated with small multipoles remain the the low $k$ regime. Under this approximation, the kernel scales asymptotically as:
\begin{equation}
S_{\ell,n}\propto \ell^{n+2},
\qquad \text{small } \ell,
\end{equation}
and 
\begin{equation}
S_{\ell,n}\propto \ell^{n},
\qquad \text{large } \ell.
\end{equation}

The angular power spectrum from Eq.~\eqref{eq: appendix factorization angular power spectrum} leads to the asymptotic scalings:
\begin{equation}
    C_{\ell} \propto \ell, \qquad \text{small } \ell,
\end{equation}
and 
\begin{equation}
    C_\ell \propto \ell^{-3}, \qquad \text{large } \ell. 
\end{equation}

Since the SNR can be schematically written as $\mathcal{I}_{\ell_1 \ell_2 \ell_3 \ell_4 }^2/C_{\ell_1} C_{\ell_2} C_{\ell_3} C_{\ell_4}$, it is natural to compare the contribution of a given multipole to the trispectrum signal against its corresponding contribution to the covariance. To build this intuition for the asymptotic hierarchy of different kernels, we introduce the simplified single-kernel diagnostic quantity
\begin{equation}
    R_{\ell, n}=\frac{A_{\ell, n}^2}{C_\ell},
\end{equation}
where 
\begin{equation}
    A_{\ell, n} = \int \mathrm{d}x \ x^2 S_{\ell, n}(x).
\end{equation}
Although this quantity does not directly correspond to the trispectrum SNR, it provides a useful asymptotic diagnostic for identifying which kernels remain least suppressed at large multipoles.

\subsection{Noise Free Asymptotic Scaling}
\label[appendix]{app:saturation noiseless}

\begin{figure}[htbp]
    \centering
    \includegraphics[width=0.9\linewidth]{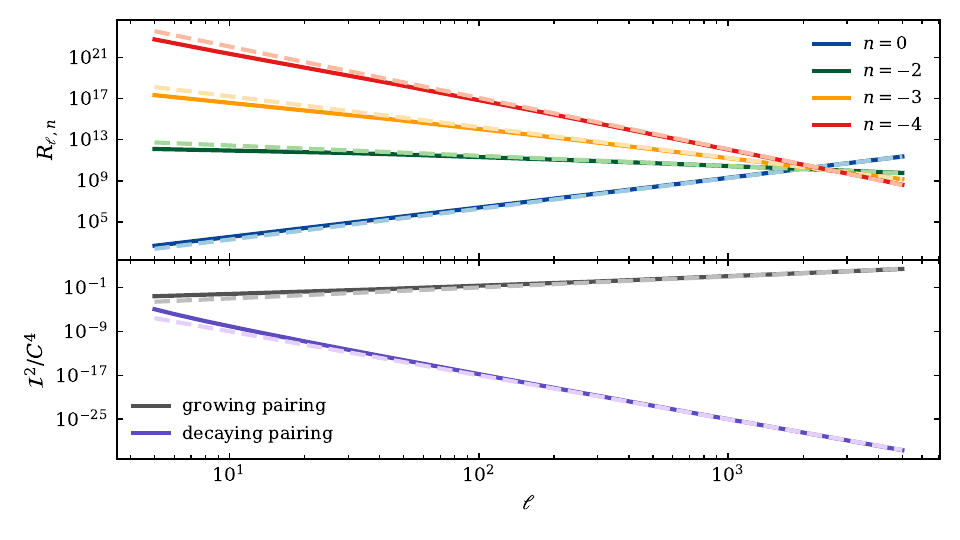}
    \caption{
    Upper panel: single-kernel diagnostic rations $R_{\ell, n}$ for different power indices $n$. Solid lines show the numerical results, while dashed lines denote the asymptotic power-law scalings derived in Eq.~\eqref{eq: appendix single ratio}. Lower panel: representative trispectrum-to-covariance scalings for the growing and decaying multipole-power-index pairings defined in Eqs.~\eqref{eq: appendix saturation growing mode noiseless} and \eqref{eq: appendix saturation decaying mode noiseless}. The growing pairing therefore gradually dominates the SNR contribution at large multipoles and does not show a clear saturation behavior within the accessible multipole range.
    }
    \label{fig: appendix saturation single kernel noiseless}
\end{figure}

In the absence of shape noise, the diagnostic ratio scales as:
\begin{equation}\label{eq: appendix single ratio}
    R_{\ell, n} \propto \ell^{2n+3}.
\end{equation}
The upper panel of \Cref{fig: appendix saturation single kernel noiseless} shows the numerical calculations together with the asymptotic power-law predictions for different power indices $n$. The hierarchy between different kernels becomes increasingly clear at large multipoles, with the $n=0$ kernel increasing with greater $\ell$ while other decreasing. 

The asymptotic behavior of the trispectrum SNR depends not only on the multipole configuration itself, but also on how the multipoles are paired with different power indices in the permutations. In the cumulative SNR calculation, configurations containing at least one small multipole often provide relatively large contributions. At the same time, there exist many configurations containing two large multipoles and two small multipoles. Motivated by this structure, we consider representative pairings between the multipole lengths and the power indices in order to understand which asymptotic configurations can continue contributing toward large multipoles. 

A simplified estimate for the trispectrum-to-covariance ratio can be written as:
\begin{equation}\label{eq: appendix saturation final ratio noiseless}
    \frac{\left[ {\mathcal{I} _{\ell _1\ell _2\ell _3\ell _4}}_{\pi}^{\left( c \right)} \right] ^2}{C_{\ell _1}C_{\ell _2}C_{\ell _3}C_{\ell _4}}\propto \frac{\int{\mathrm{d}x\,\,x^2\mathcal{S} _{\ell _1,n_1}\left( x \right) \mathcal{S} _{\ell _2,n_2}\left( x \right) \mathcal{S} _{\ell _3,n_3}\left( x \right) \mathcal{S} _{\ell _4,n_4}\left( x \right)}}{\prod_{i=1}^4{\left[ \int{\mathrm{d}x\,\,x^2\mathcal{S} _{\ell _i,0}\left( x \right) \mathcal{S} _{\ell _i,-3}\left( x \right)} \right]}} \propto \prod_{i=1}^{4} R_{\ell_i, n_i} .
\end{equation}
Among the four power indices $n=[0, -2, -3, -4]$, the $n=0$ kernel is the only contribution that increases with multipole according to Eq.~\eqref{eq: appendix single ratio}. Consequently, when the two small multipoles are held fixed and the remaining two large multipoles are taken to increasingly large values, the only pairing capable of producing a net growth is the configuration in which the large multipoles are associated with the $n=0$ and $n=-2$ kernels, while the fixed small multipoles are associated with the $n=-3$ and $n=-4$ kernels. In this case,
\begin{equation}\label{eq: appendix saturation growing mode noiseless}
    \frac{\mathcal{I}^2}{C^4} \propto \ell^2 .
\end{equation}
All other pairings are asymptotically suppressed at large multipoles. As a representative example, we additionally consider the reversed pairing, where the large multipoles are instead associated with the $n=-3$ and $n=-4$ kernels, giving
\begin{equation}\label{eq: appendix saturation decaying mode noiseless}
    \frac{\mathcal{I}^2}{C^4} \propto \ell^{-8} .
\end{equation}

The lower panel of \Cref{fig: appendix saturation single kernel noiseless} demonstrates these two representative pairing structures. Although both increasing and decreasing  contributions exist, the growing pairing gradually dominates the cumulative SNR at large multipoles. This explains why the cumulative trispectrum SNR does not exhibit a clear saturation behavior within the accessible multipole range.

\subsection{Asymptotic Scaling with Shape Noise}
\label[appendix]{app:saturation noise}

\begin{figure}
    \centering
    \includegraphics[width=0.9\linewidth]{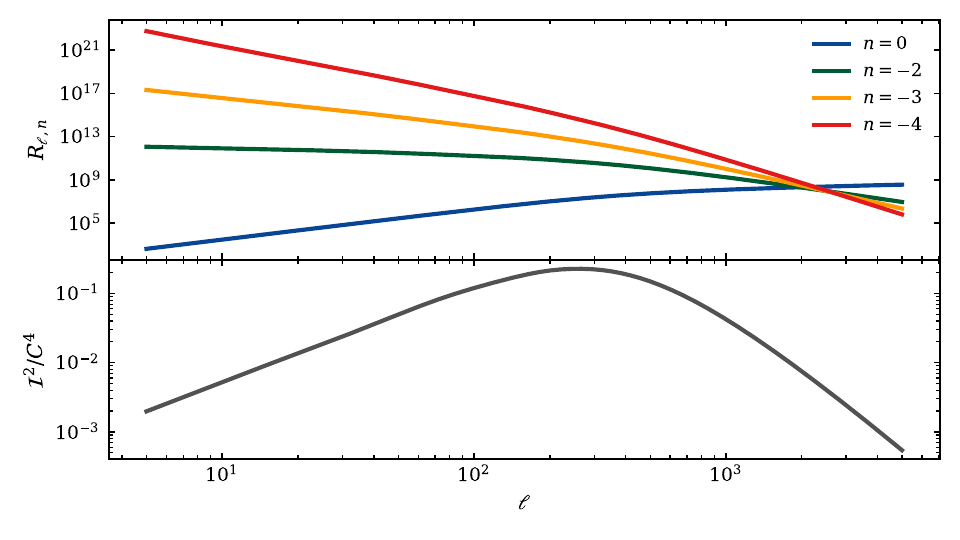}
    \caption{
    Upper panel: single kernel diagnostic ratios $R_{\ell, n}$ after including galaxy shape noise for the LSST-like high redshift lensing kernel. Lower panel: trispecrtum-to-covariance scaling for the same representative pairing shown by the gray curve in \Cref{fig: appendix saturation single kernel noiseless}, where the two varying large multipoles are associated with the $n=0$ and $n=-2$ kernels, while the $n=-3$ and $n=-4$ kernels are fixed at small $\ell=2$. The contribution initially increases at intermediate multipoles, but is eventually suppressed once the covariance becomes dominated by the approximately shape noise contribution, leading to a saturation tendency in the cumulative SNR. 
    }
    \label{fig: appendix saturation single kernel noise}
\end{figure}

When galaxy shape noise is included, the asymptotic behavior changes at sufficiently large multipoles, where the covariance becomes noise dominated. In this regime, the diagonostic ratio becomes:
\begin{equation}\label{eq: appendix saturation noise kernel ratio}
    R_{\ell, n} =\frac{A^{2}_{\ell, n}}{C_{\ell}+N_{\ell}} \simeq \frac{A^2_{\ell, n}}{N_{\ell}}
\end{equation}
As a result, the asymptotic scaling changes to 
\begin{equation}
R_{\ell,n}\propto \ell^{2n}.
\end{equation}

In the noise-free case, the asymptotic behavior is dominated by the leading power-law scaling, such that the residual logarithmic dependence of the  transfer function can be safely neglected. However, once galaxy shape noise dominates the covariance, the $n=0$ kernel scales approximately as $\ell^0$, such that the leading power-law dependence disappears. In this regime, the subleading logarithmic correction from the exact large $k$ transfer function becomes visible in the numerical calculation and produces the residual slow growth observed in the $n=0$ kernel. Nevertheless, this logarithmic enhancement remains much weaker than the power-law suppression associated with the decreasing kernels, such that the total trispectrum contribution eventually decreases at sufficiently large multipoles.

The upper panel of \Cref{fig: appendix saturation single kernel noise} shows the corresponding numerical behavior after including shape noise. In contrast to the noise free case, the large-multipole growth becomes significantly suppressed once the covariance is dominated by the approximately white noise contribution. The lower panel of \Cref{fig: appendix saturation single kernel noise} shows the corresponding pairing behavior. The growing initially increases at intermediate multipoles, but eventually decreases once the shape noise term dominates the covariance. Consequently, the cumulative trispectrum SNR develops a saturation tendency in the presence of realistic galaxy shape noise.







        



\acknowledgments

We thank Toshiki Kurita, Eiichiro Komatsu, Yong Sheng Yap, Will Coulton, Bob Cahn and Drew Jamieson for the insightful discussion. ZG acknowledges support via the funding from the U.S. Department of Energy Cosmic Frontier program, grant DE-SC002599. This project is funded by Deutsche Forschungsgemeinschaft (DFG) -- Project number 554476934. We also thank organizers and participants of \textit{Parity Violation from Home 2025} and \textit{Cambridge-LMU Cosmology Meeting}.

SC and ZG are deeply indebted to Stella Seitz for introducing and inspiring them to pursue research in the field of gravitational lensing, in which she made several foundational contributions. She recognized early on the potential of wide-area weak lensing surveys, advanced statistical tools, all of which are central to this work. Sadly, she did not live to see its final form. May she rest in peace, knowing that her legacy will live on through this and future works.



\bibliography{bibliography}

@article{Komatsu:2001rj,
    author = "Komatsu, Eiichiro and Spergel, David N.",
    title = "{Acoustic signatures in the primary microwave background bispectrum}",
    eprint = "astro-ph/0005036",
    archivePrefix = "arXiv",
    doi = "10.1103/PhysRevD.63.063002",
    journal = "Phys. Rev. D",
    volume = "63",
    pages = "063002",
    year = "2001"
}

@article{Noumi:2012vr,
    author = "Noumi, Toshifumi and Yamaguchi, Masahide and Yokoyama, Daisuke",
    title = "{Effective field theory approach to quasi-single field inflation and effects of heavy fields}",
    eprint = "1211.1624",
    archivePrefix = "arXiv",
    primaryClass = "hep-th",
    reportNumber = "UT-KOMABA-12-9, TIT-HEP-625",
    doi = "10.1007/JHEP06(2013)051",
    journal = "JHEP",
    volume = "06",
    pages = "051",
    year = "2013"
}

@article{Mandelstam:1958xc,
    author = "Mandelstam, S.",
    title = "{Determination of the pion - nucleon scattering amplitude from dispersion relations and unitarity. General theory}",
    doi = "10.1103/PhysRev.112.1344",
    journal = "Phys. Rev.",
    volume = "112",
    pages = "1344--1360",
    year = "1958"
}

@article{Hou:2025cey,
    author = "Hou, Jiamin and others",
    title = "{Parity-odd four-point correlation function from the DESI data release 1 luminous red galaxy sample}",
    eprint = "2512.20132",
    archivePrefix = "arXiv",
    primaryClass = "astro-ph.CO",
    reportNumber = "FERMILAB-PUB-25-0977-PPD",
    doi = "10.1103/2dmd-hyt1",
    journal = "Phys. Rev. D",
    volume = "113",
    number = "10",
    pages = "103502",
    year = "2026"
}

@article{DES:2020ebm,
    author = "Myles, J. and others",
    collaboration = "DES",
    title = "{Dark Energy Survey Year 3 results: redshift calibration of the weak lensing source galaxies}",
    eprint = "2012.08566",
    archivePrefix = "arXiv",
    primaryClass = "astro-ph.CO",
    reportNumber = "FERMILAB-PUB-20-659-AE",
    doi = "10.1093/mnras/stab1515",
    journal = "Mon. Not. Roy. Astron. Soc.",
    volume = "505",
    number = "3",
    pages = "4249--4277",
    year = "2021"
}

@article{Bartolo:2015dga,
    author = "Bartolo, Nicola and Matarrese, Sabino and Peloso, Marco and Shiraishi, Maresuke",
    title = "{Parity-violating CMB correlators with non-decaying statistical anisotropy}",
    eprint = "1505.02193",
    archivePrefix = "arXiv",
    primaryClass = "astro-ph.CO",
    reportNumber = "UMN-TH-3435-15, IPMU15-0067",
    doi = "10.1088/1475-7516/2015/07/039",
    journal = "JCAP",
    volume = "07",
    pages = "039",
    year = "2015"
}

@article{Jeong:2012df,
    author = "Jeong, Donghui and Kamionkowski, Marc",
    title = "{Clustering Fossils from the Early Universe}",
    eprint = "1203.0302",
    archivePrefix = "arXiv",
    primaryClass = "astro-ph.CO",
    doi = "10.1103/PhysRevLett.108.251301",
    journal = "Phys. Rev. Lett.",
    volume = "108",
    pages = "251301",
    year = "2012"
}

@article{Bao:2025onc,
    author = "Bao, Yunjia and Wang, Lian-Tao and Xianyu, Zhong-Zhi and Zhong, Yi-Ming",
    title = "{Anatomy of parity-violating trispectra in galaxy surveys}",
    eprint = "2504.02931",
    archivePrefix = "arXiv",
    primaryClass = "astro-ph.CO",
    doi = "10.1103/wzrq-8r8m",
    journal = "Phys. Rev. D",
    volume = "112",
    number = "10",
    pages = "103536",
    year = "2025"
}

@article{Cabass:2022oap,
    author = "Cabass, Giovanni and Ivanov, Mikhail M. and Philcox, Oliver H. E.",
    title = "{Colliders and ghosts: Constraining inflation with the parity-odd galaxy four-point function}",
    eprint = "2210.16320",
    archivePrefix = "arXiv",
    primaryClass = "astro-ph.CO",
    doi = "10.1103/PhysRevD.107.023523",
    journal = "Phys. Rev. D",
    volume = "107",
    number = "2",
    pages = "023523",
    year = "2023"
}

@article{Lee:2023jby,
    author = "Lee, Mang Hei Gordon and McCulloch, Ciaran and Pajer, Enrico",
    title = "{Leading loops in cosmological correlators}",
    eprint = "2305.11228",
    archivePrefix = "arXiv",
    primaryClass = "hep-th",
    doi = "10.1007/JHEP11(2023)038",
    journal = "JHEP",
    volume = "11",
    pages = "038",
    year = "2023"
}

@article{Reinhard:2024evr,
    author = "Reinhard, Matthew and Slepian, Zachary and Hou, Jiamin and Greco, Alessandro",
    title = "{Full Parity-Violating Trispectrum in Axion Inflation: Reduction to Low-D Integrals}",
    eprint = "2412.16037",
    archivePrefix = "arXiv",
    primaryClass = "astro-ph.CO",
    month = "12",
    year = "2024"
}

@article{Fujita:2023inz,
    author = "Fujita, Tomohiro and Murata, Tomoaki and Obata, Ippei and Shiraishi, Maresuke",
    title = "{Parity-violating scalar trispectrum from a rolling axion during inflation}",
    eprint = "2310.03551",
    archivePrefix = "arXiv",
    primaryClass = "astro-ph.CO",
    reportNumber = "RUP-23-20",
    doi = "10.1088/1475-7516/2024/05/127",
    journal = "JCAP",
    volume = "05",
    pages = "127",
    year = "2024"
}

@article{Niu:2022fki,
    author = "Niu, Xuce and Rahat, Moinul Hossain and Srinivasan, Karthik and Xue, Wei",
    title = "{Parity-odd and even trispectrum from axion inflation}",
    eprint = "2211.14324",
    archivePrefix = "arXiv",
    primaryClass = "hep-ph",
    doi = "10.1088/1475-7516/2023/05/018",
    journal = "JCAP",
    volume = "05",
    pages = "018",
    year = "2023"
}

@article{Stefanyszyn:2023qov,
    author = "Stefanyszyn, David and Tong, Xi and Zhu, Yuhang",
    title = "{Cosmological correlators through the looking glass: reality, parity, and factorisation}",
    eprint = "2309.07769",
    archivePrefix = "arXiv",
    primaryClass = "hep-th",
    doi = "10.1007/JHEP05(2024)196",
    journal = "JHEP",
    volume = "05",
    pages = "196",
    year = "2024"
}

@article{Jazayeri:2023kji,
    author = "Jazayeri, Sadra and Renaux-Petel, S{\'e}bastien and Tong, Xi and Werth, Denis and Zhu, Yuhang",
    title = "{Parity violation from emergent nonlocality during inflation}",
    eprint = "2308.11315",
    archivePrefix = "arXiv",
    primaryClass = "hep-th",
    doi = "10.1103/PhysRevD.108.123523",
    journal = "Phys. Rev. D",
    volume = "108",
    number = "12",
    pages = "123523",
    year = "2023"
}

@article{Thavanesan:2025kyc,
    author = "Thavanesan, Ayngaran",
    title = "{No-go Theorem for Cosmological Parity Violation}",
    eprint = "2501.06383",
    archivePrefix = "arXiv",
    primaryClass = "hep-th",
    month = "1",
    year = "2025"
}

@article{Sinde:2026adb,
    author = "Sinde, T. and Fonseca, J.",
    title = "{From photometric surveys to HI intensity mapping: Improving constraints on magnification biases while testing gravity}",
    eprint = "2603.28840",
    archivePrefix = "arXiv",
    primaryClass = "astro-ph.CO",
    month = "3",
    year = "2026"
}

@article{Ma:2005rc,
    author = "Ma, Zhao-Ming and Hu, Wayne and Huterer, Dragan",
    title = "{Effect of photometric redshift uncertainties on weak lensing tomography}",
    eprint = "astro-ph/0506614",
    archivePrefix = "arXiv",
    doi = "10.1086/497068",
    journal = "Astrophys. J.",
    volume = "636",
    pages = "21--29",
    year = "2005"
}

@article{LSSTDarkEnergyScience:2018jkl,
    author = "Mandelbaum, Rachel and others",
    collaboration = "LSST Dark Energy Science",
    title = "{The LSST Dark Energy Science Collaboration (DESC) Science Requirements Document}",
    eprint = "1809.01669",
    archivePrefix = "arXiv",
    primaryClass = "astro-ph.CO",
    reportNumber = "FERMILAB-PUB-18-465-A",
    doi = "10.2172/1471560",
    month = "9",
    year = "2018"
}

@article{Liu:2019fag,
    author = "Liu, Tao and Tong, Xi and Wang, Yi and Xianyu, Zhong-Zhi",
    title = "{Probing P and CP Violations on the Cosmological Collider}",
    eprint = "1909.01819",
    archivePrefix = "arXiv",
    primaryClass = "hep-ph",
    doi = "10.1007/JHEP04(2020)189",
    journal = "JHEP",
    volume = "04",
    pages = "189",
    year = "2020"
}

@article{Cabass:2022rhr,
    author = "Cabass, Giovanni and Jazayeri, Sadra and Pajer, Enrico and Stefanyszyn, David",
    title = "{Parity violation in the scalar trispectrum: no-go theorems and yes-go examples}",
    eprint = "2210.02907",
    archivePrefix = "arXiv",
    primaryClass = "hep-th",
    doi = "10.1007/JHEP02(2023)021",
    journal = "JHEP",
    volume = "02",
    pages = "021",
    year = "2023"
}

@misc{Mathematica,
  author = {Wolfram Research{,} Inc.},
  title = {Mathematica, {V}ersion 14.0},
  note = {Champaign, IL, 2024}
}

@article{Diemer:2017bwl,
    author = "Diemer, Benedikt",
    title = "{COLOSSUS: A python toolkit for cosmology, large-scale structure, and dark matter halos}",
    eprint = "1712.04512",
    archivePrefix = "arXiv",
    primaryClass = "astro-ph.CO",
    doi = "10.3847/1538-4365/aaee8c",
    journal = "Astrophys. J. Suppl.",
    volume = "239",
    number = "2",
    pages = "35",
    year = "2018"
}

@article{Planck:2018vyg,
    author = "Aghanim, N. and others",
    collaboration = "Planck",
    title = "{Planck 2018 results. VI. Cosmological parameters}",
    eprint = "1807.06209",
    archivePrefix = "arXiv",
    primaryClass = "astro-ph.CO",
    doi = "10.1051/0004-6361/201833910",
    journal = "Astron. Astrophys.",
    volume = "641",
    pages = "A6",
    year = "2020",
    note = "[Erratum: Astron.Astrophys. 652, C4 (2021)]"
}

@article{Kurita:2025hmp,
    author = "Kurita, Toshiki and Jamieson, Drew and Komatsu, Eiichiro and Schmidt, Fabian",
    title = "{Parity violation in galaxy shapes: Primordial non-Gaussianity}",
    eprint = "2509.08787",
    archivePrefix = "arXiv",
    primaryClass = "astro-ph.CO",
    doi = "10.1103/fxh6-hpmk",
    journal = "Phys. Rev. D",
    volume = "113",
    number = "6",
    pages = "063557",
    year = "2026"
}

@article{Okamoto:2002ik,
    author = "Okamoto, Takemi and Hu, Wayne",
    title = "{The angular trispectra of CMB temperature and polarization}",
    eprint = "astro-ph/0206155",
    archivePrefix = "arXiv",
    doi = "10.1103/PhysRevD.66.063008",
    journal = "Phys. Rev. D",
    volume = "66",
    pages = "063008",
    year = "2002"
}

@article{Wright:2024qvd,
    author = "Wright, Angus H. and others",
    title = "{The fifth data release of the Kilo Degree Survey: Multi-epoch optical/NIR imaging covering wide and legacy-calibration fields}",
    eprint = "2503.19439",
    archivePrefix = "arXiv",
    primaryClass = "astro-ph.GA",
    doi = "10.1051/0004-6361/202346730",
    journal = "Astron. Astrophys.",
    volume = "686",
    pages = "A170",
    year = "2024"
}

@article{DES:2020ekd,
    author = "Gatti, M. and others",
    collaboration = "DES",
    title = "{Dark energy survey year 3 results: weak lensing shape catalogue}",
    eprint = "2011.03408",
    archivePrefix = "arXiv",
    primaryClass = "astro-ph.CO",
    reportNumber = "FERMILAB-PUB-20-545-AE, DES-2015-0048",
    doi = "10.1093/mnras/stab918",
    journal = "Mon. Not. Roy. Astron. Soc.",
    volume = "504",
    number = "3",
    pages = "4312--4336",
    year = "2021"
}

@article{DES:2020aks,
    author = "Sevilla-Noarbe, I. and others",
    collaboration = "DES",
    title = "{Dark Energy Survey Year 3 Results: Photometric Data Set for Cosmology}",
    eprint = "2011.03407",
    archivePrefix = "arXiv",
    primaryClass = "astro-ph.CO",
    reportNumber = "FERMILAB-PUB-20-569-AE, DES-2019-0451",
    doi = "10.3847/1538-4365/abeb66",
    journal = "Astrophys. J. Suppl.",
    volume = "254",
    number = "2",
    pages = "24",
    year = "2021"
}

@article{DES:2016jjg,
    author = "Abbott, T. and others",
    collaboration = "DES",
    title = "{The Dark Energy Survey: more than dark energy {\textendash} an overview}",
    eprint = "1601.00329",
    archivePrefix = "arXiv",
    primaryClass = "astro-ph.CO",
    reportNumber = "FERMILAB-PUB-16-003-AE, DES-2015-0085",
    doi = "10.1093/mnras/stw641",
    journal = "Mon. Not. Roy. Astron. Soc.",
    volume = "460",
    number = "2",
    pages = "1270--1299",
    year = "2016"
}

@article{Kuijken:2015vca,
    author = "Kuijken, Konrad and others",
    title = "{Gravitational Lensing Analysis of the Kilo Degree Survey}",
    eprint = "1507.00738",
    archivePrefix = "arXiv",
    primaryClass = "astro-ph.CO",
    doi = "10.1093/mnras/stv2140",
    journal = "Mon. Not. Roy. Astron. Soc.",
    volume = "454",
    number = "4",
    pages = "3500--3532",
    year = "2015"
}

@article{DES:2015gax,
    author = "Abbott, T. and others",
    collaboration = "DES",
    title = "{Cosmology from cosmic shear with Dark Energy Survey Science Verification data}",
    eprint = "1507.05552",
    archivePrefix = "arXiv",
    primaryClass = "astro-ph.CO",
    reportNumber = "DES-2015-0076, FERMILAB-PUB-15-285-AE",
    doi = "10.1103/PhysRevD.94.022001",
    journal = "Phys. Rev. D",
    volume = "94",
    number = "2",
    pages = "022001",
    year = "2016"
}

@article{Aihara_2017,
   title={The Hyper Suprime-Cam SSP Survey: Overview and survey design},
   volume={70},
   ISSN={2053-051X},
   url={http://dx.doi.org/10.1093/pasj/psx066},
   DOI={10.1093/pasj/psx066},
   number={SP1},
   journal={Publications of the Astronomical Society of Japan},
   publisher={Oxford University Press (OUP)},
   author={Aihara, Hiroaki and Arimoto, Nobuo and Armstrong, Robert and Arnouts, Stéphane and Bahcall, Neta A and Bickerton, Steven and Bosch, James and Bundy, Kevin and Capak, Peter L and Chan, James H H and Chiba, Masashi and Coupon, Jean and Egami, Eiichi and Enoki, Motohiro and Finet, Francois and Fujimori, Hiroki and Fujimoto, Seiji and Furusawa, Hisanori and Furusawa, Junko and Goto, Tomotsugu and Goulding, Andy and Greco, Johnny P and Greene, Jenny E and Gunn, James E and Hamana, Takashi and Harikane, Yuichi and Hashimoto, Yasuhiro and Hattori, Takashi and Hayashi, Masao and Hayashi, Yusuke and Hełminiak, Krzysztof G and Higuchi, Ryo and Hikage, Chiaki and Ho, Paul T P and Hsieh, Bau-Ching and Huang, Kuiyun and Huang, Song and Ikeda, Hiroyuki and Imanishi, Masatoshi and Inoue, Akio K and Iwasawa, Kazushi and Iwata, Ikuru and Jaelani, Anton T and Jian, Hung-Yu and Kamata, Yukiko and Karoji, Hiroshi and Kashikawa, Nobunari and Katayama, Nobuhiko and Kawanomoto, Satoshi and Kayo, Issha and Koda, Jin and Koike, Michitaro and Kojima, Takashi and Komiyama, Yutaka and Konno, Akira and Koshida, Shintaro and Koyama, Yusei and Kusakabe, Haruka and Leauthaud, Alexie and Lee, Chien-Hsiu and Lin, Lihwai and Lin, Yen-Ting and Lupton, Robert H and Mandelbaum, Rachel and Matsuoka, Yoshiki and Medezinski, Elinor and Mineo, Sogo and Miyama, Shoken and Miyatake, Hironao and Miyazaki, Satoshi and Momose, Rieko and More, Anupreeta and More, Surhud and Moritani, Yuki and Moriya, Takashi J and Morokuma, Tomoki and Mukae, Shiro and Murata, Ryoma and Murayama, Hitoshi and Nagao, Tohru and Nakata, Fumiaki and Niida, Mana and Niikura, Hiroko and Nishizawa, Atsushi J and Obuchi, Yoshiyuki and Oguri, Masamune and Oishi, Yukie and Okabe, Nobuhiro and Okamoto, Sakurako and Okura, Yuki and Ono, Yoshiaki and Onodera, Masato and Onoue, Masafusa and Osato, Ken and Ouchi, Masami and Price, Paul A and Pyo, Tae-Soo and Sako, Masao and Sawicki, Marcin and Shibuya, Takatoshi and Shimasaku, Kazuhiro and Shimono, Atsushi and Shirasaki, Masato and Silverman, John D and Simet, Melanie and Speagle, Joshua and Spergel, David N and Strauss, Michael A and Sugahara, Yuma and Sugiyama, Naoshi and Suto, Yasushi and Suyu, Sherry H and Suzuki, Nao and Tait, Philip J and Takada, Masahiro and Takata, Tadafumi and Tamura, Naoyuki and Tanaka, Manobu M and Tanaka, Masaomi and Tanaka, Masayuki and Tanaka, Yoko and Terai, Tsuyoshi and Terashima, Yuichi and Toba, Yoshiki and Tominaga, Nozomu and Toshikawa, Jun and Turner, Edwin L and Uchida, Tomohisa and Uchiyama, Hisakazu and Umetsu, Keiichi and Uraguchi, Fumihiro and Urata, Yuji and Usuda, Tomonori and Utsumi, Yousuke and Wang, Shiang-Yu and Wang, Wei-Hao and Wong, Kenneth C and Yabe, Kiyoto and Yamada, Yoshihiko and Yamanoi, Hitomi and Yasuda, Naoki and Yeh, Sherry and Yonehara, Atsunori and Yuma, Suraphong},
   year={2017},
   month=sep }

@article{Slepian:2025kbb,
    author = "Slepian, Zachary and others",
    title = "{Measurement of Parity-Violating Modes of the Dark Energy Spectroscopic Instrument (DESI) Year 1 Luminous Red Galaxies' 4-Point Correlation Function}",
    eprint = "2508.09133",
    archivePrefix = "arXiv",
    primaryClass = "astro-ph.CO",
    reportNumber = "FERMILAB-PUB-25-0581-PPD",
    month = "8",
    year = "2025"
}

@article{Philcox:2024mmz,
    author = "Philcox, Oliver H. E. and Ereza, Julia",
    title = "{Could Sample Variance be Responsible for the Parity-Violating Signal Seen in the BOSS Galaxy Survey?}",
    eprint = "2401.09523",
    archivePrefix = "arXiv",
    primaryClass = "astro-ph.CO",
    month = "1",
    year = "2024"
}

@article{Krolewski:2024paz,
    author = "Krolewski, Alex and May, Simon and Smith, Kendrick and Hopkins, Hans",
    title = "{No evidence for parity violation in BOSS}",
    eprint = "2407.03397",
    archivePrefix = "arXiv",
    primaryClass = "astro-ph.CO",
    doi = "10.1088/1475-7516/2024/08/044",
    journal = "JCAP",
    volume = "08",
    pages = "044",
    year = "2024"
}

@article{Hou:2022wfj,
    author = "Hou, Jiamin and Slepian, Zachary and Cahn, Robert N.",
    title = "{Measurement of parity-odd modes in the large-scale 4-point correlation function of Sloan Digital Sky Survey Baryon Oscillation Spectroscopic Survey twelfth data release CMASS and LOWZ galaxies}",
    eprint = "2206.03625",
    archivePrefix = "arXiv",
    primaryClass = "astro-ph.CO",
    doi = "10.1093/mnras/stad1062",
    journal = "Mon. Not. Roy. Astron. Soc.",
    volume = "522",
    number = "4",
    pages = "5701--5739",
    year = "2023"
}

@article{Philcox:2022hkh,
    author = "Philcox, Oliver H. E.",
    title = "{Probing parity violation with the four-point correlation function of BOSS galaxies}",
    eprint = "2206.04227",
    archivePrefix = "arXiv",
    primaryClass = "astro-ph.CO",
    doi = "10.1103/PhysRevD.106.063501",
    journal = "Phys. Rev. D",
    volume = "106",
    number = "6",
    pages = "063501",
    year = "2022"
}

@article{Sorbo:2011rz,
    author = "Sorbo, Lorenzo",
    title = "{Parity violation in the Cosmic Microwave Background from a pseudoscalar inflaton}",
    eprint = "1101.1525",
    archivePrefix = "arXiv",
    primaryClass = "astro-ph.CO",
    doi = "10.1088/1475-7516/2011/06/003",
    journal = "JCAP",
    volume = "06",
    pages = "003",
    year = "2011"
}

@article{Bartolo:2004if,
    author = "Bartolo, N. and Komatsu, E. and Matarrese, Sabino and Riotto, A.",
    title = "{Non-Gaussianity from inflation: Theory and observations}",
    eprint = "astro-ph/0406398",
    archivePrefix = "arXiv",
    reportNumber = "DFPD-04-A-12",
    doi = "10.1016/j.physrep.2004.08.022",
    journal = "Phys. Rept.",
    volume = "402",
    pages = "103--266",
    year = "2004"
}

@article{Arkani-Hamed:2015bza,
    author = "Arkani-Hamed, Nima and Maldacena, Juan",
    title = "{Cosmological Collider Physics}",
    eprint = "1503.08043",
    archivePrefix = "arXiv",
    primaryClass = "hep-th",
    month = "3",
    year = "2015"
}

@article{Zhu:2013fja,
    author = "Zhu, Tao and Zhao, Wen and Huang, Yongqing and Wang, Anzhong and Wu, Qiang",
    title = "{Effects of parity violation on non-gaussianity of primordial gravitational waves in Ho{\v{r}}ava-Lifshitz gravity}",
    eprint = "1305.0600",
    archivePrefix = "arXiv",
    primaryClass = "hep-th",
    doi = "10.1103/PhysRevD.88.063508",
    journal = "Phys. Rev. D",
    volume = "88",
    pages = "063508",
    year = "2013"
}

@article{Wang:2012fi,
    author = "Wang, Anzhong and Wu, Qiang and Zhao, Wen and Zhu, Tao",
    title = "{Polarizing primordial gravitational waves by parity violation}",
    eprint = "1208.5490",
    archivePrefix = "arXiv",
    primaryClass = "astro-ph.CO",
    doi = "10.1103/PhysRevD.87.103512",
    journal = "Phys. Rev. D",
    volume = "87",
    number = "10",
    pages = "103512",
    year = "2013"
}

@article{Dyda:2012rj,
    author = "Dyda, Sergei and Flanagan, Eanna E. and Kamionkowski, Marc",
    title = "{Vacuum Instability in Chern-Simons Gravity}",
    eprint = "1208.4871",
    archivePrefix = "arXiv",
    primaryClass = "gr-qc",
    doi = "10.1103/PhysRevD.86.124031",
    journal = "Phys. Rev. D",
    volume = "86",
    pages = "124031",
    year = "2012"
}

@article{Shiraishi:2011st,
    author = "Shiraishi, Maresuke and Nitta, Daisuke and Yokoyama, Shuichiro",
    title = "{Parity Violation of Gravitons in the CMB Bispectrum}",
    eprint = "1108.0175",
    archivePrefix = "arXiv",
    primaryClass = "astro-ph.CO",
    doi = "10.1143/PTP.126.937",
    journal = "Prog. Theor. Phys.",
    volume = "126",
    pages = "937--959",
    year = "2011"
}

@article{Soda:2011am,
    author = "Soda, Jiro and Kodama, Hideo and Nozawa, Masato",
    title = "{Parity Violation in Graviton Non-gaussianity}",
    eprint = "1106.3228",
    archivePrefix = "arXiv",
    primaryClass = "hep-th",
    reportNumber = "KUNS-2346, KEK-TH-1467, KEK-COSMO-76",
    doi = "10.1007/JHEP08(2011)067",
    journal = "JHEP",
    volume = "08",
    pages = "067",
    year = "2011"
}

@article{Satoh:2010ep,
    author = "Satoh, Masaki",
    title = "{Slow-roll Inflation with the Gauss-Bonnet and Chern-Simons Corrections}",
    eprint = "1008.2724",
    archivePrefix = "arXiv",
    primaryClass = "astro-ph.CO",
    doi = "10.1088/1475-7516/2010/11/024",
    journal = "JCAP",
    volume = "11",
    pages = "024",
    year = "2010"
}

@article{Alexander:2009tp,
    author = "Alexander, Stephon and Yunes, Nicolas",
    title = "{Chern-Simons Modified General Relativity}",
    eprint = "0907.2562",
    archivePrefix = "arXiv",
    primaryClass = "hep-th",
    doi = "10.1016/j.physrep.2009.07.002",
    journal = "Phys. Rept.",
    volume = "480",
    pages = "1--55",
    year = "2009"
}

@article{Takahashi:2009wc,
    author = "Takahashi, Tomohiro and Soda, Jiro",
    title = "{Chiral Primordial Gravitational Waves from a Lifshitz Point}",
    eprint = "0904.0554",
    archivePrefix = "arXiv",
    primaryClass = "hep-th",
    doi = "10.1103/PhysRevLett.102.231301",
    journal = "Phys. Rev. Lett.",
    volume = "102",
    pages = "231301",
    year = "2009"
}

@article{Lyth:2005jf,
    author = "Lyth, David H. and Quimbay, Carlos and Rodriguez, Yeinzon",
    title = "{Leptogenesis and tensor polarisation from a gravitational Chern-Simons term}",
    eprint = "hep-th/0501153",
    archivePrefix = "arXiv",
    doi = "10.1088/1126-6708/2005/03/016",
    journal = "JHEP",
    volume = "03",
    pages = "016",
    year = "2005"
}

@article{Alexander:2004wk,
    author = "Alexander, Stephon and Martin, Jerome",
    title = "{Birefringent gravitational waves and the consistency check of inflation}",
    eprint = "hep-th/0410230",
    archivePrefix = "arXiv",
    reportNumber = "SLAC-PUB-10816",
    doi = "10.1103/PhysRevD.71.063526",
    journal = "Phys. Rev. D",
    volume = "71",
    pages = "063526",
    year = "2005"
}

@article{Carroll:1998zi,
    author = "Carroll, Sean M.",
    title = "{Quintessence and the rest of the world}",
    eprint = "astro-ph/9806099",
    archivePrefix = "arXiv",
    reportNumber = "NSF-ITP-98-063",
    doi = "10.1103/PhysRevLett.81.3067",
    journal = "Phys. Rev. Lett.",
    volume = "81",
    pages = "3067--3070",
    year = "1998"
}

@article{Harari:1992ea,
    author = "Harari, Diego and Sikivie, Pierre",
    title = "{Effects of a Nambu-Goldstone boson on the polarization of radio galaxies and the cosmic microwave background}",
    reportNumber = "UFIFT-HEP-92-9",
    doi = "10.1016/0370-2693(92)91363-E",
    journal = "Phys. Lett. B",
    volume = "289",
    pages = "67--72",
    year = "1992"
}

@article{Carroll:1991zs,
    author = "Carroll, Sean M. and Field, George B.",
    title = "{The Einstein equivalence principle and the polarization of radio galaxies}",
    reportNumber = "PRINT-95-220",
    doi = "10.1103/PhysRevD.43.3789",
    journal = "Phys. Rev. D",
    volume = "43",
    pages = "3789",
    year = "1991"
}

@article{Carroll:1989vb,
    author = "Carroll, Sean M. and Field, George B. and Jackiw, Roman",
    title = "{Limits on a Lorentz and Parity Violating Modification of Electrodynamics}",
    reportNumber = "MIT-CTP-1782",
    doi = "10.1103/PhysRevD.41.1231",
    journal = "Phys. Rev. D",
    volume = "41",
    pages = "1231",
    year = "1990"
}

@article{Kamionkowski:1996ks,
    author = "Kamionkowski, Marc and Kosowsky, Arthur and Stebbins, Albert",
    title = "{Statistics of cosmic microwave background polarization}",
    eprint = "astro-ph/9611125",
    archivePrefix = "arXiv",
    reportNumber = "FERMILAB-PUB-96-426-A, CU-TP-787, CAL-617",
    doi = "10.1103/PhysRevD.55.7368",
    journal = "Phys. Rev. D",
    volume = "55",
    pages = "7368--7388",
    year = "1997"
}

@article{Zaldarriaga:1996xe,
    author = "Zaldarriaga, Matias and Seljak, Uros",
    title = "{An all sky analysis of polarization in the microwave background}",
    eprint = "astro-ph/9609170",
    archivePrefix = "arXiv",
    doi = "10.1103/PhysRevD.55.1830",
    journal = "Phys. Rev. D",
    volume = "55",
    pages = "1830--1840",
    year = "1997"
}

@article{Diego-Palazuelos:2025dmh,
    author = "Diego-Palazuelos, Patricia and Komatsu, Eiichiro",
    title = "{Cosmic birefringence from the Atacama Cosmology Telescope Data Release 6}",
    eprint = "2509.13654",
    archivePrefix = "arXiv",
    primaryClass = "astro-ph.CO",
    doi = "10.1103/pbc3-t52s",
    journal = "Phys. Rev. D",
    volume = "113",
    number = "10",
    pages = "L101302",
    year = "2026"
}

@article{Eskilt:2022cff,
    author = "Eskilt, Johannes R. and Komatsu, Eiichiro",
    title = "{Improved constraints on cosmic birefringence from the WMAP and Planck cosmic microwave background polarization data}",
    eprint = "2205.13962",
    archivePrefix = "arXiv",
    primaryClass = "astro-ph.CO",
    doi = "10.1103/PhysRevD.106.063503",
    journal = "Phys. Rev. D",
    volume = "106",
    number = "6",
    pages = "063503",
    year = "2022"
}

@article{Diego-Palazuelos:2022dsq,
    author = "Diego-Palazuelos, P. and others",
    title = "{Cosmic Birefringence from the Planck Data Release 4}",
    eprint = "2201.07682",
    archivePrefix = "arXiv",
    primaryClass = "astro-ph.CO",
    doi = "10.1103/PhysRevLett.128.091302",
    journal = "Phys. Rev. Lett.",
    volume = "128",
    number = "9",
    pages = "091302",
    year = "2022"
}

@article{Minami:2020odp,
    author = "Minami, Yuto and Komatsu, Eiichiro",
    title = "{New Extraction of the Cosmic Birefringence from the Planck 2018 Polarization Data}",
    eprint = "2011.11254",
    archivePrefix = "arXiv",
    primaryClass = "astro-ph.CO",
    doi = "10.1103/PhysRevLett.125.221301",
    journal = "Phys. Rev. Lett.",
    volume = "125",
    number = "22",
    pages = "221301",
    year = "2020"
}

@article{Shiraishi:2010kd,
    author = "Shiraishi, Maresuke and Nitta, Daisuke and Yokoyama, Shuichiro and Ichiki, Kiyotomo and Takahashi, Keitaro",
    title = "{CMB Bispectrum from Primordial Scalar, Vector and Tensor non-Gaussianities}",
    eprint = "1012.1079",
    archivePrefix = "arXiv",
    primaryClass = "astro-ph.CO",
    doi = "10.1143/PTP.125.795",
    journal = "Prog. Theor. Phys.",
    volume = "125",
    pages = "795--813",
    year = "2011"
}

@article{Kamionkowski:2010rb,
    author = "Kamionkowski, Marc and Souradeep, Tarun",
    title = "{The Odd-Parity CMB Bispectrum}",
    eprint = "1010.4304",
    archivePrefix = "arXiv",
    primaryClass = "astro-ph.CO",
    doi = "10.1103/PhysRevD.83.027301",
    journal = "Phys. Rev. D",
    volume = "83",
    pages = "027301",
    year = "2011"
}

@article{Hou:2024udn,
    author = "Hou, Jiamin and Slepian, Zachary and Jamieson, Drew",
    title = "{Can Baryon Acoustic Oscillations Illuminate the Parity-Violating Galaxy 4PCF?}",
    eprint = "2410.05230",
    archivePrefix = "arXiv",
    primaryClass = "astro-ph.CO",
    doi = "10.1103/zpmc-pqzk",
    journal = "Phys. Rev. D",
    volume = "112",
    number = "12",
    pages = "123502",
    year = "2025"
}

@article{Zhu:2024wme,
    author = "Zhu, Hong-Ming and Pen, Ue-Li",
    title = "{Systematic Analysis of Parity-Violating Modes}",
    eprint = "2409.11400",
    archivePrefix = "arXiv",
    primaryClass = "astro-ph.CO",
    doi = "10.1103/8s52-x3r2",
    journal = "Phys. Rev. Lett.",
    volume = "135",
    number = "11",
    pages = "111003",
    year = "2025"
}

@article{Shim:2024tue,
    author = "Shim, Junsup and Pen, Ue-Li and Yu, Hao-Ran and Okumura, Teppei",
    title = "{Probing Vector Chirality in the Early Universe}",
    eprint = "2406.06080",
    archivePrefix = "arXiv",
    primaryClass = "astro-ph.CO",
    doi = "10.1103/ym2n-lzts",
    journal = "Phys. Rev. Lett.",
    volume = "135",
    number = "14",
    pages = "141002",
    year = "2025"
}

@article{Philcox:2023uor,
    author = {Philcox, Oliver H. E. and K{\"o}nig, Morgane J. and Alexander, Stephon and Spergel, David N.},
    title = "{What can galaxy shapes tell us about physics beyond the standard model?}",
    eprint = "2309.08653",
    archivePrefix = "arXiv",
    primaryClass = "astro-ph.CO",
    doi = "10.1103/PhysRevD.109.063541",
    journal = "Phys. Rev. D",
    volume = "109",
    number = "6",
    pages = "063541",
    year = "2024"
}

@article{Lue:1998mq,
    author = "Lue, Arthur and Wang, Li-Min and Kamionkowski, Marc",
    title = "{Cosmological signature of new parity violating interactions}",
    eprint = "astro-ph/9812088",
    archivePrefix = "arXiv",
    reportNumber = "CU-TP-926, CAL-675",
    doi = "10.1103/PhysRevLett.83.1506",
    journal = "Phys. Rev. Lett.",
    volume = "83",
    pages = "1506--1509",
    year = "1999"
}

@article{Cahn:2021ltp,
    author = "Cahn, Robert N. and Slepian, Zachary and Hou, Jiamin",
    title = "{Test for Cosmological Parity Violation Using the 3D Distribution of Galaxies}",
    eprint = "2110.12004",
    archivePrefix = "arXiv",
    primaryClass = "astro-ph.CO",
    doi = "10.1103/PhysRevLett.130.201002",
    journal = "Phys. Rev. Lett.",
    volume = "130",
    number = "20",
    pages = "201002",
    year = "2023"
}

@article{Komatsu:2022nvu,
    author = "Komatsu, Eiichiro",
    title = "{New physics from the polarized light of the cosmic microwave background}",
    eprint = "2202.13919",
    archivePrefix = "arXiv",
    primaryClass = "astro-ph.CO",
    doi = "10.1038/s42254-022-00452-4",
    journal = "Nature Rev. Phys.",
    volume = "4",
    number = "7",
    pages = "452--469",
    year = "2022"
}

@article{Garwin:1957hc,
    author = "Garwin, R. L. and Lederman, L. M. and Weinrich, Marcel",
    title = "{Observations of the Failure of Conservation of Parity and Charge Conjugation in Meson Decays: The Magnetic Moment of the Free Muon}",
    doi = "10.1103/PhysRev.105.1415",
    journal = "Phys. Rev.",
    volume = "105",
    pages = "1415--1417",
    year = "1957"
}

@article{Wu:1957my,
    author = "Wu, C. S. and Ambler, E. and Hayward, R. W. and Hoppes, D. D. and Hudson, R. P.",
    title = "{Experimental Test of Parity Conservation in $\beta$ Decay}",
    doi = "10.1103/PhysRev.105.1413",
    journal = "Phys. Rev.",
    volume = "105",
    pages = "1413--1414",
    year = "1957"
}

@article{Lee:1956qn,
    author = "Lee, T. D. and Yang, Chen-Ning",
    title = "{Question of Parity Conservation in Weak Interactions}",
    doi = "10.1103/PhysRev.104.254",
    journal = "Phys. Rev.",
    volume = "104",
    pages = "254--258",
    year = "1956"
}

@article{Smith:2015uia,
    author = "Smith, Kendrick M. and Senatore, Leonardo and Zaldarriaga, Matias",
    title = "{Optimal analysis of the CMB trispectrum}",
    eprint = "1502.00635",
    archivePrefix = "arXiv",
    primaryClass = "astro-ph.CO",
    month = "2",
    year = "2015"
}

@ARTICLE{2019JOSS....4.1397M,
       author = {{Murray}, Steven and {Poulin}, Francis},
        title = "{hankel: A Python library for performing simple and accurate Hankel transformations}",
      journal = {The Journal of Open Source Software},
         year = 2019,
        month = may,
       volume = {4},
       number = {37},
          eid = {1397},
        pages = {1397},
          doi = {10.21105/joss.01397},
archivePrefix = {arXiv},
       eprint = {1906.01088},
 primaryClass = {astro-ph.IM},
       adsurl = {https://ui.adsabs.harvard.edu/abs/2019JOSS....4.1397M}
}

@article{Bartelmann:1999yn,
    author = "Bartelmann, Matthias and Schneider, Peter",
    title = "{Weak gravitational lensing}",
    eprint = "astro-ph/9912508",
    archivePrefix = "arXiv",
    doi = "10.1016/S0370-1573(00)00082-X",
    journal = "Phys. Rept.",
    volume = "340",
    pages = "291--472",
    year = "2001"
}

@article{Mandelbaum:2017jpr,
    author = "Mandelbaum, Rachel",
    title = "{Weak lensing for precision cosmology}",
    eprint = "1710.03235",
    archivePrefix = "arXiv",
    primaryClass = "astro-ph.CO",
    doi = "10.1146/annurev-astro-081817-051928",
    journal = "Ann. Rev. Astron. Astrophys.",
    volume = "56",
    pages = "393--433",
    year = "2018"
}

@inproceedings{Narayan:1996ba,
    author = "Narayan, Ramesh and Bartelmann, Matthias",
    title = "{Lectures on gravitational lensing}",
    booktitle = "{13th Jerusalem Winter School in Theoretical Physics: Formation of Structure in the Universe}",
    eprint = "astro-ph/9606001",
    archivePrefix = "arXiv",
    month = "6",
    year = "1996"
}

@inproceedings{Schneider:2005ka,
    author = "Schneider, Peter",
    title = "{Weak gravitational lensing}",
    booktitle = "{33rd Advanced Saas Fee Course on Gravitational Lensing: Strong, Weak, and Micro}",
    eprint = "astro-ph/0509252",
    archivePrefix = "arXiv",
    doi = "10.1007/978-3-540-30310-7_3",
    pages = "269--451",
    year = "2006"
}

@article{Mehrem:2009ip,
    author = "Mehrem, Rami",
    title = "{The Plane Wave Expansion, Infinite Integrals and Identities involving Spherical Bessel Functions}",
    eprint = "0909.0494",
    archivePrefix = "arXiv",
    primaryClass = "math-ph",
    month = "9",
    year = "2009"
}

@book{Varshalovich:1988ifq,
    author = "Varshalovich, D. A. and Moskalev, A. N. and Khersonskii, V. K.",
    title = "{Quantum Theory of Angular Momentum}: {Irreducible Tensors, Spherical Harmonics, Vector Coupling Coefficients, 3nj Symbols}",
    doi = "10.1142/0270",
    isbn = "978-981-4415-49-1, 978-9971-5-0107-5",
    publisher = "World Scientific Publishing Company",
    year = "1988"
}

@article{Eisenstein:1997ik,
    author = "Eisenstein, Daniel J. and Hu, Wayne",
    title = "{Baryonic features in the matter transfer function}",
    eprint = "astro-ph/9709112",
    archivePrefix = "arXiv",
    reportNumber = "IASSNS-AST-97-51",
    doi = "10.1086/305424",
    journal = "Astrophys. J.",
    volume = "496",
    pages = "605",
    year = "1998"
}

@article{Eisenstein:1997jh,
    author = "Eisenstein, Daniel J. and Hu, Wayne",
    title = "{Power spectra for cold dark matter and its variants}",
    eprint = "astro-ph/9710252",
    archivePrefix = "arXiv",
    reportNumber = "IASSNS-AST-97-61",
    doi = "10.1086/306640",
    journal = "Astrophys. J.",
    volume = "511",
    pages = "5",
    year = "1997"
}

@article{Borges:2007bh,
    author = "Borges, H. A. and Carneiro, S. and Fabris, J. C. and Pigozzo, C.",
    title = "{Evolution of density perturbations in decaying vacuum cosmology}",
    eprint = "0711.2689",
    archivePrefix = "arXiv",
    primaryClass = "astro-ph",
    doi = "10.1103/PhysRevD.77.043513",
    journal = "Phys. Rev. D",
    volume = "77",
    pages = "043513",
    year = "2008"
}

@article{Slepian:2015zra,
    author = "Slepian, Zachary and Eisenstein, Daniel J.",
    title = "{A Simple Analytic Treatment of Linear Growth of Structure with Baryon Acoustic Oscillations}",
    eprint = "1509.08199",
    archivePrefix = "arXiv",
    primaryClass = "astro-ph.CO",
    doi = "10.1093/mnras/stv2889",
    journal = "Mon. Not. Roy. Astron. Soc.",
    volume = "457",
    number = "1",
    pages = "24--37",
    year = "2016"
}

@book{Dodelson:2020bqr,
    author = "Dodelson, Scott and Schmidt, Fabian",
    title = "{Modern Cosmology}",
    doi = "10.1016/C2017-0-01943-2",
    publisher = "Academic Press",
    year = "2020"
}

@article{Limber:1954zz,
    author = "Limber, D. Nelson",
    title = "{The Analysis of Counts of the Extragalactic Nebulae in Terms of a Fluctuating Density Field. II}",
    doi = "10.1086/145870",
    journal = "Astrophys. J.",
    volume = "119",
    pages = "655",
    year = "1954"
}

@article{Lemos:2017arq,
    author = "Lemos, Pablo and Challinor, Anthony and Efstathiou, George",
    title = "{The effect of Limber and flat-sky approximations on galaxy weak lensing}",
    eprint = "1704.01054",
    archivePrefix = "arXiv",
    primaryClass = "astro-ph.CO",
    doi = "10.1088/1475-7516/2017/05/014",
    journal = "JCAP",
    volume = "05",
    pages = "014",
    year = "2017"
}

@article{Hu:2001fa,
    author = "Hu, Wayne",
    title = "{Angular trispectrum of the CMB}",
    eprint = "astro-ph/0105117",
    archivePrefix = "arXiv",
    doi = "10.1103/PhysRevD.64.083005",
    journal = "Phys. Rev. D",
    volume = "64",
    pages = "083005",
    year = "2001"
}

@article{Jamieson:2024mau,
    author = "Jamieson, Drew and Caravano, Angelo and Hou, Jiamin and Slepian, Zachary and Komatsu, Eiichiro",
    title = "{Parity-odd power spectra: concise statistics for cosmological parity violation}",
    eprint = "2406.15683",
    archivePrefix = "arXiv",
    primaryClass = "astro-ph.CO",
    doi = "10.1093/mnras/stae1924",
    journal = "Mon. Not. Roy. Astron. Soc.",
    volume = "533",
    number = "3",
    pages = "2582--2598",
    year = "2024"
}

@article{Coulton:2023oug,
    author = "Coulton, William R. and Philcox, Oliver H. E. and Villaescusa-Navarro, Francisco",
    title = "{Signatures of a parity-violating universe}",
    eprint = "2306.11782",
    archivePrefix = "arXiv",
    primaryClass = "astro-ph.CO",
    doi = "10.1103/PhysRevD.109.023531",
    journal = "Phys. Rev. D",
    volume = "109",
    number = "2",
    pages = "023531",
    year = "2024"
}

@article{Greco:2025xtt,
    author = "Greco, Alessandro and Slepian, Zachary and Hou, Jiamin and Krolewski, Alex",
    title = "{CMB Lensing Trispectrum as a Probe of Parity Violation in LSS}",
    eprint = "2505.15789",
    archivePrefix = "arXiv",
    primaryClass = "astro-ph.CO",
    month = "5",
    year = "2025"
}

@article{Planck:2018jri,
    author = "Akrami, Y. and others",
    collaboration = "Planck",
    title = "{Planck 2018 results. X. Constraints on inflation}",
    eprint = "1807.06211",
    archivePrefix = "arXiv",
    primaryClass = "astro-ph.CO",
    doi = "10.1051/0004-6361/201833887",
    journal = "Astron. Astrophys.",
    volume = "641",
    pages = "A10",
    year = "2020"
}

@article{Cahn:2020axu,
    author = "Cahn, Robert N. and Slepian, Zachary",
    title = "{Isotropic N-point basis functions and their properties}",
    eprint = "2010.14418",
    archivePrefix = "arXiv",
    primaryClass = "astro-ph.CO",
    doi = "10.1088/1751-8121/acdfc4",
    journal = "J. Phys. A",
    volume = "56",
    number = "32",
    pages = "325204",
    year = "2023"
}

@article{Newman:1966ub,
    author = "Newman, E. T. and Penrose, R.",
    title = "{Note on the Bondi-Metzner-Sachs group}",
    doi = "10.1063/1.1931221",
    journal = "J. Math. Phys.",
    volume = "7",
    pages = "863--870",
    year = "1966"
}

@ARTICLE{1927ZPhy...43..624W,
       author = {{Wigner}, E.},
        title = "{Einige Folgerungen aus der Schr{\"o}dingerschen Theorie f{\"u}r die Termstrukturen}",
      journal = {Zeitschrift fur Physik},
         year = 1927,
        month = sep,
       volume = {43},
       number = {9-10},
        pages = {624-652},
          doi = {10.1007/BF01397327},
       adsurl = {https://ui.adsabs.harvard.edu/abs/1927ZPhy...43..624W}
}

@article{Eckart:1930utp,
    author = "Eckart, Carl",
    title = "{The Application of Group Theory to the Quantum Dynamics of Monatomic Systems}",
    doi = "10.1103/RevModPhys.2.305",
    journal = "Rev. Mod. Phys.",
    volume = "2",
    pages = "305--380",
    year = "1930"
}

@article{Xiang:2021mzd,
    author = "Xiang, Shaohui and Wang, Liming and Yan, Zong-Chao and Qiao, Haoxue",
    title = "{A program for simplifying summation of Wigner 3j-symbols}",
    doi = "10.1016/j.cpc.2021.107880",
    journal = "Comput. Phys. Commun.",
    volume = "264",
    pages = "107880",
    year = "2021"
}

@article{Shiraishi:2016mok,
    author = "Shiraishi, Maresuke",
    title = "{Parity violation in the CMB trispectrum from the scalar sector}",
    eprint = "1608.00368",
    archivePrefix = "arXiv",
    primaryClass = "astro-ph.CO",
    reportNumber = "IPMU16-0111",
    doi = "10.1103/PhysRevD.94.083503",
    journal = "Phys. Rev. D",
    volume = "94",
    number = "8",
    pages = "083503",
    year = "2016"
}

@article{Azyzy2025,
    author = "Azyzy, Sha and Jamieson, Drew and Komatsu, Eiichiro and Kurita, Toshiki",
    title = "{Nonlinear evolution of the matter trispectrum with primordial parity violation}",
    eprint = "2510.06164",
    archivePrefix = "arXiv",
    primaryClass = "astro-ph.CO",
    doi = "10.1088/1475-7516/2026/06/054",
    journal = "JCAP",
    volume = "06",
    pages = "054",
    year = "2026"
}

@article{Deshpande:2020jjs,
    author = "Deshpande, Anurag C. and Kitching, Thomas D.",
    title = "{Post-Limber Weak Lensing Bispectrum, Reduced Shear Correction, and Magnification Bias Correction}",
    eprint = "2004.01666",
    archivePrefix = "arXiv",
    primaryClass = "astro-ph.CO",
    doi = "10.1103/PhysRevD.101.103531",
    journal = "Phys. Rev. D",
    volume = "101",
    number = "10",
    pages = "103531",
    year = "2020"
}






\end{document}